# Assessing Technical Performance in Differential Gene Expression Experiments with External Spike-in RNA Control Ratio Mixtures

Sarah A. Munro[1*], Steve P. Lund[1], P. Scott Pine[1], Hans Binder[2], Djork-Arné Clevert[3], Ana Conesa[4], Joaquin Dopazo[4,5], Mario Fasold[2], Sepp Hochreiter[3], Huixiao Hong[6], Nederah Jafari[7], David P. Kreil[8,9], Paweł P. Łabaj[8], Sheng Li[10], Yang Liao[11,12], Simon Lin[13], Joseph Meehan[6], Christopher E. Mason[10], Javier Santoyo[4,14], Robert A. Setterquist[15], Leming Shi[16], Wei Shi[11,12], Gordon K. Smyth[11,17], Nancy Stralis-Pavese[8], Zhenqiang Su[6], Weida Tong[6], Charles Wang[18], Jian Wang[19], Joshua Xu[6], Zhan Ye[13], Yong Yang[19], Ying Yu[16], Marc Salit[1*]

[*] Corresponding authors
[1] National Institute of Standards and Technology 100 Bureau Dr Gaithersburg, Maryland 20899 USA
[2] Interdisciplinary Centre for Bioinformatics, Härtelstr. 16 - 18, 04107 Leipzig, Germany
[3] Institute of Bioinformatics, Johannes Kepler University Linz, Linz, Austria
[4] Institute of Computational Medicine, Principe Felipe Research Center, Avd. Eduardo Primo Yúfera 3, 46012 Valencia, Spain
[5] CIBER de Enfermedades Raras (CIBERER), Valencia, Spain
[6] National Center for Toxicological Research Food and Drug Administration 3900 NCTR Road Jefferson, Arkansas 72079 USA
[7] Genomics Core Facility, Feinberg School of Medicine, Northwestern University Ward building 9-150, 303 E. Chicago Ave. Chicago, Illinois, 60611 USA
[8] Chair of Bioinformatics, Boku University Vienna, Muthgasse 18, 1190 Vienna, Austria
[9] University of Warwick, U.K.
[10] Department of Physiology and Biophysics and the Institute for Computational Biomedicine, Weill Cornell Medical College, 1305 York Ave., Rm. Y13-04, Box 140, New York, NY 10021 USA
[11] Bioinformatics Division, The Walter and Eliza Hall Institute of Medical Research, 1G Royal Parade, Parkville, VIC 3052, Australia
[12] Department of Computing and Information Systems, The University of Melbourne, Parkville, VIC 3010 Australia
[13] Marshfield Clinic Research Foundation, 1000 N Oak Ave, Marshfield, Wisconsin 54449 USA
[14] Medical Genome Project, Andalusian Center for Human Genomic Sequencing, c/ Albert Einstein s/n. Plta. Baja, Sevilla, 41092, Spain
[15] Thermo Fisher Scientific, Research & Development, 2170 Woodward St, Austin, TX, 78744 USA
[16] State Key Laboratory of Genetic Engineering and MOE Key Laboratory of Contemporary Anthropology, Schools of Life Sciences and Pharmacy, Fudan University, Shanghai 201203, China
[17] Department of Mathematics and Statistics, The University of Melbourne, Parkville, VIC 3010, Australia
[18] Center for Genomics and Division of Microbiology & Molecular Genetics, School of Medicine, Loma Linda University, Loma Linda, California 92350 USA
[19] Research Informatics, Eli Lilly and Company, Lilly Corporate Center, Indianapolis, Indiana 46285, USA

*There is a critical need for standard approaches to assess, report, and compare the technical performance of genome-scale differential gene expression experiments. We assess technical performance with a proposed standard "dashboard" of metrics derived from analysis of external spike-in RNA control ratio mixtures. These control ratio mixtures with defined abundance ratios enable assessment of diagnostic performance of differentially expressed transcript lists, limit of detection of ratio (LODR) estimates, and expression ratio variability and measurement bias. The performance metrics suite is applicable to analysis of a typical experiment, and here we also apply these metrics to evaluate technical performance among laboratories. An interlaboratory study using identical samples shared amongst 12 laboratories with three different measurement processes demonstrated generally consistent diagnostic power across 11 laboratories. Ratio measurement variability and bias were also comparable amongst laboratories for the same measurement process. Different biases were observed for measurement processes using different mRNA enrichment protocols.*

Ratios of mRNA transcript abundance between sample types are measures of biological activity. These measurements of differential gene expression are important to underpin new biological hypotheses and to support critical applications such as selection of disease classifiers and regulatory oversight of drug therapies. Controls and associated ratio performance metrics are essential to understand the reproducibility and validity of differential expression experimental results.

External RNA spike-in controls developed by the External RNA Controls Consortium (ERCC)[1] can serve as technology-independent positive and negative controls for differential expression experiments. Method validation based on these ERCC controls supports comparisons between experiments, laboratories, technology platforms, and data analysis methods[2-6]. In any differential expression experiment, with any technology platform, a pair of ERCC control ratio mixtures can be added ("spiked") into total RNA samples such that for each ERCC control the relative abundance of the control between samples is either of known difference (a true positive control) or the same (a true negative control).

To enable rapid, reproducible, and automated analysis of any differential expression experiment we present a new software tool, the *erccdashboard* R package, which produces ERCC ratio performance metrics from expression values (e.g. sequence counts or microarray signal intensities). These ratio performance measures include diagnostic performance of differential expression detection with Receiver Operating Characteristic (ROC) curves and Area Under the Curve (AUC) statistics, limit of detection of ratio estimates, and expression ratio technical variability and bias.

Ratio performance measures provided by the erccdashboard package do not supersede other quality control (QC) measures, such as the QC methods recommended to evaluate sequence data both before and after alignment to a reference sequence in RNA-Seq experiments[7-11]. Sequence-level QC methods are important for evaluating the quality of data in both transcript-discovery and differential expression RNA-Seq experiments, but do not provide the additional analysis of positive and negative controls to fully evaluate differential expression experiment technical performance.

Analysis of ERCC ratio mixtures with the erccdashboard package provides technology-independent ratio performance metrics (applicable to RNA-Seq, microarrays, or any future gene expression measurement technologies). These metrics are a significant extension beyond previous work with ERCC transcripts in RNA-Seq measurements[12]. In this earlier work, a single mixture of ERCC transcripts was used to assess dynamic range and precision in individual transcript discovery RNA-Seq measurements. This earlier work did not assess differential expression experiments using ratio performance metrics from ERCC control ratio mixtures.

The source to create ERCC ratio mixtures is a plasmid DNA library of ERCC sequences that is available as a standard reference material from NIST (SRM 2374[13]). This library of 96 sequences is intended for use as controls in commercial products, such as the pair of ERCC ratio mixtures used in this analysis. In these commercially available mixtures (Mix 1 and Mix 2), 92 ERCC RNA molecule species were pooled to create mixes with true positive and true negative relative abundance differences. The two ERCC ratio mixtures are each composed of 4 subpools (23 ERCC controls per subpool) with defined abundance ratios between the mixes (Fig. 1a). Three of the subpools have different ERCC abundances in Mix 1 and Mix 2 (4:1, 1:2, and 1.5:1 ratios), and one subpool has identical ERCC abundances in the two mixes (a 1:1 ratio). Within each subpool ERCC abundances span a $2^{20}$ dynamic range. Figure 1b illustrates the ratio-abundance relationship of the 92 controls in the pair of mixtures.

Ratio mixture analysis with the erccdashboard is demonstrated for two types of differential expression studies: (1) rat toxicogenomics experiments with different treatments conducted at a single laboratory[14] and (2) interlaboratory analysis of the samples used in the MicroArray Quality Control (MAQC) study[15], Universal Human Reference RNA[16] (UHRR) and Human Brain Reference RNA (HBRR). The rat toxicogenomics study design consists of biological replicates for treatment and control conditions and illustrates a canonical RNA-Seq differential expression experiment with biological sample replication (Fig. 1c). In the interlaboratory study of the reference RNA samples, design library replicates are compared in lieu of biological replicates (Fig. 1d). The interlaboratory study design offers a valuable opportunity to evaluate performance of experiments at individual laboratories and reproducibility between laboratories, even in the absence of biological replication due to the use of reference samples.

Aliquots from a pair of spiked reference RNA samples were distributed to multiple labs for the Sequencing Quality Control (SEQC) project[17] and the Association of Biomolecular Resource Facilities (ABRF) interlaboratory study[18]. Both studies measured the same samples on multiple measurement platforms. Subsets of experiments from these studies are analyzed here with the erccdashboard package. These experiments include RNA-Seq experiments from the SEQC study using the Illumina HiSeq platform (ILM SEQC Lab 1-6) and the Life Technologies 5500 platform (LIF SEQC Lab 7-9) and ABRF study Illumina HiSeq platform (ILM ABRF Lab 10-12). Three laboratories in the SEQC project also performed microarray experiments with these same samples, discussed in supplementary material (Illumina BeadArray experiments at Lab 13 and 14 and a custom Agilent 1M array at Lab 15).

**Results**

***Reproducible Assessment of Technical Performance with Spike-In Controls***

Analysis of experiment expression values using the erccdashboard package provides four main technical performance figures. In Figures 2-5 two arbitrarily-selected example experiments from the large SEQC experiment cohort are evaluated (for all results see Fig. S1-S20). These two examples are a rat toxicogenomics methimazole-treated (MET) and control (CTL) sample RNA-Seq experiment with biological replication (panel a in Fig. 2-5 and Fig. S2) and the Lab 5 RNA-Seq reference sample experiment (panel b in Fig. 2-5 and Fig. S10).

***Dynamic Range of Control Measurements***

The $2^{20}$ range of RNA abundance in ERCC Mix 1 and Mix 2 (Fig. 1b) is used to assess the dynamic range of an experiment. The rat RNA-Seq experiment has a ~$2^{15}$ dynamic range (Fig. 2a) and the reference sample RNA-Seq experiment dynamic range spans the $2^{20}$ design dynamic range (Fig. 2b). This difference is due to increased sequencing depth in the reference sample experiment.

Particular ERCC controls consistently deviate from the expected signal-abundance relationship. These ERCC-specific effects were quantified with linear models (see Methods), and were observed at each site that used a poly-A selection protocol for mRNA enrichment, but not at sites using ribosomal RNA depletion for mRNA enrichment (see labeled ERCC controls in Fig. S21). The ERCC-specific effects were particularly strong for the Illumina labs that used poly-A selection in sample preparation (Lab 1-6, Fig. S22).

RNA-Seq of pure ERCC mixtures (Lab 2, 3, and 5) provides more evidence supporting a hypothesis that a poly-A selection bias is responsible for the observed ERCC-specific deviations. At Lab 5 library preparation of pure ERCC samples included poly-A selection and the results showed ERCC-specific effects. No such effects were seen in results from pure samples without poly-A selection at Lab 2 and 3 (Fig. S23). As noted previously[12] the short, ~20–26 nt, poly-A tails on the controls were designed for oligo-dT-primed microarray target preparation, and not intended for use in oligo-dT separation protocols. The effects appear to be correlated with ERCC transcript length (Fig. S24).

### *Diagnostic Performance of Control Ratios*

When true differences in expression exist between samples in an experiment, those differences should be detected in differential expression tests; where no differences exist no difference should be detected. The true positive and true negative ERCC control ratios can be used in a Receiver Operator Characteristic (ROC) curve analysis of rank-ordered differential expression test p-values (Fig. 3). ROC curves and the corresponding area under the curve (AUC) statistics[19] change based on the discrimination of true positive values and true negative values in this rank-ordered list. Perfect diagnostic performance is represented by AUC = 1 and a diagnostic failure is indicated by AUC = 0.5, meaning that discriminatory power of an experiment is equivalent to a random guess.

Within each experiment there is a predictable increase in diagnostic performance with increasing ERCC ratio differences (Fig. 3a,b). This relationship between design ratio and diagnostic performance relies on balanced, matched distributions of positive and negative control abundances. This design requirement is a critical consideration for preparation of any set of ERCC ratio mixtures for diagnostic performance evaluation.

In the rat experiment, all AUC statistics were > 0.9, indicating good diagnostic power (Fig. 3a). For the reference RNA experiment (Fig. 3b) diagnostic performance from ROC curves as AUC statistics is slightly lower. This is explained by the greater sequencing depth in these experiments, resulting in detection of more ERCC controls and a more stringent ROC analysis. This highlights a limitation of ROC curve analysis; it does not directly assess diagnostic performance as a function of abundance. To address this shortcoming, we introduce a new performance measure, limit of detection of ratio (LODR) estimates.

### *Limit of Detection of Ratio (LODR) Estimates*

Identifying differentially expressed transcripts is the objective of differential expression experiments, but how much information (signal) is needed to have confidence that a given fold change in expression of transcripts will be detected?

With limit of detection of ratio (LODR) estimates, empirical ERCC control ratio measurements can inform researchers of diagnostic power at a given fold change relative to transcript abundance for an experiment.

An LODR estimate for a particular fold change is the minimum signal above which differentially expressed transcripts can be detected with a specified degree of confidence. LODR offers a statistically derived, objective alternative to other methods of parsing gene lists.

An LODR estimate is obtained for a specified ratio by modeling the relationship between differential expression test p-values and signal. An acceptable false discovery rate (FDR) must be chosen to estimate an LODR. For the selected FDR (q-value) a threshold p-value can be selected from the population of p-values from the experiment. An LODR estimate for each differential ratio is found based on the intersection of the model confidence interval upper bound (90%) with the p-value threshold. A recommended default for erccdashboard analysis is FDR = 0.05, but this input parameter may be adjusted. For all rat RNA-Seq experiments (Fig. 4a and Fig. S1–S5) FDR = 0.1, because in these sequencing experiments the differential expression testing yields p-value distributions which do not contain strong evidence for differences between the samples. A smaller FDR for these experiments would decrease the threshold p-value and increase the LODR estimates. A much lower threshold, FDR = 0.01, is used in the reference sample experiments (Fig. 4b and Fig. S6-S20), because large differences in reference sample transcript abundances yield a large number of small p-values. See Methods for more guidance and detail on LODR estimation. In supplementary material we also describe a way to assess validity of the ERCC control data for LODR estimation, and an alternative model-based approach for LODR estimation (Fig. S25–S26).

Detection of differential expression improves with increasing signal for all experiments (Fig. 4ab); this cannot be discerned with ROC analysis. The AUC results for the rat experiment (Fig. 3a) had very similar diagnostic performance for all ratios (all ratios have AUC > 0.95), but the LODR estimates for each ratio are significantly different (Fig. 4a). This analysis demonstrates that although AUC statistics can be a good summary of overall diagnostic performance, LODR estimates provide valuable evidence of diagnostic performance with respect to transcript abundance.

ERCC results that are above each LODR estimate are annotated with filled points on MA plots[20] (Fig. 5a,b); such annotated MA plots can be used to design future experiments to achieve balance between cost and the desired diagnostic power (e.g. changing sequencing depth). Spike-in control LODR estimates provide an objective expectation for detection of differentially expressed endogenous transcripts, but will not substitute for careful experimental design with appropriate biological replication.

### Bias and Variability in Control Ratio Measurements

Bias and variability of control ratio measurements are evaluated graphically with MA plots. A bias is observed for the control ratio measurements in the reference RNA experiment attributable to the documented difference in mRNA fraction between the two reference samples[21]. Following mRNA enrichment, the relative amount of ERCC mix to endogenous RNA in HBRR is greater than the amount in UHRR. This contributes to a consistent bias in ERCC control ratio measurements from the nominal ratios (see Fig. 5b nominal ratio annotation).

Correcting this bias is critical for accurate differential expression testing. A model to describe this bias in control ratios, $r_m$, is:

$$R_S = r_m \left(\frac{E_1}{E_2}\right)_S$$

where $R_S$ is the nominal ratio of controls in subpool $S$ of a pure ERCC mixture and $(E_1/E_2)_S$ is the observed ratio of measured ERCC expression values in subpool $S$ in sample 1 and sample 2. In the absence of bias $r_m = 1$ ($\log(r_m) = 0$).

Given this model, $r_m$ should be a property of the samples. An empirical $r_m$ value is calculated using the previously reported mRNA fractions of these samples[21]. Deviation from this empirical $r_m$ is likely due to bias contributed during sample handling and library preparation procedures (which include mRNA enrichment procedures). Estimates of $\log(r_m)$ for these samples are consistent with this empirical $\log(r_m)$ estimate (Fig. 5b, S6-S20), but with large measurement uncertainties.

Evidence of bias from mRNA fraction differences calls for a different normalization approach to be applied to the data. Recent work has shown that simple normalization approaches can be insufficient for experiments where mRNA fractions are significantly different[22]. Systematic deviation of ERCC control ratios is indicative of a batch effect of some sort. Although normalization can address this, it may be prudent to repeat the experiment.

In both RNA-Seq experiments (Fig. 5a,b) the measured ratios showed convergence to the $r_m$ corrected ratios (dashed lines Fig. 5) with increasing signal. ERCC ratio measurements in the reference sample experiment have smaller variability compared to the rat experiment measurements. This difference in ratio variability can be attributed to lower sequencing depth in the rat experiment as well as variability in spiking the biological samples (reference samples were spiked once in bulk and then aliquoted).

### Application of the erccdashboard for Interlaboratory Analysis

Interlaboratory reproducibility of RNA-Seq experiments are evaluated by comparing erccdashboard performance measures using the spiked reference RNA samples. Three different measurement processes (sample preparation and sequencing platform) were used at different laboratories: Illumina SEQC sequencing sites (ILM SEQC Lab 1-6), Life Technologies SEQC sequencing sites (LIF SEQC Lab 7-9), and Illumina ABRF sequencing sites (ILM ABRF Lab 10-12).

At the ILM ABRF sites ribosomal RNA depletion was used for mRNA enrichment. At ILM SEQC and LIF SEQC sites reference sample total RNA went through two rounds of poly-A selection, but a different type of kit and experimental protocol was used for each platform. Poly-A selection was done independently for each library replicate at ILM SEQC sites and at LIF SEQC sites poly-A selection was done for each sample type.

Strong conclusions regarding performance of particular laboratories, measurement processes, or sequencing platforms (these factors are confounded) would require a more systematic study design repeated over time.

LODR estimates complement AUC statistics for each interlaboratory site (Fig. 6a,b), supporting the use of the more informative LODR as a new performance metric. For the ILM SEQC experiments (Lab 1-6) although the AUC statistics for all ratios at Lab 2 indicate slightly decreased diagnostic performance, the LODR estimates showed similar performance across all six sites. LODR estimates from the ILM ABRF experiments were consistent with ILM SEQC experiments despite lower AUC statistics for the ILM ABRF experiments (Lab 10 -12). For the LIF SEQC experiments (Lab 7-9) both the AUC statistics and LODR estimates indicated reduced diagnostic performance at Lab 7. For 1:1.5 ratio measurements in this experiment diagnostic performance is very poor, AUC < 0.7, and an LODR estimate could not be obtained for the specified FDR.

Weighted mean estimates of the mRNA fraction difference between the UHRR and HBRR samples for the ILM SEQC experiments generally show agreement with the previously reported $r_m$ measurement (Fig. 6c) with the exception of Lab 3 This lab also had an increased standard error for $r_m$ compared to the other ILM SEQC labs. This difference is echoed in other upstream QC analysis of the ILM SEQC data that showed decreased sequencing read quality at Lab 3[17, 23].

Large standard errors for $r_m$ were obtained for the laboratories in both the ILM ABRF and LIF SEQC experiments. This increased variability in the $r_m$ estimates is echoed in violin plots of ratio standard deviations at each site (Fig. 6d). In the ILM ABRF experiments Lab 10 had particularly high ratio measurement variability suggesting the presence of a batch effect at this site. At Lab 7 in the LIF SEQC experiments, the $r_m$ estimate standard errors and overall ratio measurement variability were very high (Fig. 6c,d), and this site also showed poor diagnostic performance (Fig. 6a,b).

***Assessment of Within-Platform Differences Using QC Metrics from Mapped Reads Is Consistent with erccdashboard Results***

Analysis of the ERCC control ratio "truth-set" provides evidence of the poor ratio measurement performance in the Lab 7 differential expression experiment, but technology-specific QC measures are needed to link observations of poor ratio measurement performance to upstream causes such as sample preparation issues. QC assessment of the mapped read data for the three Life Technologies sites was used to identify possible reasons for performance differences in these experiments.

Lab 7 performance is not an artifact of read mapping and quantification; similar results were obtained for LIF SEQC data using both the Life Technologies LifeScope analysis pipeline (Fig. S12-S14) and the Subread and featureCounts analysis pipeline (Fig. S27–S29). Mapped read QC metrics from RNASeQC for a small subset of UHRR data from Lab 7-9 mapped with LifeScope show possible reasons for performance differences. Lab 7 had an increased percentage of duplicated reads in the libraries they prepared; a fifth library prepared at an independent site (and then shared amongst the three laboratories for sequencing) showed a lower duplication rate (Fig. S30). These results suggest that libraries prepared at Lab 7 had low complexity. Evidence for 3' coverage bias at this laboratory is seen in coverage plots for the 1000 middle and the 1000 top expressed transcripts (Fig. S31-32). There were no significant differences in ribosomal RNA mapping fractions for libraries 1-4 at the three laboratories (Fig. S33).

**Discussion**

The erccdashboard R package is a method validation tool for standard analysis of differential gene expression experiments. Key technical performance parameters from ERCC control ratio mixtures are evaluated with four main analysis figures produced by the software. These technology-agnostic performance measures include dynamic range, diagnostic performance, limit of detection of ratio (LODR) estimates, and expression ratio bias and technical variability. Method validation can be accomplished with these performance measures for any gene expression measurement technology, including both RNA-Seq and microarrays, which can give comparable differential expression results with appropriate experimental design and analysis.

Individual experiments and reproducibility between experiments can be assessed with erccdashboard performance measures. Rat toxicogenomics experiments are presented as examples of canonical differential expression experiments with individually spiked biological replicates. Interlaboratory experiments used technical replicates from a single pair of spiked reference samples shared across multiple laboratories to understand reproducibility of experiments across laboratories with three different measurement processes.

LODR estimates are a new type of performance measure that summarize diagnostic performance with respect to abundance in any experiment and can be informative for experimental design. With the exception of one experiment that showed poor performance for both the LODR and AUC metrics, the interlaboratory experiments showed good diagnostic power with both performance measures.

Most reference RNA experiments had a ratio measurement bias that could be explained by the known mRNA fraction difference for the reference RNA samples, but several experiments had standard errors that did not overlap with the reported mRNA fraction difference, and some had very large standard errors. For these experiments there may be other batch effects that shifted the ratio measurement bias or contributed to the large standard errors. These differences between experiments highlights the utility of ERCC ratio measurements as a truth set to identify sources of bias, such as mRNA fraction differences or batch effects.

Reproducible research calls for standard approaches to assess, report, and compare the technical performance of genome-scale differential expression experiments. These erccdashboard performance measures are a standard method to enable the enormous scientific community conducting differential expression experiments to critically assess the performance of single experiments, performance of a given laboratory over time, or performance among laboratories. As measurement technology costs decrease, differential expression measurements are increasing in scope and complexity, including experimental designs with large sample cohorts, measured over time, at multiple laboratories. Even a single canonical differential expression experiment can involve the effort of multiple investigators, from the experimentalist generating the samples and eventually reporting the conclusions to the many scientists performing sample preparation, sequencing, bioinformatics, and statistical analysis. Consistent, standard method validation of experiments with erccdashboard analysis will provide scientists with confidence in the technical performance of their experiments at any scale.

**Methods**
***Reference RNA Sample Preparation and RNA-Seq***

The two ERCC spike-in RNA transcript mixtures (Ambion, Life Technologies) were produced from plasmid DNA templates (NIST Standard Reference Material 2374). The reference RNA samples, Universal Human Reference RNA[16] (Agilent Technologies) and Human Brain Reference RNA (Ambion, Life Technologies) were spiked with the two ERCC spike-in RNA transcript mixtures (Ambion, Life Technologies) by FDA National Center for Toxicological Research (NCTR) and distributed to SEQC site laboratories for sequencing on Illumina, Life Technologies, and Roche platforms as described in the main SEQC project manuscript[17] and these samples were also used in the ABRF interlaboratory study[18]. In brief, 50 μL of ERCC

Mix 1 was spiked into 2500 μL UHRR (Universal Human Reference RNA) total RNA and 50 μL ERCC Mix 2 was spiked into 2500 μL HBRR (Human Brain Reference RNA) total RNA. Single aliquots (10 μL each) of these two samples were sent to each participating laboratory to produce replicate library preparations of samples.

For the SEQC study there were separate library preparation protocols for the Illumina and Life Technologies platforms including different poly-A selection protocols for mRNA enrichment. Replicate library preparations (n=4) were prepared at every laboratory and then at each laboratory all library preparations were barcoded, pooled, and sequenced with 2 x 100 paired-end sequencing chemistry for Illumina and 50 x 35 paired-end sequencing chemistry for Life Technologies using the full fluidic capacity of an instrument (all lanes and flow cells). Experiments for SEQC interlaboratory analysis from six Illumina sites and three Life Technologies sites were compared in this analysis.

In addition to the SEQC data we also evaluated three Illumina sequencing experiments from the ABRF study that used ribo-depletion for mRNA enrichment instead of poly-A selection. In these experiments replicate library preparations (n = 3) were sequenced at each lab with 2 x 50 paired-end sequencing chemistry.

For the Illumina SEQC reference RNA libraries the mean number of reads per library was 260 098 869 reads, for the Life Technologies SEQC reference RNA libraries the mean number of reads per library was 109 307 746 reads, and for the ABRF Illumina reference RNA libraries the mean number of reads per library was 257 451 483 reads.

***Rat Toxicogenomics Sample Preparation and RNA-Seq***

Library preparation for rat toxicogenomics study samples was performed at a single laboratory with sequencing runs on Illumina HiScanSQ and HiSeq 2000 instruments as described in the companion rat toxicogenomics manuscript.[14] A subset of the data, measured with the HiScanSQ, was analyzed here. Rats in the MET, 3ME, and NAP sample sets were treated orally with methimazole, 3-methylcholanthrene, and betanapthoflavone, and compared to the same set of control rats. Rats in the THI and NIT sample sets were treated by injection with thioacetamide and n-nitrosodimethylamine. RNA samples from treated rat replicates were spiked with ERCC Mix 1 (per treatment type n = 3). We retained the match control (CTL) samples that were spiked with ERCC Mix 2; for the MET, 3ME, and NAP experiments there were n = 3 CTL samples and for the THI and NIT experiments the three CTL samples with the highest RIN numbers were used from a set of five CTL samples. For the five rat toxicogenomics experiments (21 samples), the mean number of total reads per library was 40 281 946 reads.

### Bioinformatic Analysis of RNA-Seq Experiments

Rat toxicogenomics sample data were mapped at NCTR against rat and ERCC reference sequences using Tophat[24]. Sequence reads from the SEQC interlaboratory study were aligned to human (hg19) and ERCC reference sequences. SEQC study Illumina platform data were mapped with BWA[25] and gene level counts corresponding to human and ERCC nucleic acid features were quantified using reference annotations for the ERCC controls and hg19 (NCBI RefSeq, Release 52). SEQC study Life Technologies platform data were mapped with LifeScope (Life Technologies, Foster City, CA) and reference annotations from UCSC and NIST. Life Technologies platform data were also mapped with the Subread aligner[26] and summarized using the featureCounts program[27]. ABRF Illumina data used in this analysis were mapped with the STAR aligner using the hg19 genome assembly and the Gencode v12 annotation was used for read counting with the Rmake pipeline (http://physiology.med.cornell.edu/faculty/mason/lab/r-make/). Count data from these experiments were used in the erccdashboard analysis. The default normalization for all RNA-Seq experiment sample replicates was 75$^{th}$ percentile normalization of count data.

### Reference RNA Microarray Analysis

In the SEQC study there were three microarray experiments. Two experiments used Illumina Bead Arrays (Lab 13 and 14). For Lab 13 and 14 triplicate arrays were prepared for each reference RNA sample. Microarray signal intensity data were not background-corrected or normalized using Illumina software. The unnormalized data was processed to keep only the results in all sample replicates (n = 6) that had probe detection p-vals that were ≤ 0.05. In erccdashboard analysis the replicates in these array experiments were normalized using the 75$^{th}$ percentile intensity for each replicate array. At Lab 15 custom Agilent 1M microarrays (n = 4 per sample) with a variance stabilizing normalization[28] were used in erccdashboard analysis. For the Agilent arrays probe sequence specific signals were modeled using established methods, saturation effects detrended, and outlier probes downweighted.[29-31]

### Analysis of Expression Values with erccdashboard

Expression values were analyzed using the erccdashboard R package (https://github.com/munrosa/erccdashboard) which was developed in the R language[32]. All diagnostic plots were generated based on tools available in the ggplot2[33] and gridExtra[34] R packages.

A negative binomial generalized linear model (GLM) was fit to counts for individual ERCC controls from each replicate of the treatment and control samples to estimate the bias in the empirical ERCC ratios ($r_m$). These individual ERCC $r_m$ estimates and standard errors were used to produce an overall weighted mean $r_m$ estimate with a weighted standard error estimate. The $r_m$ estimate must be applied as a correction factor to ERCC data prior to further analysis.

The relationship between normalized ERCC spike-in signal data and the design ERCC spike-in abundances can be used to evaluate dynamic range. To estimate ERCC-specific effects for RNA-Seq experiments a linear model was fit to the signal-abundance data, providing a global slope ($\beta_{global}$) and global intercept ($\alpha_{global}$). A second model was fit allowing an intercept for each ERCC, but fixing the slope as $\beta_{global}$. ERCC-specific effects were then estimated as the difference between the global intercept ($\alpha_{global}$) and the ERCC-specific intercepts. For each ERCC the resulting intercept difference from the fits was divided by the corresponding standard error to give a unitless indicator of deviance from the expected signal-abundance relationship for each control.

Differential expression testing of ERCCs and endogenous genes was performed with QuasiSeq[35], using a negative binomial dispersion trend estimated from edgeR[36, 37], to generate p-values for all endogenous and ERCC features. ROC curves and AUC statistics were produced using the ROCR package[38]. To construct the ROC curves, the 1:1 subpool p-values were the true negative group for each differential ratio ROC curve.

Estimation of LODR requires the parameters: fold change, *fold*; probability, *prob*; and p-value threshold, $p_{thresh}$. An LODR estimate is defined as the minimum count above which a transcript with an absolute log fold change, |log(fold)|, has at least a *prob*\*100% chance of obtaining a q-value of *FDR* or less. The choice of $p_{thresh}$ is based on specification of an acceptable false discovery rate (*FDR*), typically this may be *FDR* = 0.05, but for samples with higher or lower populations of differentially expressed genes one can be more or less conservative in this choice. In our analysis, *FDR* = 0.1 was used to compare all rat data sets and *FDR* = 0.01 was used for all human reference RNA data sets. For each p-value obtained from differential expression testing of the population of transcripts a q-value (estimated false discovery rate) is computed. The maximum p-value that has a corresponding q-value less than or equal to *FDR* is defined as $p_{thresh}$.

LODR estimates for each of the differential ERCC ratios were made using locfit[39] regression trends (including a pointwise 80% prediction interval) of the relationship between abundance (log10(average count)) and strength of differential expression (log10(p-value)). For a given *fold* (ratio) the LODR is the average count where the upper bound of ratio prediction interval intersects with a chosen $p_{thresh}$. This method of estimating LODR is annotated with colored arrows in Figure 4. For each LODR estimate 90% confidence intervals were obtained via bootstrapping (residuals from the corresponding locfit curve were repeatedly resampled to estimate LODR).

For evaluating ratio measurement variability for the pair of samples in an experiment, ratios of ERCC control signals for the samples were examined with respect to the average of the sample ERCC control signals. MA plots of this data were annotated to indicate ERCC ratio measurements above and below the LODR estimates for each ratio. Violin plots of the density distribution of ERCC control ratio standard deviations (with the upper 10th percentile trimmed) are used to evaluate ratio measurement variability for multiple experiments.

### *Mapped Read QC metrics analysis*

Mapped read QC metrics were produced for Life Technologies data from Lab 7-9. The percentage of rRNA mapped in all UHRR Libraries (1-5) technical replicates (all lanes and flow cells) at Lab 7-9 were extracted from LifeScope mapping filter reports that result from sample alignment to a reference file of filter reference sequences. A subset of UHRR Library 1-5 bam files that were each downsampled to ~1 million read pairs using the downSampleSam function in Picard[40] were analyzed using the RNASeQC analysis tool[8] to assess duplicate read rates and coverage bias across transcripts.

### *Data Accession Codes*

Sequence data used in this analysis are from the companion SEQC manuscripts[14,17] and the ABRF study manuscript[18]. The full SEQC project data set has been deposited in GEO and is accessible by the code GSE47792 and the full ABRF study data set is accessible by the code GSE46876. Expression measure tables derived from the RNA-Seq and microarray data are available as a supplementary data file, so that the analysis presented here may be reproduced in R with the erccdashboard package.

### *Acknowledgements*

We thank David L. Duewer for review of the manuscript, Cecelie Boysen for discussion of results, and all other members of the SEQC consortium who supported this work. PPL, NSP and DPK acknowledge support from the Vienna Scientific Cluster (VSC), the Vienna Science and Technology Fund (WWTF), Baxter AG, Austrian Research Centres (ARC) Seibersdorf, and the Austrian Centre of Biopharmaceutical Technology (ACBT).

### *Disclaimer*

This work includes contributions from, and was reviewed by, the FDA. This work has been approved for publication by this agency, but it does not necessarily reflect official agency policy. Certain commercial equipment, instruments, or materials are identified in this paper in order to specify the experimental procedure adequately. Such identification is not intended to imply recommendation or endorsement by the National Institute of Standards and Technology (NIST) or the Food and Drug Administration (FDA), nor is it intended to imply that the materials or equipment identified are necessarily the best available for the purpose.

**Figure Captions:**

**Figure 1 (a)** Two mixtures of the same 92 ERCC RNA transcripts are prepared such that 4 subpools with 23 transcripts per subpool are in 4 defined abundance ratios between the 2 mixtures. **(b)** Within each ratio subpool the 23 controls (several points overlap) span a broad dynamic range of transcript concentrations. **(c)** In a typical single laboratory RNA-Seq experiment biological replicates would be prepared for treatment and control samples. Rat toxicogenomics experimental samples represent this experimental design **(d)** In the SEQC experimental design UHRR and HBRR samples have no biological replicates, but have extensive technical replicates including multiple library preparation replicates that are analyzed for the interlaboratory assessment of reproducibility instead of biological replicates.

**Figure 2** The relationship between signal and abundance for ERCC spike-in controls is shown to assess dynamic range of three different experiments (**a**) biological replicates (n=3) of control (CTL) and methimazole treated (MET) from a rat toxicogenomics experiment **(b)** an RNA-Seq measurement of reference samples (UHRR and HBRR) with library preparation technical replicates (n = 4) from Lab 5 of an interlaboratory study. In each figure points are colored by ratio subpool, errorbars represent the standard deviations of replicates, and shape represents sample type. In the RNA-Seq results ERCC controls that did not have at least 1 count in three libraries for either sample were not included in the signal-abundance plot.

**Figure 3** ROC curves and AUC statistics for the three differential ratio subpools are shown for (**a**) biological replicates (n=3) of control (CTL) and methimazole treated (MET) from a rat toxicogenomics experiment **(b)** an RNA-Seq measurement of reference samples (UHRR and HBRR) with library preparation technical replicates (n = 4) from Lab 5 of an interlaboratory study. Annotation tables include AUC statistics for each group of true positive ERCC controls along with the number of controls that were used in this analysis ("detected") and the total number that were included in the ERCC control mixtures ("spiked").

**Figure 4** P-values and modeled LODR as a function of average counts for the 4 different ERCC ratios. The black dashed line annotates the p-value threshold derived

form the FDR chosen for each experiment **(a)** FDR = 0.1 for biological replicates (n=3) of control (CTL) and methimazole treated (MET) rats **(b)** FDR = 0.01 for the reference sample RNA-seq experiment technical replicates (n = 4) from Lab 5 of an interlaboratory study. Colored arrows indicate the LODR estimate (average counts) for each fold change estimate that crosses the line indicating $p_{thresh}$ and the upper boundary of the model confidence interval. LODR results and bootstrap confidence intervals are in the annotation table below the plot.

**Figure 5** MA plots of ratio measurements of ERCC controls as a function of abundance (points colored based on ratio) and endogenous transcript measurements (grey points) for **(a)** biological replicates (n=3) of control (CTL) and methimazole treated (MET) from a rat toxicogenomics experiment **(b)** an RNA-Seq measurement of reference samples (UHRR and HBRR) with library preparation technical replicates (n = 4) from Lab 5 of an interlaboratory study.
ERCC data points represent mean ratio measurements per ERCC and error bars represent standard deviation of replicates. Filled circles indicate ERCC ratios above the LODR estimate for 4:1, 1:1.5, and 1:2 ratios. The estimate of mRNA fraction differences between the samples, $r_m$, is provided in an inset table and used to adjust the nominal ERCC ratios. The nominal ratios are annotated with solid colored lines for each ratio subpool and the adjusted ratios are annotated with dashed colored lines.

**Figure 6** Interlaboratory comparison of ERCC dashboard performance measures for reference samples with two different platforms at nine laboratories. The legend is common to all figures, color indicates measurement process and transparency of each color is used to indicate results for different ratio **(a)** Area Under the Curve (AUC) statistics for the ERCC controls at the three differential ratios. **(b)** Limit of Detection of Ratio (LODR) count estimates (on a log scale) for the ERCC controls at the three differential ratios. **(c)** Weighted mean estimates of mRNA fraction differences for the sample set with error bars representing weighted standard errors. The solid black line represents the measurement of $r_m$ from previous work[21] and dashed black lines show the confidence interval from the standard deviation of this estimate **(d)** Violin plots of showing distributions of ERCC control ratio standard deviations at each laboratory.

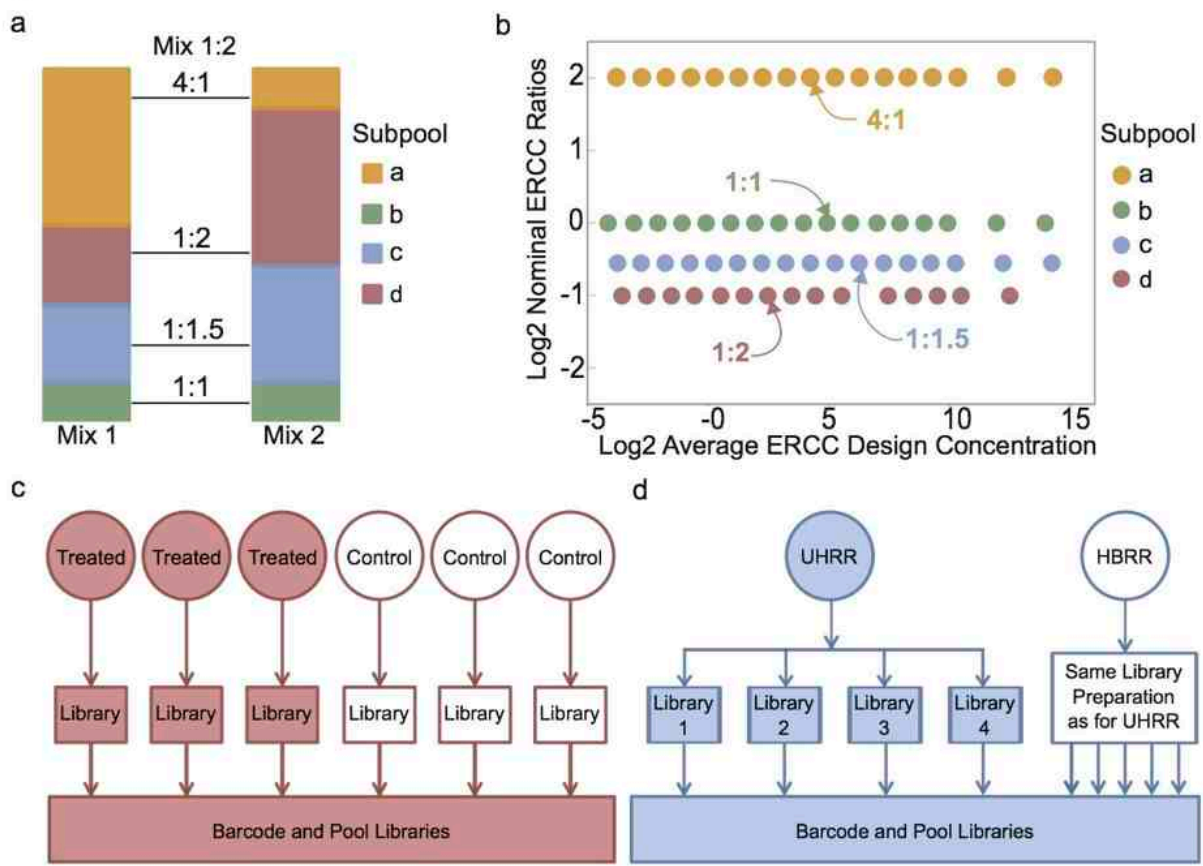

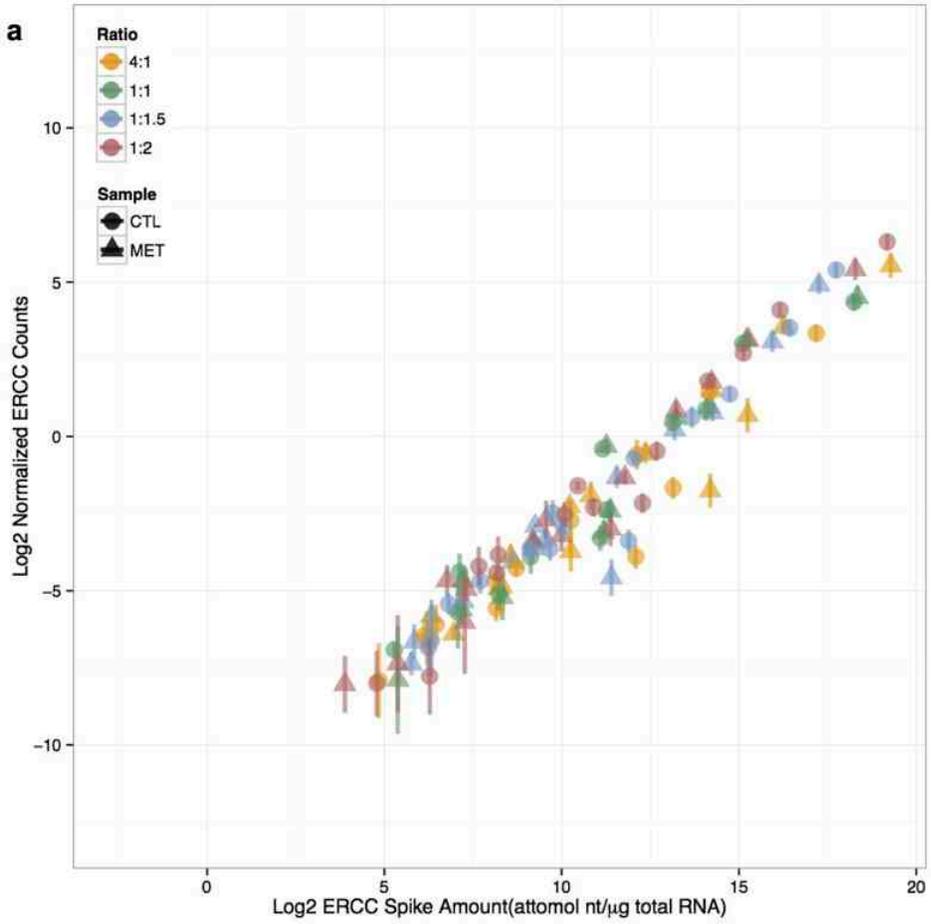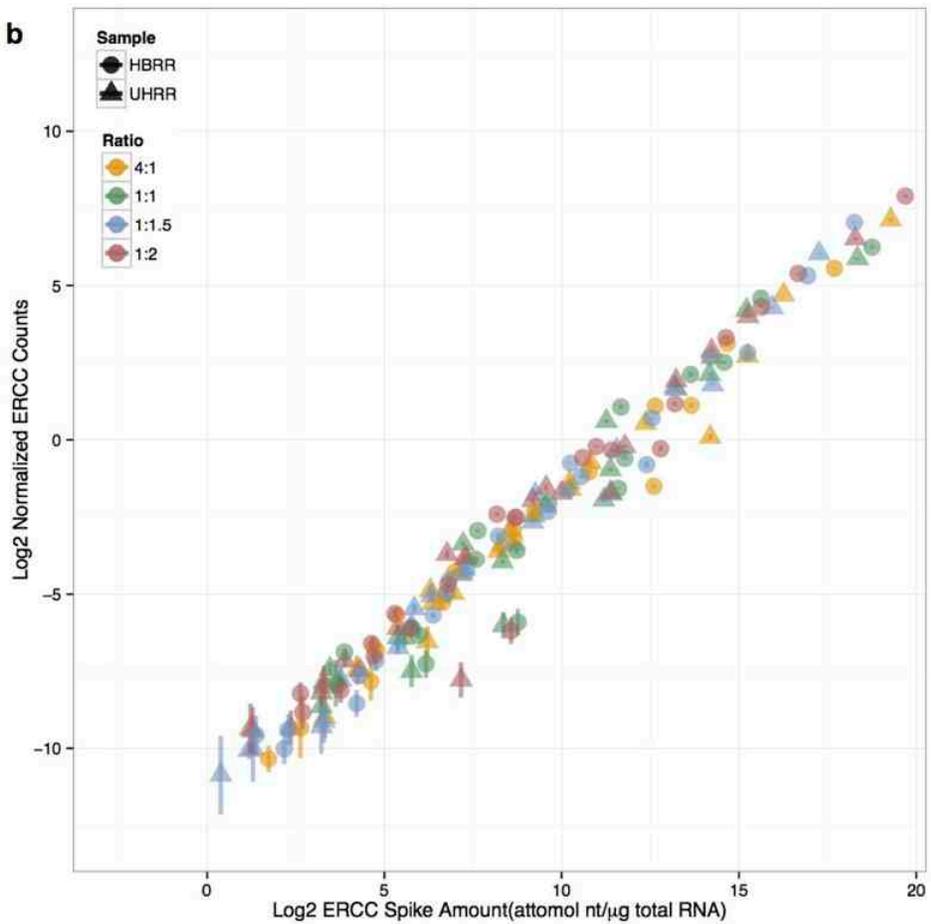

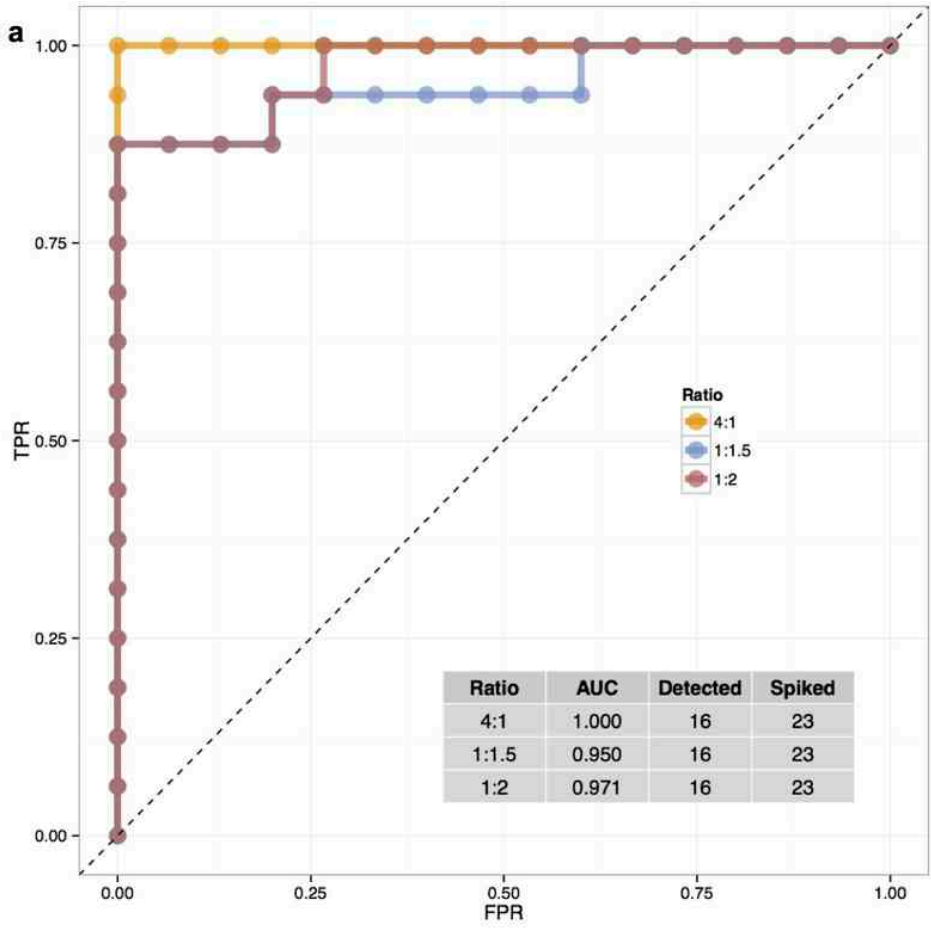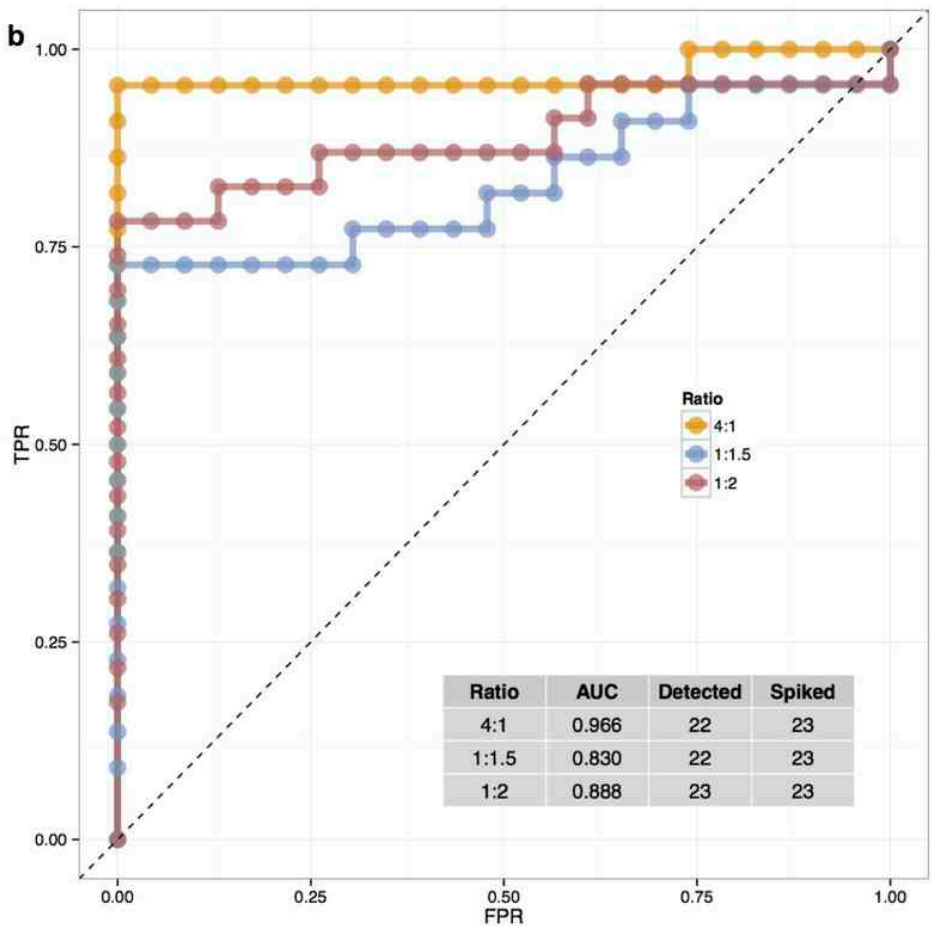

## Figure 4 a

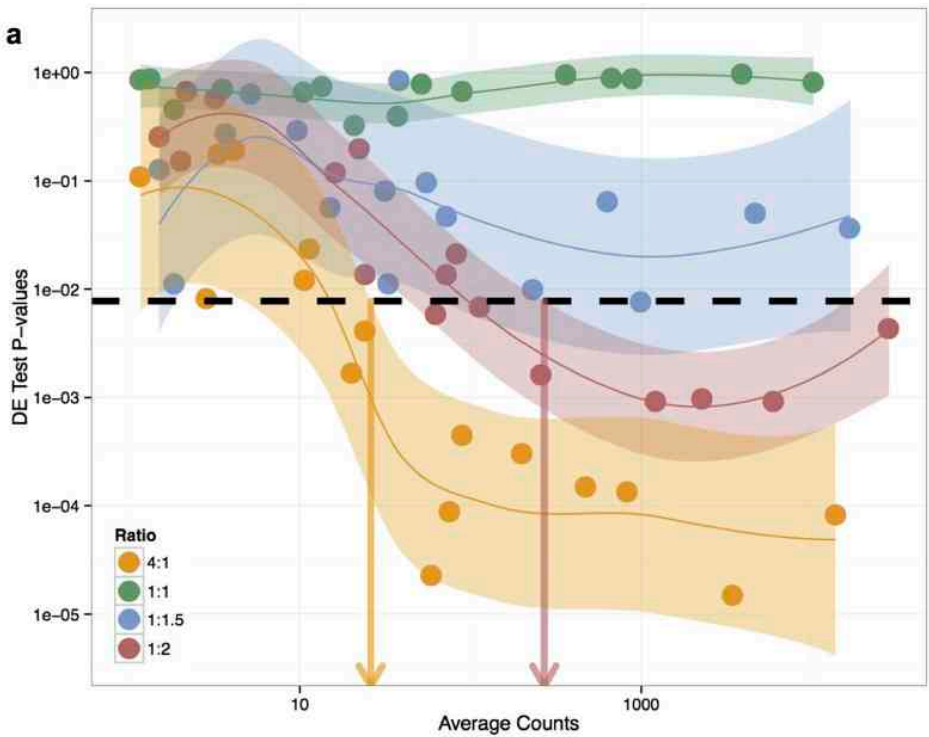

| Ratio | LODR Estimate | 90% CI Lower Bound | 90% CI Upper Bound |
|---|---|---|---|
| 4:1 | 26 | 19 | 32 |
| 1:1.5 | Inf | NA | NA |
| 1:2 | 270 | 140 | 390 |

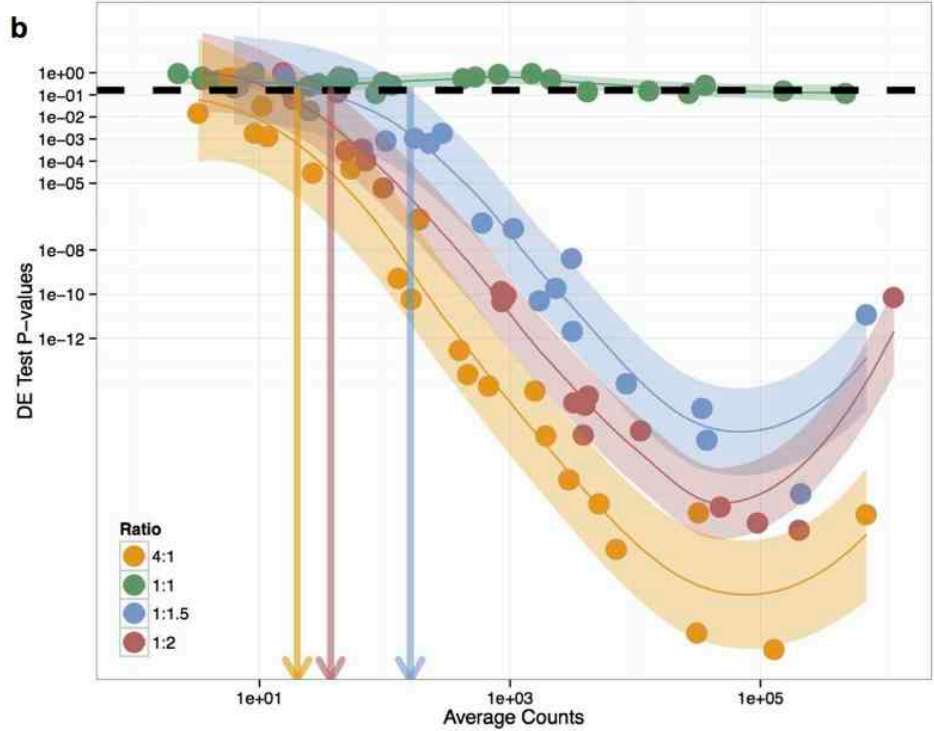

| Ratio | LODR Estimate | 90% CI Lower Bound | 90% CI Upper Bound |
|---|---|---|---|
| 4:1 | 20 | 6.2 | 26 |
| 1:1.5 | 160 | 69 | 190 |
| 1:2 | 37 | 23 | 41 |

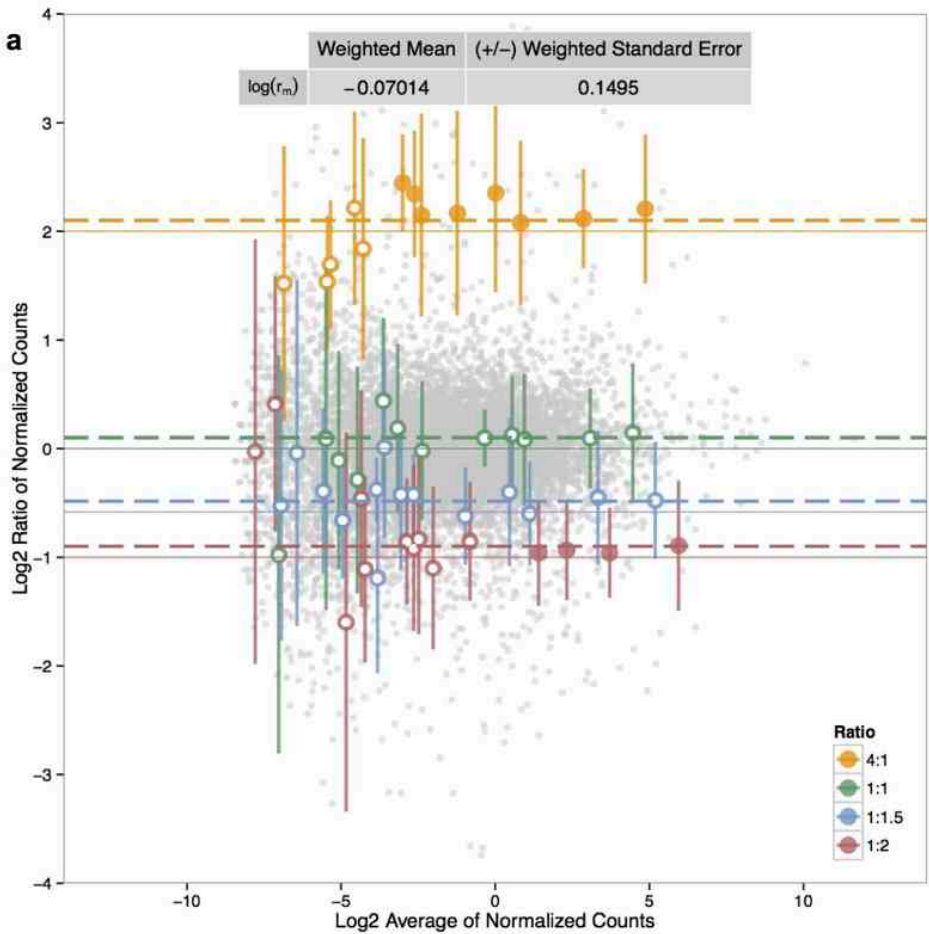

Figure 5 a

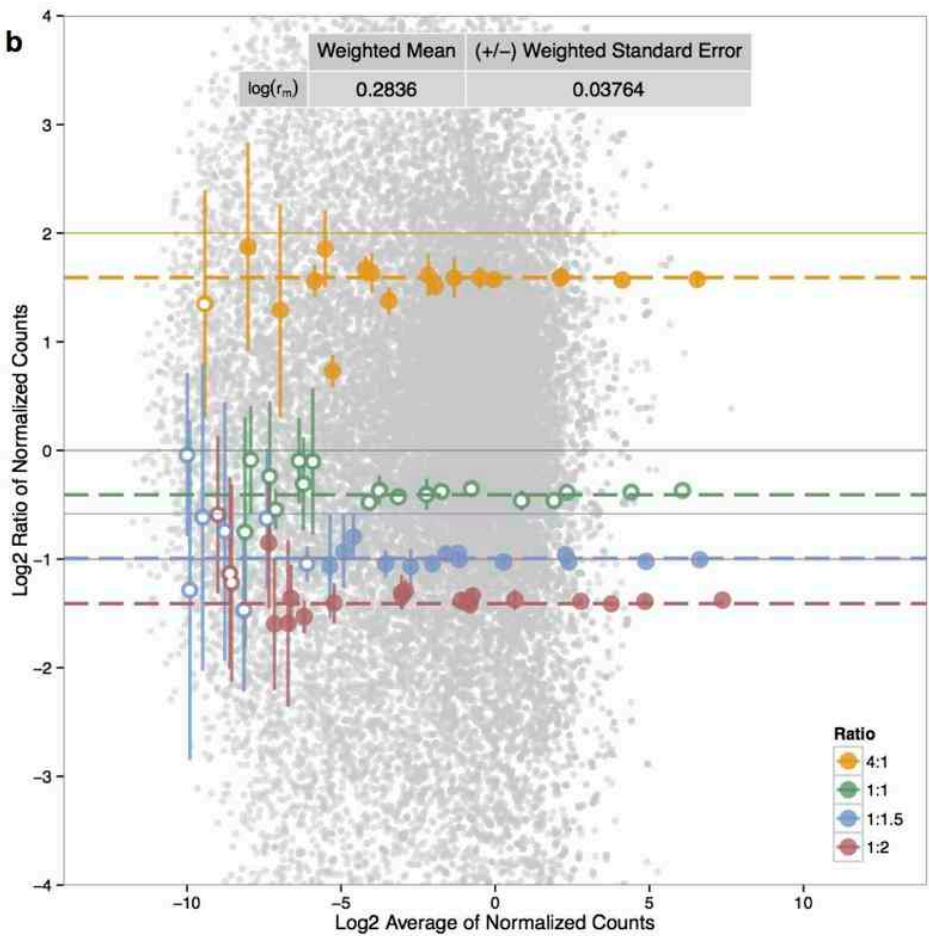

Figure 6 a

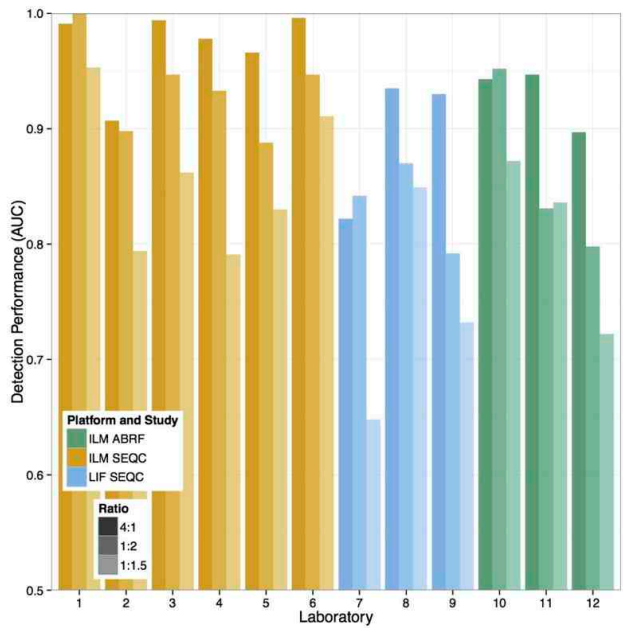
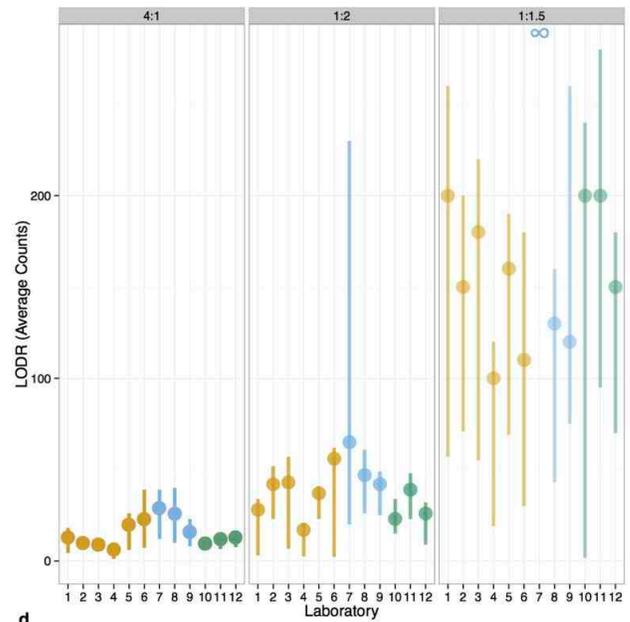
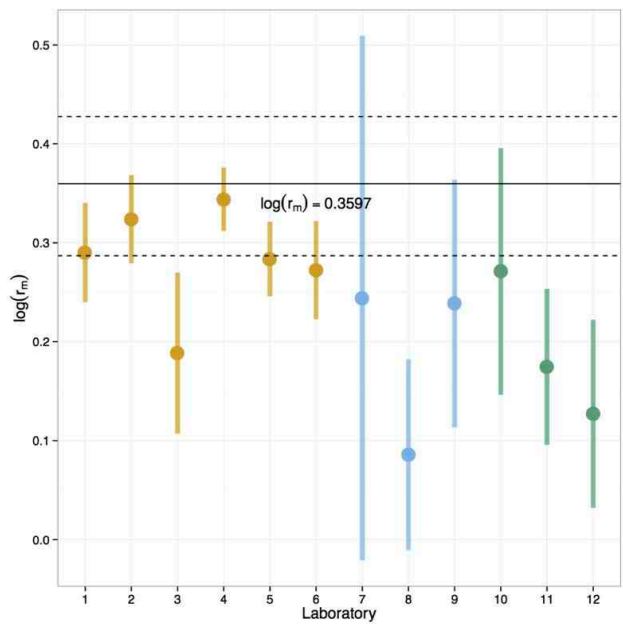
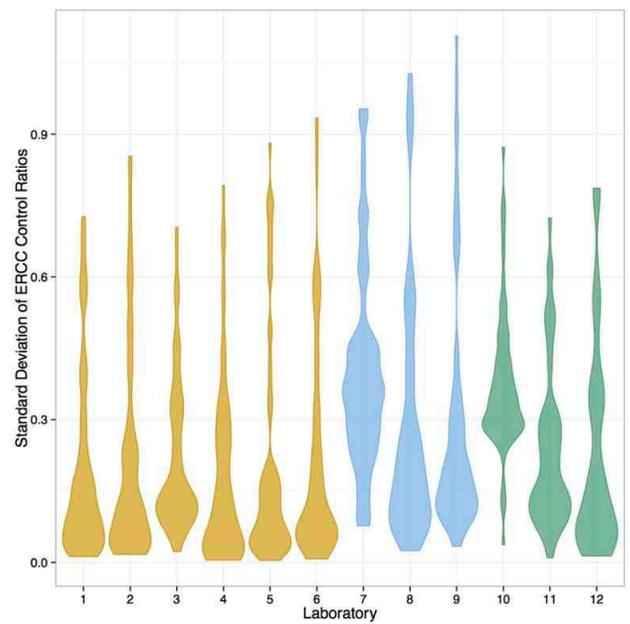

**Supplementary Material**

I. Figures from erccdashboard for rat toxicogenomics (Fig. S1-S5) and interlaboratory UHRR/HBRR analysis (Fig. S6–S20)

II. ERCC specific effects across sites and platforms (Fig. S21–S22)

III. Comparison of ERCC control ratio mixtures with and without poly-A selection (Fig. S23–S24)

IV. Qualitative dispersion estimate analysis to assess validity of control data (Fig. S25)

V. LODR estimation from endogenous transcript simulation (Fig. S26)

VI. Figures from erccdashboard for Subread and featureCounts analysis of Lab 7-9 data (Fig. S27-S29)

VII. Mapped read QC Metrics (Fig. S30–S33)

I.  **Figures from erccdashboard for rat toxicogenomics and interlaboratory UHRR/HBRR analysis (Fig. S1–S20)**

The following figures are the erccdashboard main technical performance figures for 20 different experiments. See main text and captions for Fig. 2-5 for detailed descriptions of these four performance figures.

**Rat Toxicogenomics Experiments**
Fig. S1   erccdashboard results for 3-methylcholanthrene (3ME) and control (CTL, set 1) rat experiment
Fig. S2   erccdashboard results for methimazole (MET) and control (CTL, set 1) rat experiment
Fig. S3   erccdashboard results for betanapthoflavone (NAP) and control (CTL, set 1) rat experiment
Fig. S4   erccdashboard results for n-nitrosodimethylamine(NIT) and control (CTL, set 2) rat experiment
Fig. S5   erccdashboard results for yhioacetamide (THI) and Control (CTL, set 2) rat experiment

**Reference RNA (UHRR/HBRR) RNA-Seq Experiments**
Fig. S6-S11     erccdashboard results for SEQC Illumina platform experiments at Lab 1-6
Fig. S12-S14    erccdashboard results for SEQC Life Technologies platform experiments at Lab 7-9
Fig. S15-S17    erccdashboard results for ABRF Illumina platform experiments at Lab 10-12

**Reference RNA (UHRR/HBRR) Microarray Experiments**
Fig. S18-S19    erccdashboard results for SEQC Illumina Beadarray microarray platform experiments at Lab 13 – 14
Fig. S20        erccdashboard results for SEQC Agilent 1M custom microarray platform experiments at Lab 15

In typical microarray results saturation effects are observed (Fig. S18-19). Signal processing techniques (see Methods) were applied to Agilent array data (Fig. S20) to reduce these effects and the resulting microarray ratio-abundance plots are comparable to those generated with RNA-Seq data.

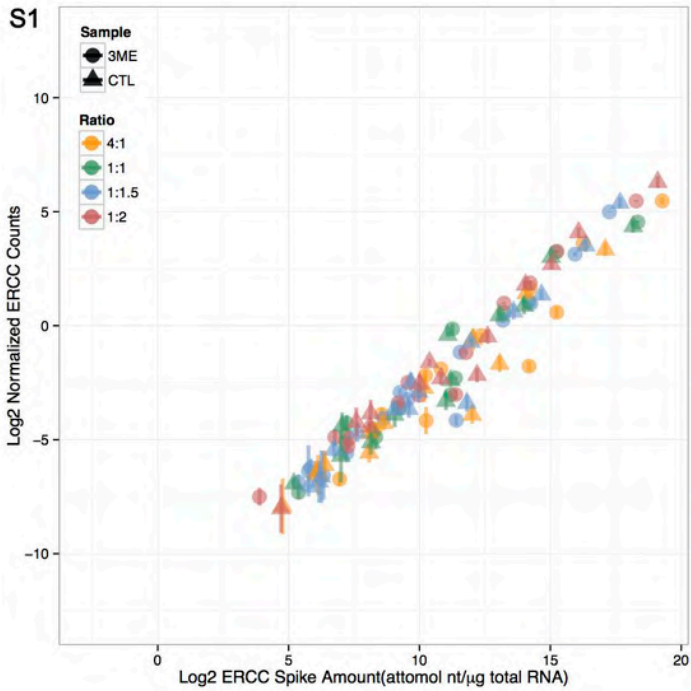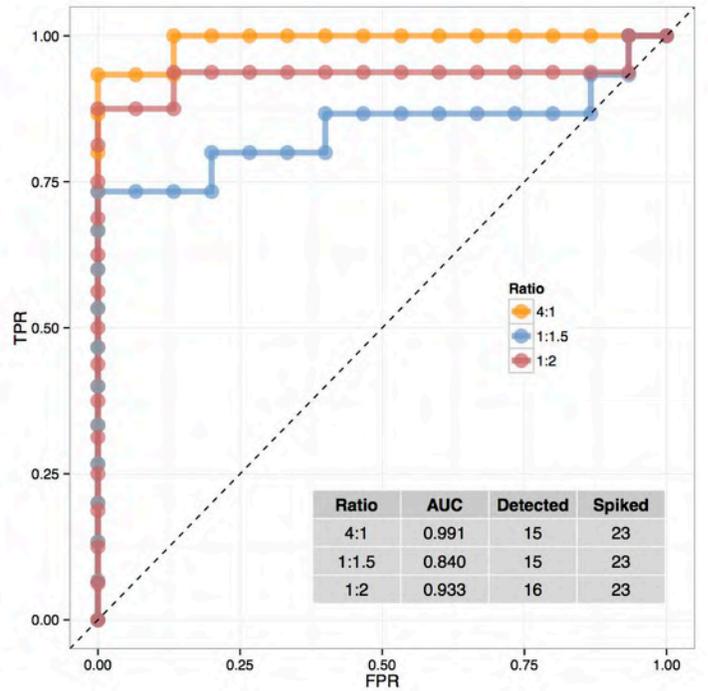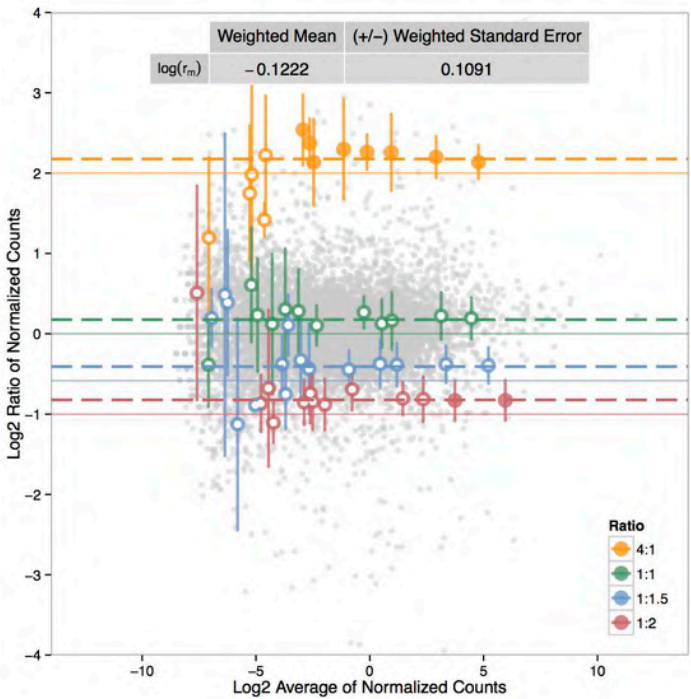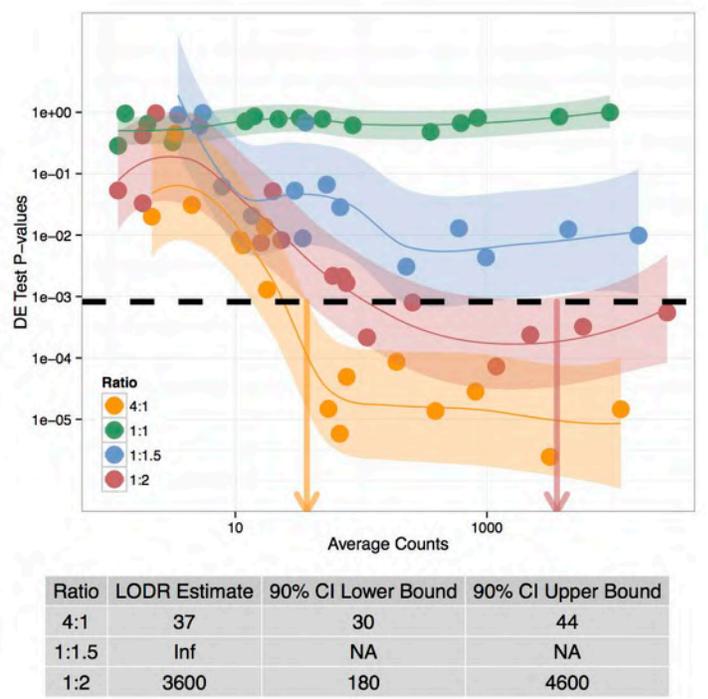

S2

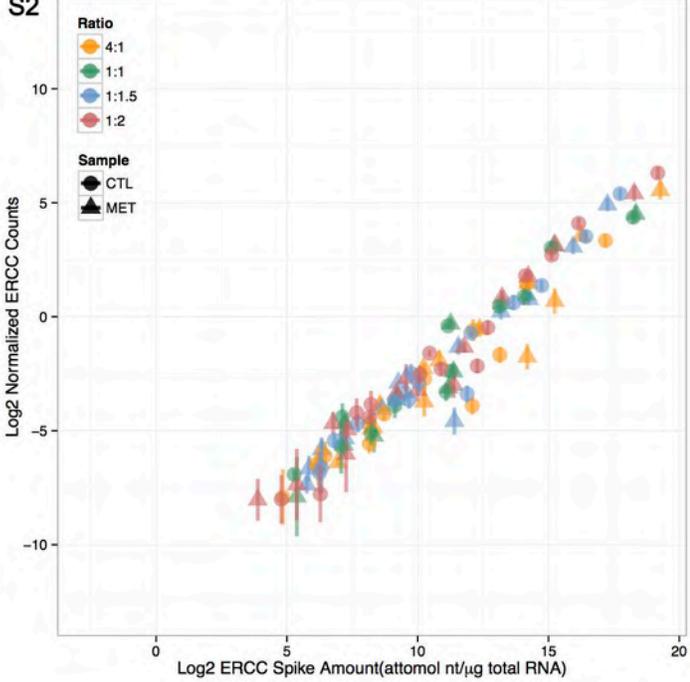
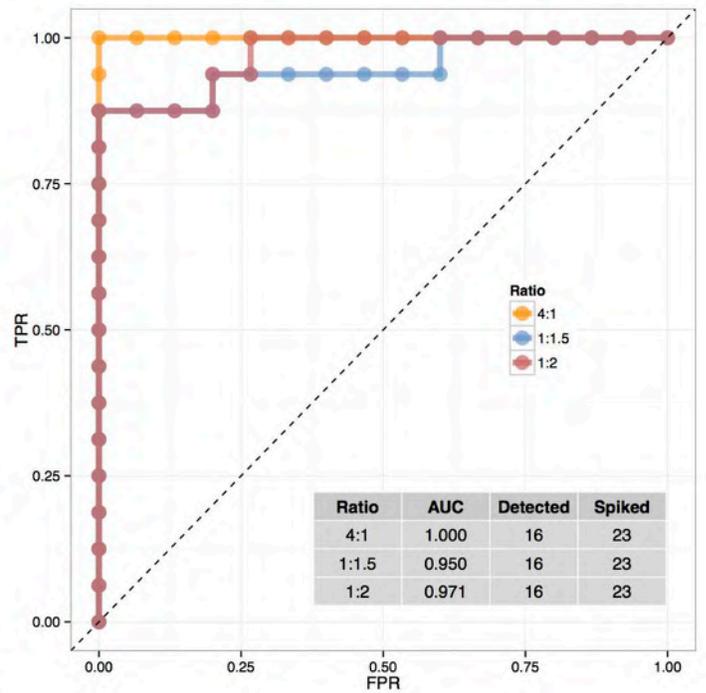
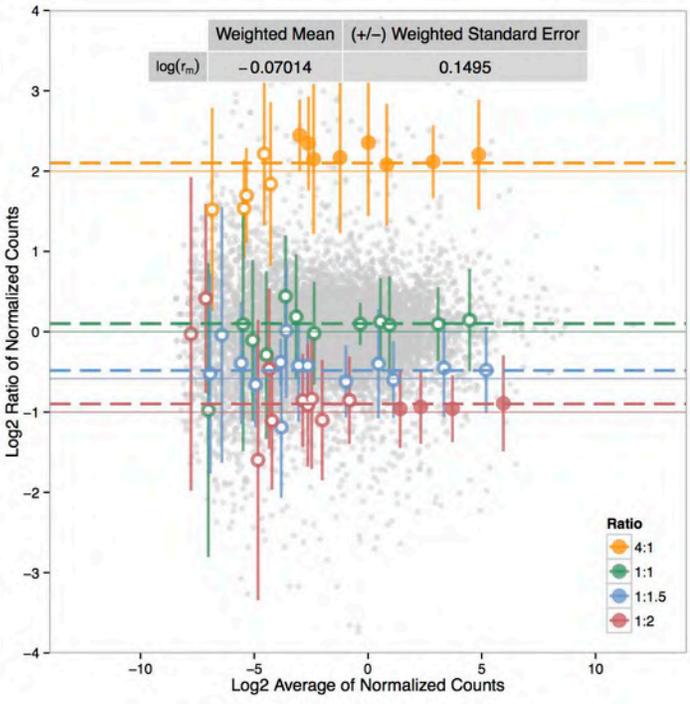
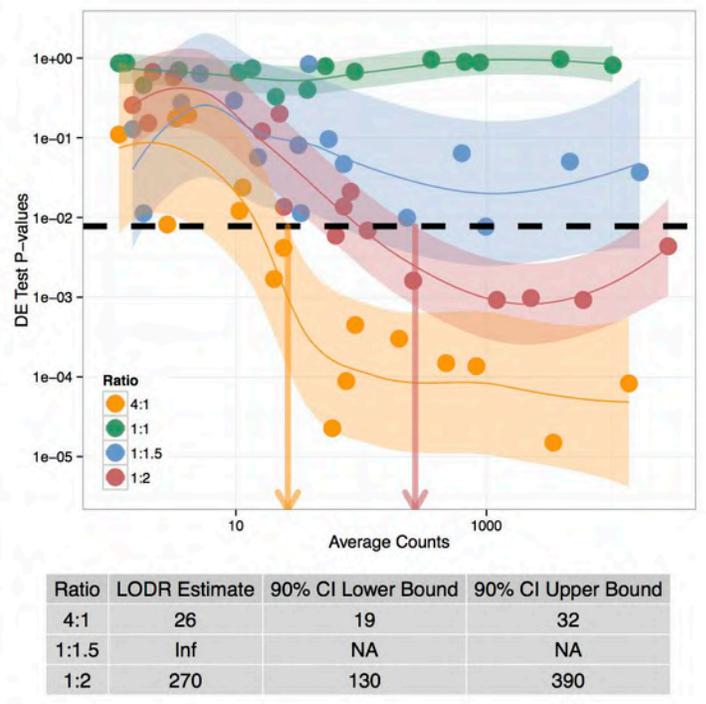

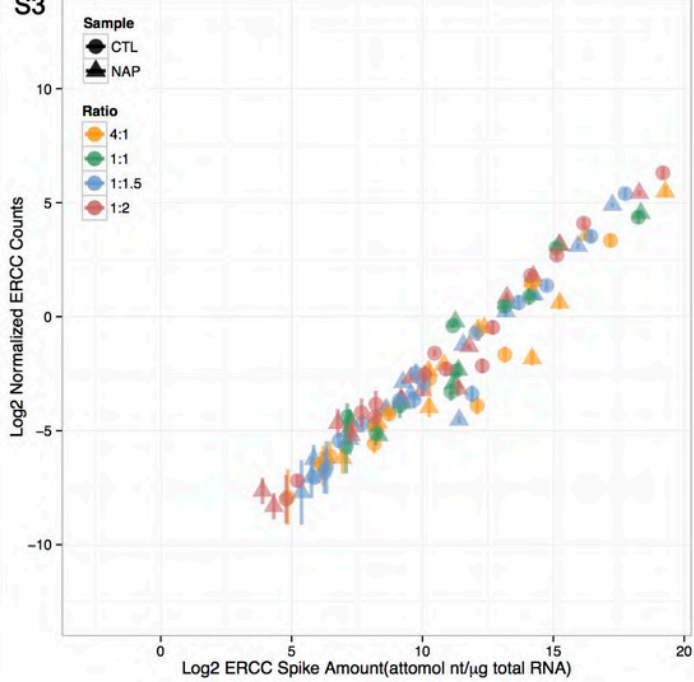
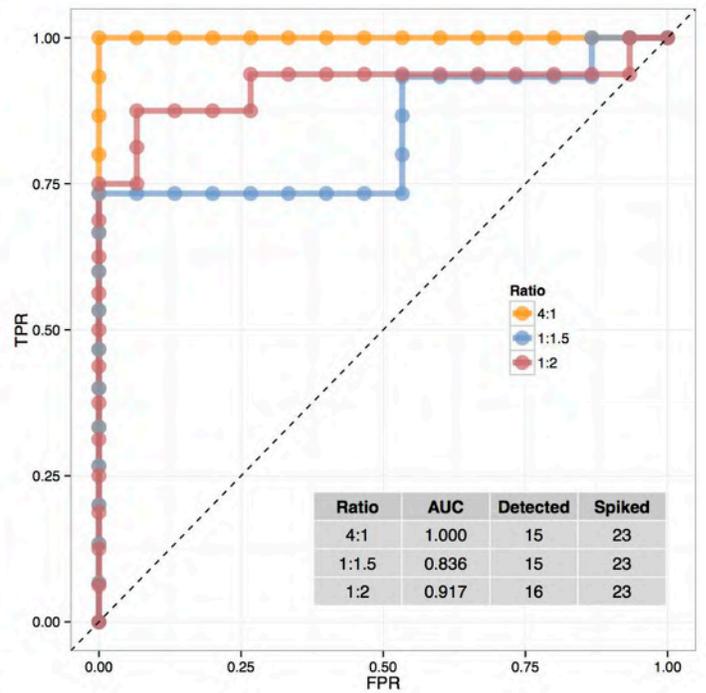
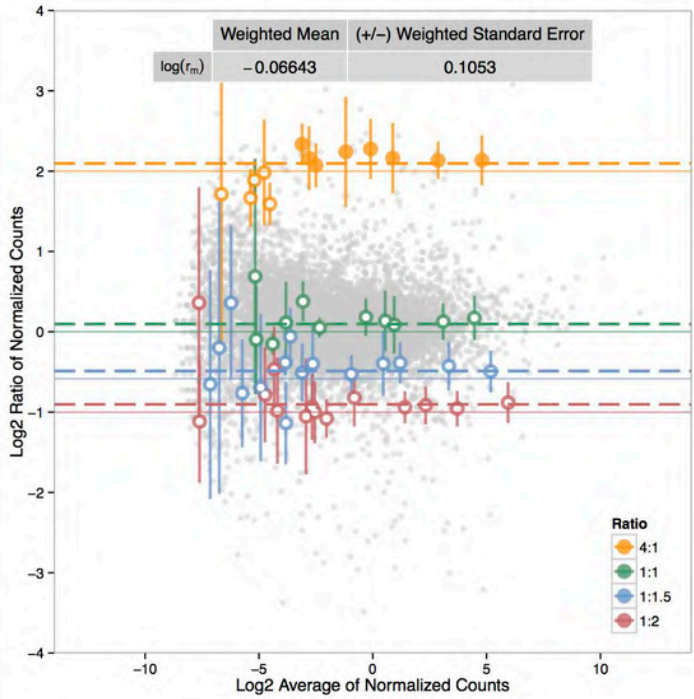
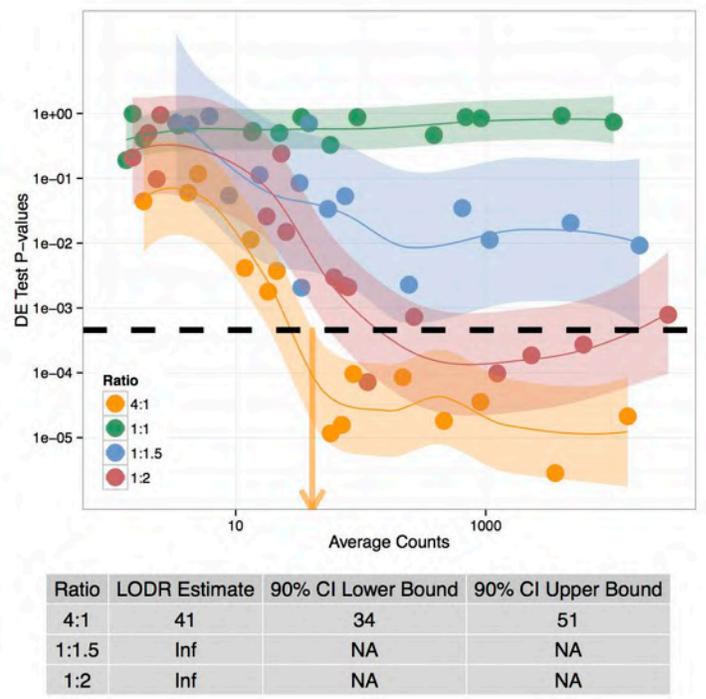

S4

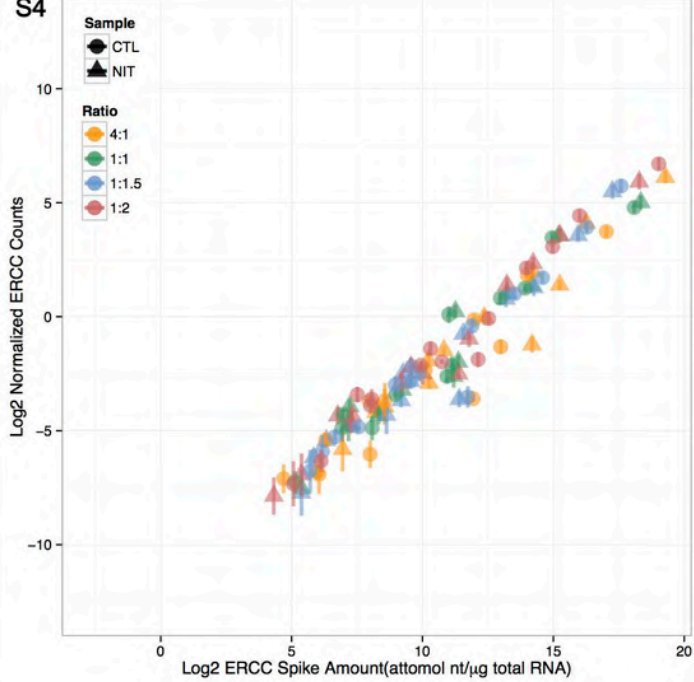
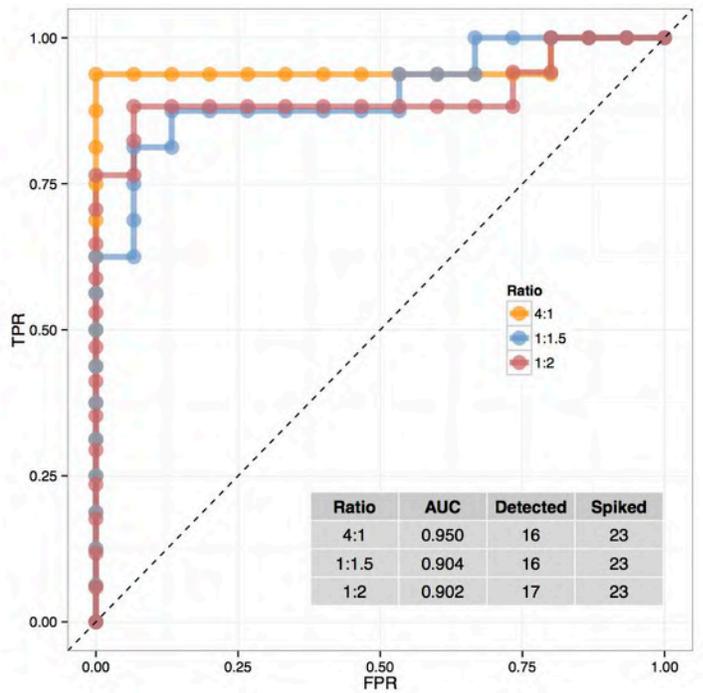
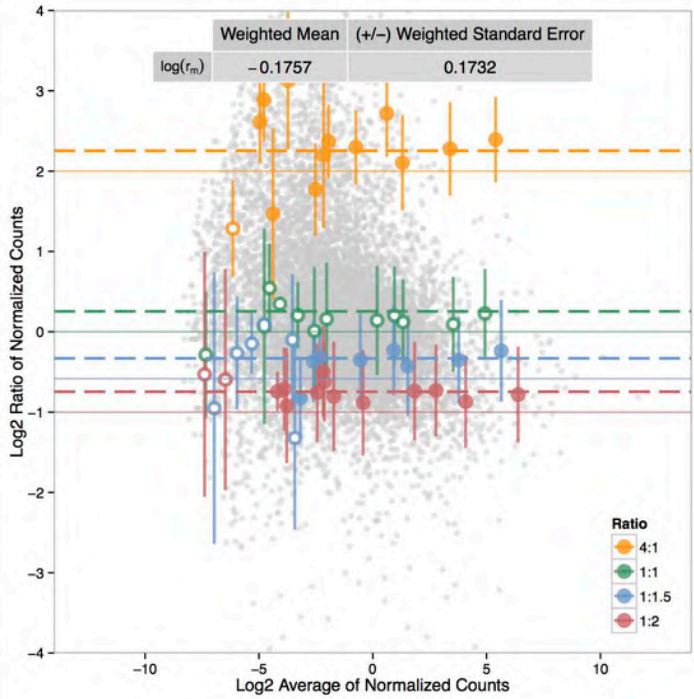
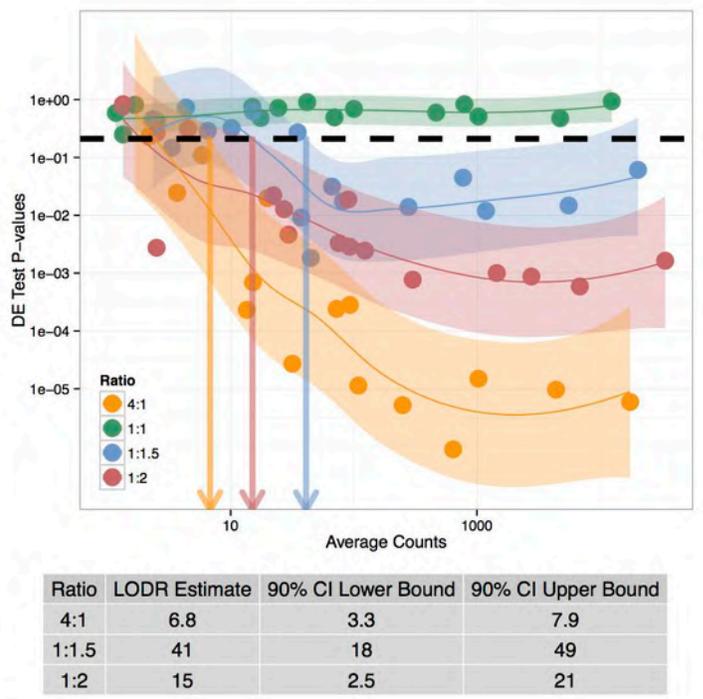

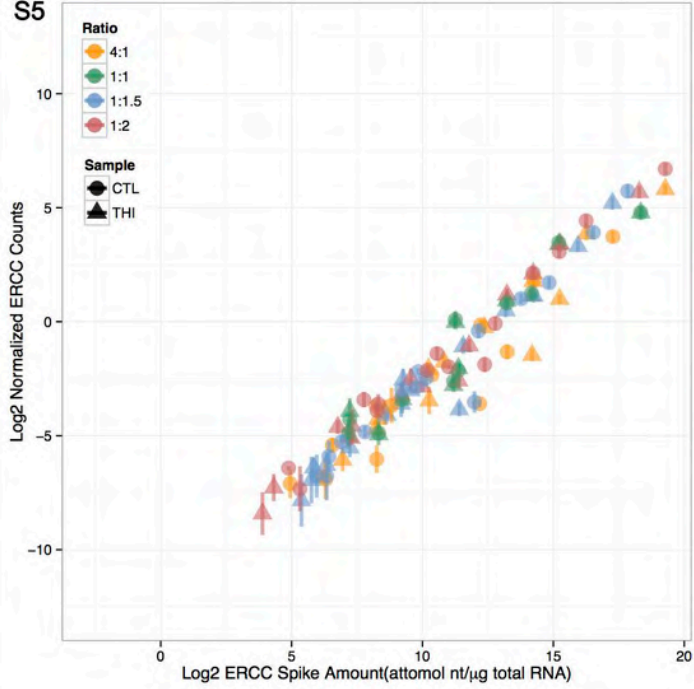
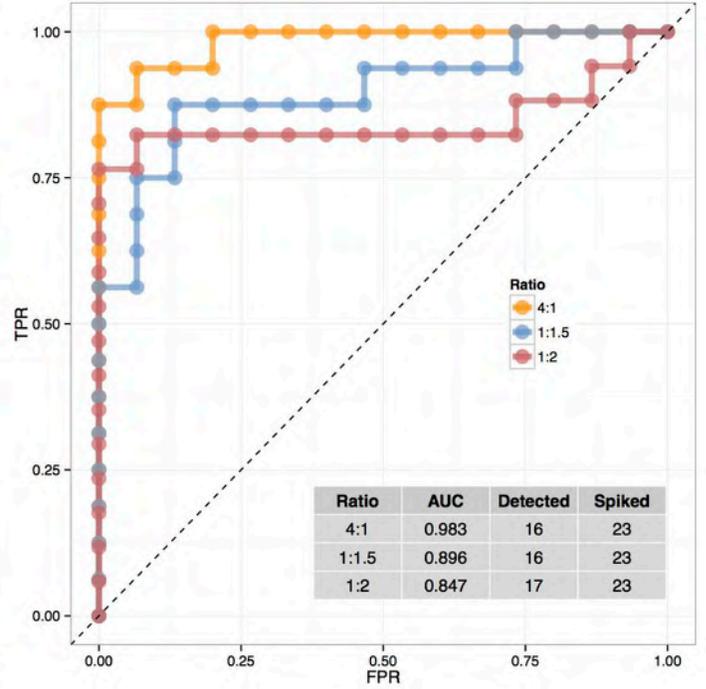
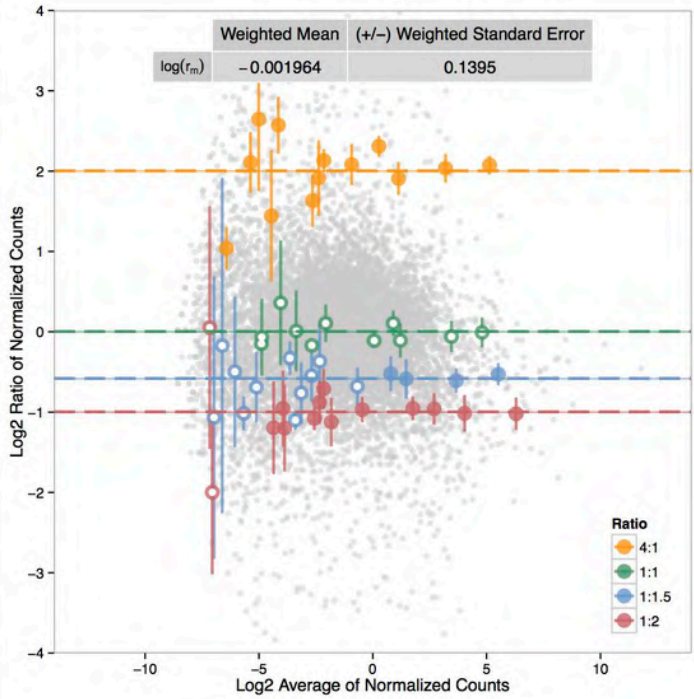
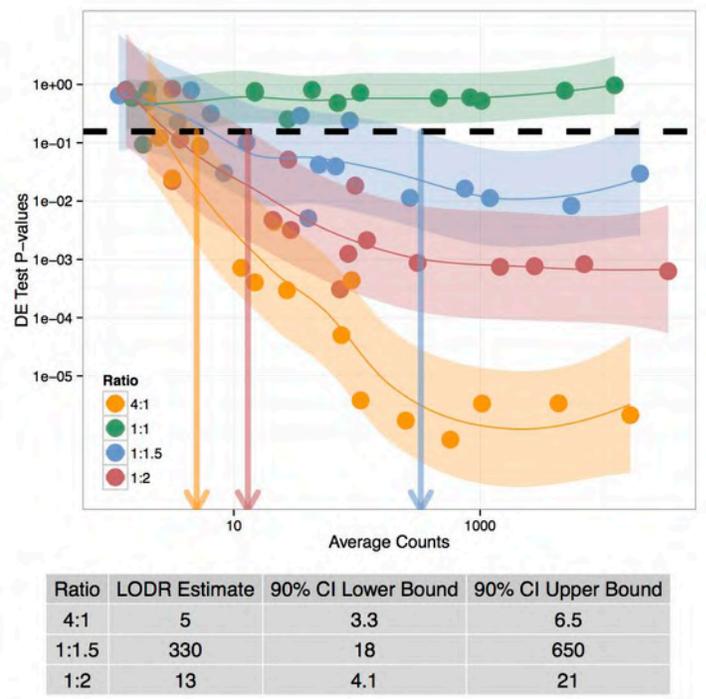

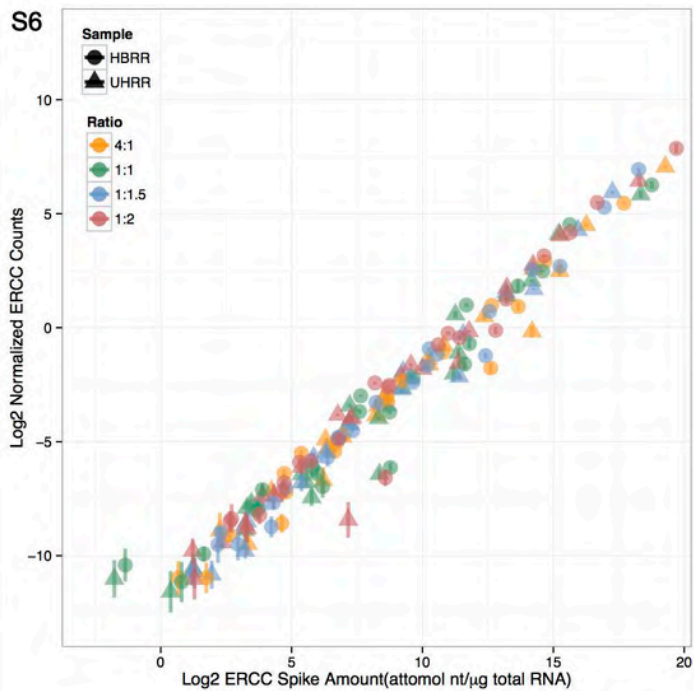
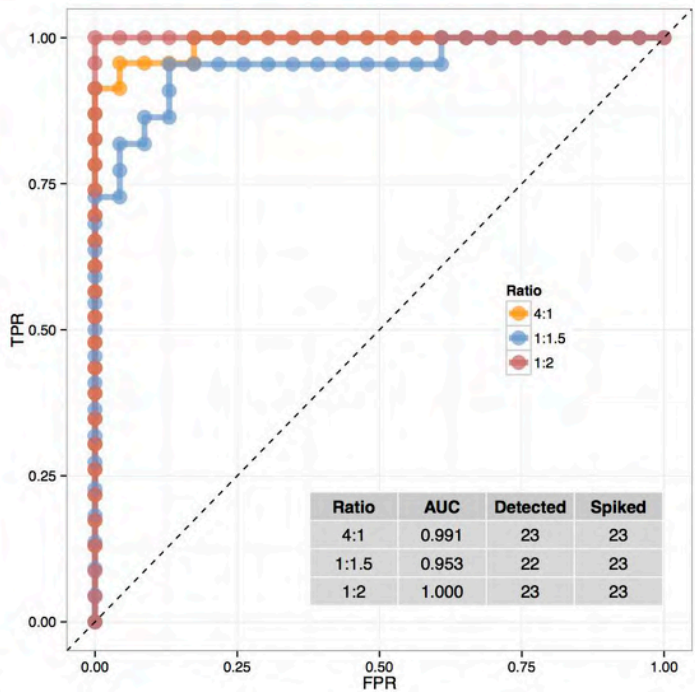
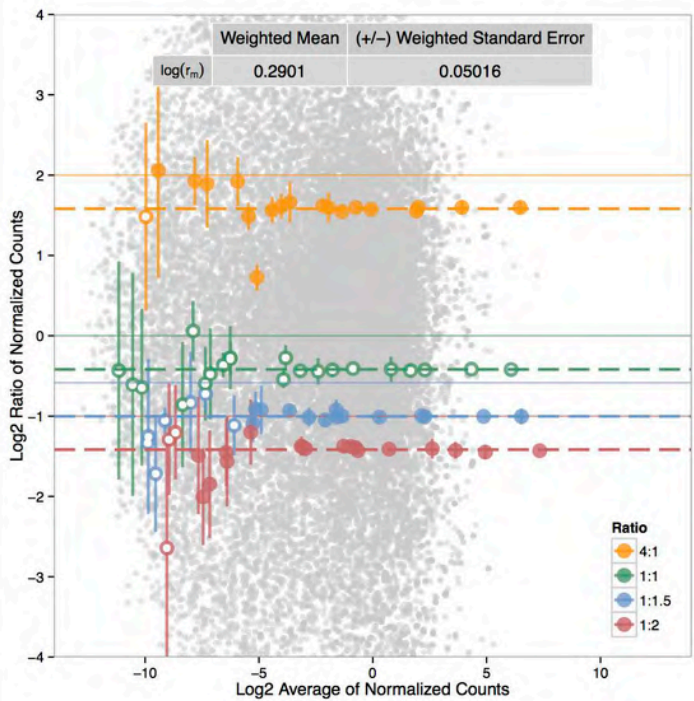
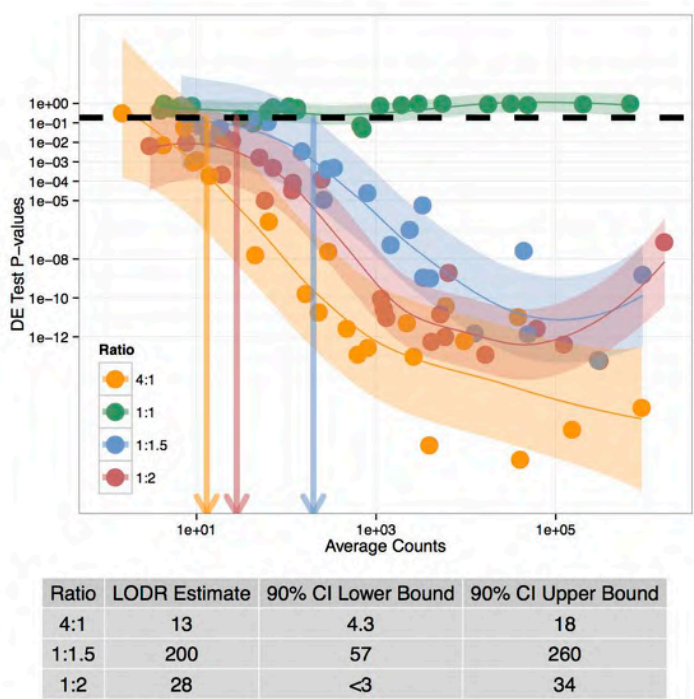

S7

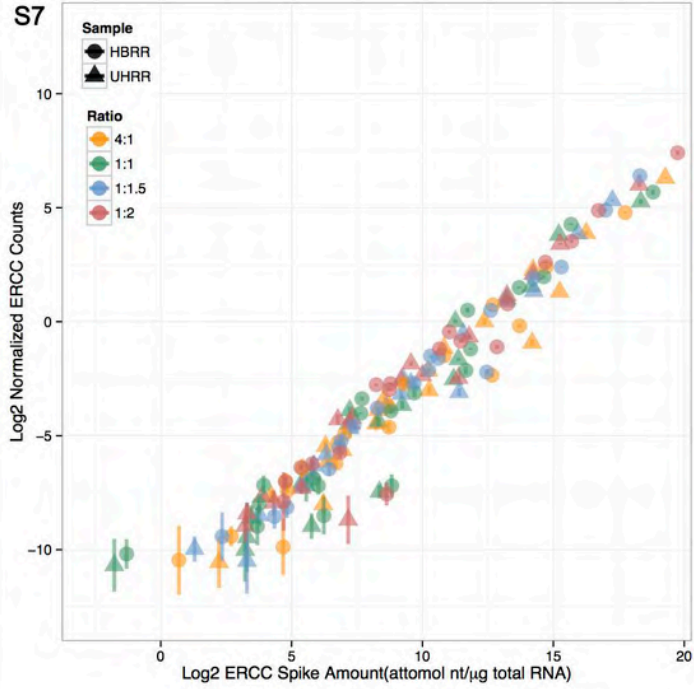
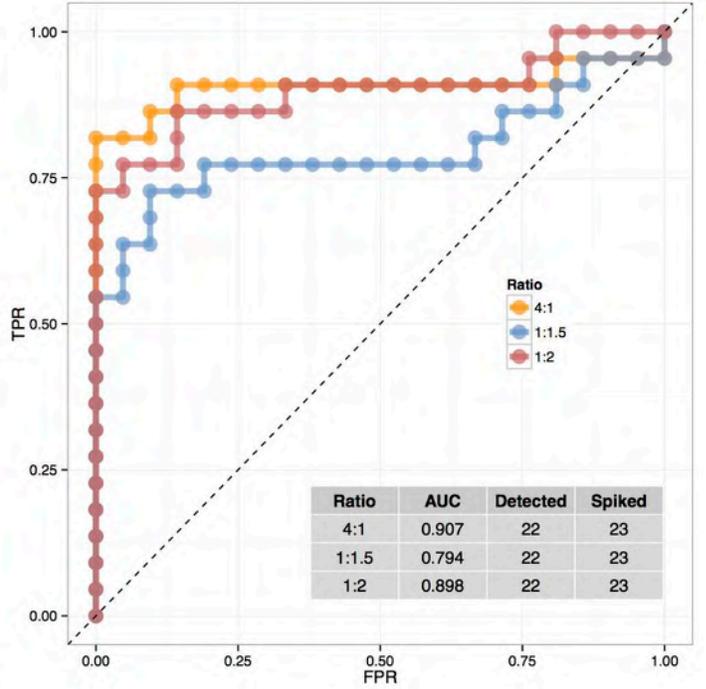
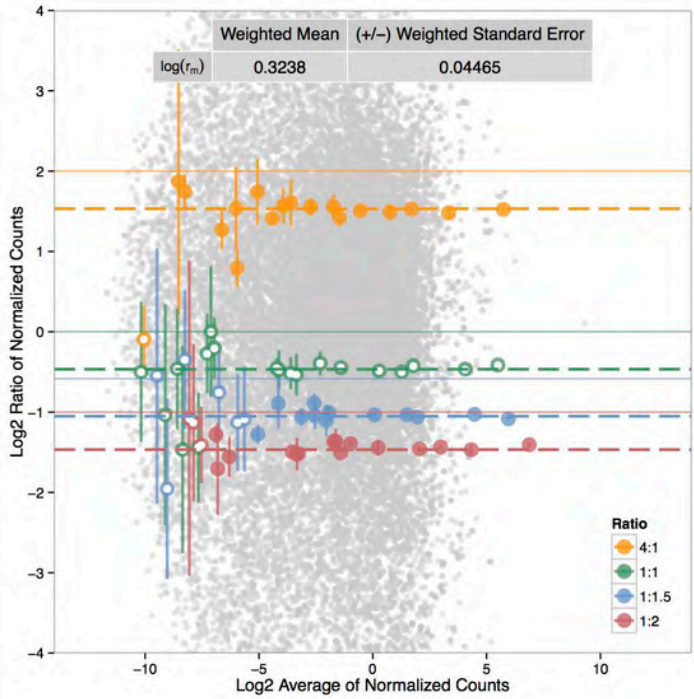
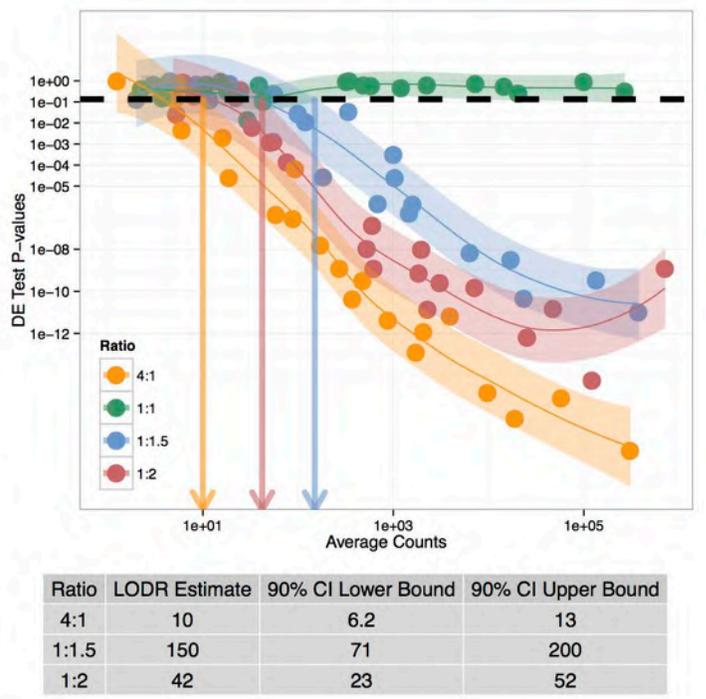

S8

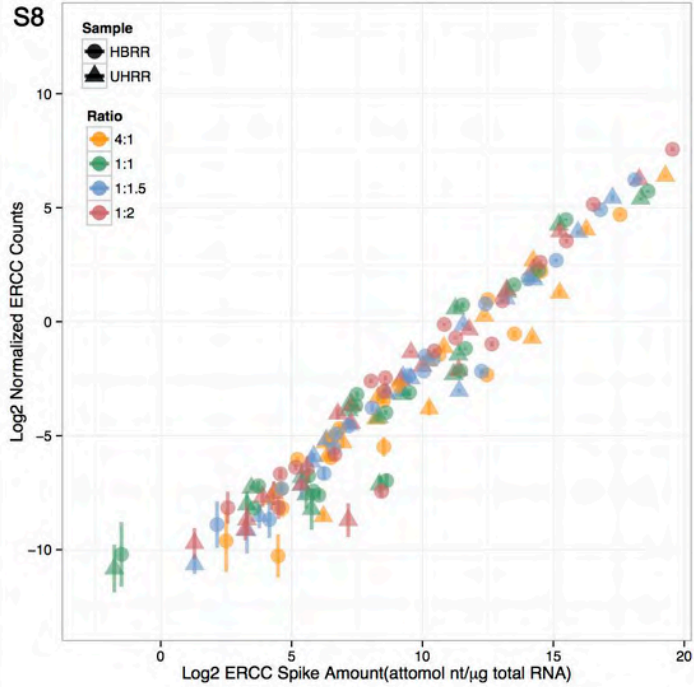
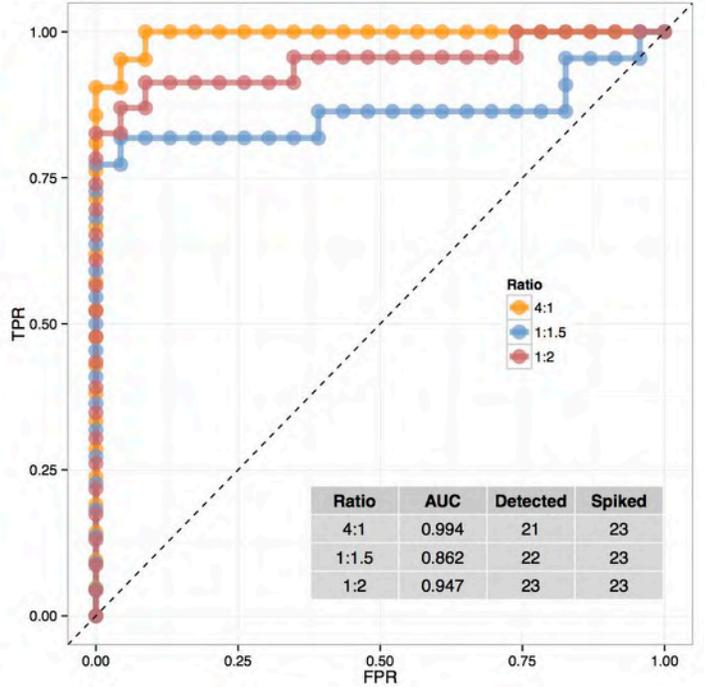
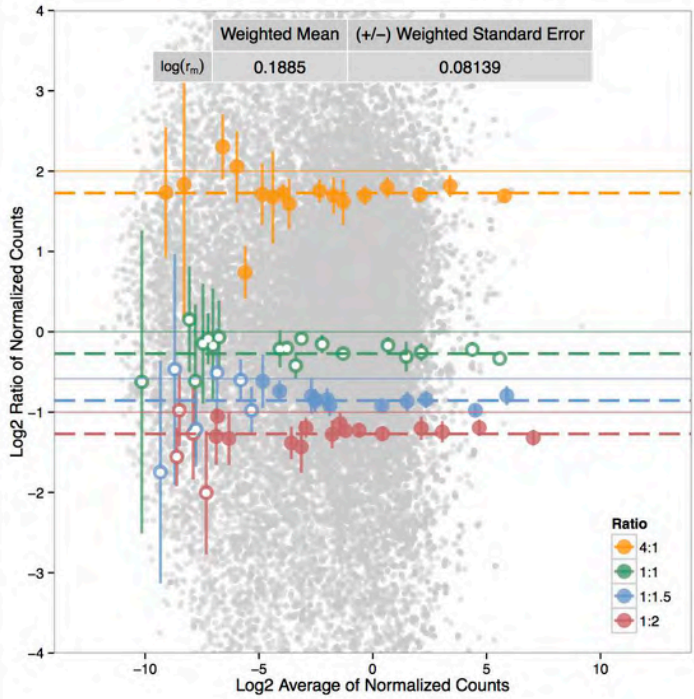
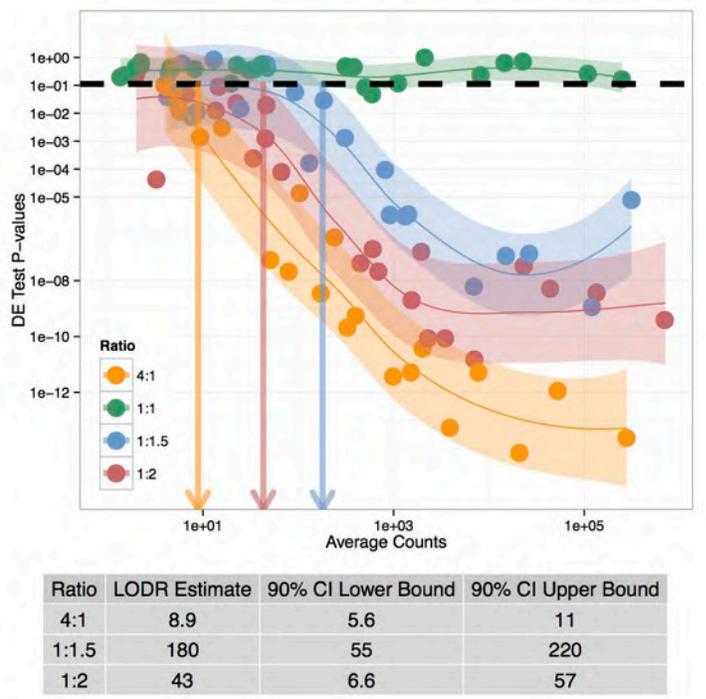

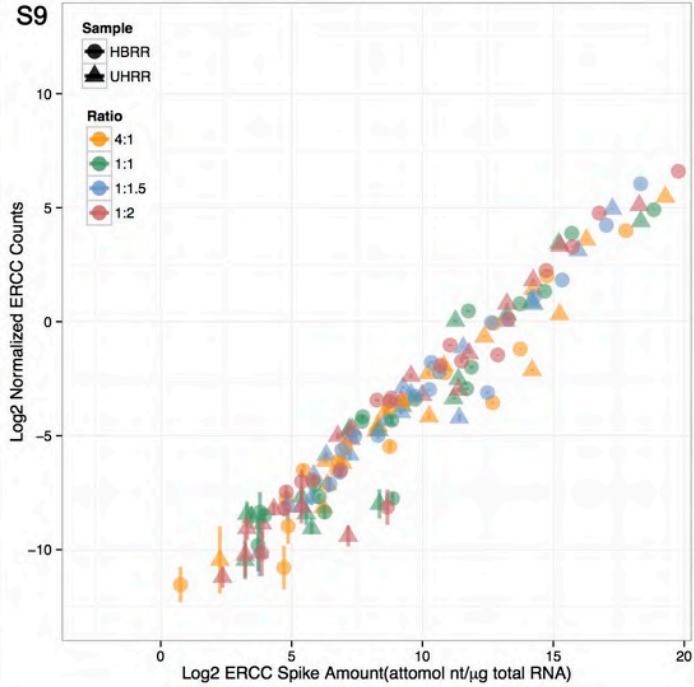
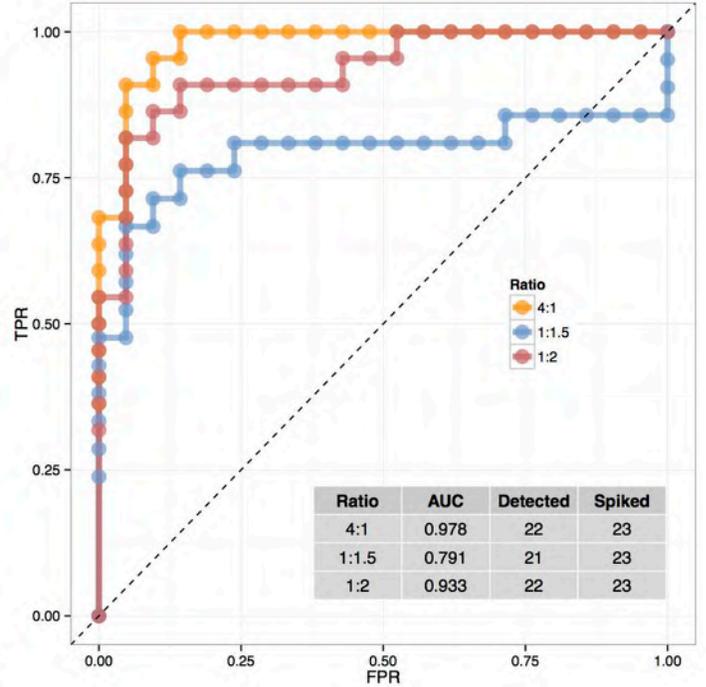
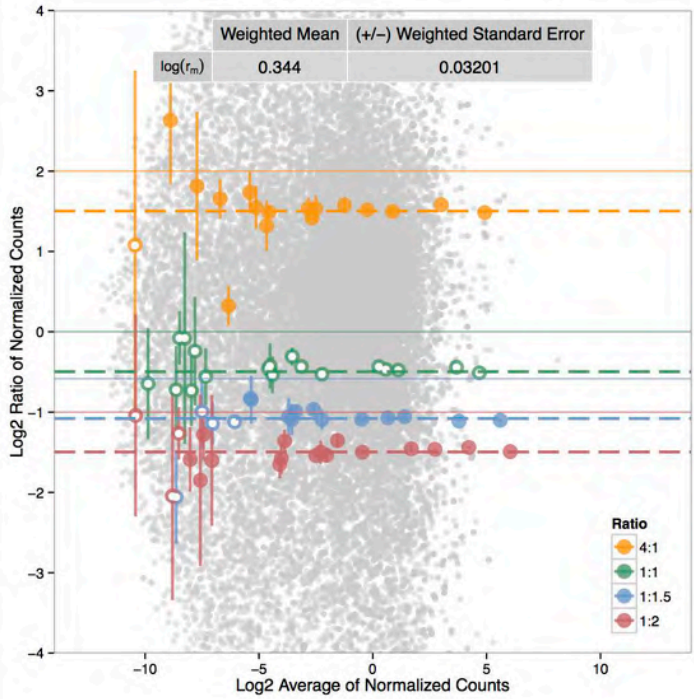
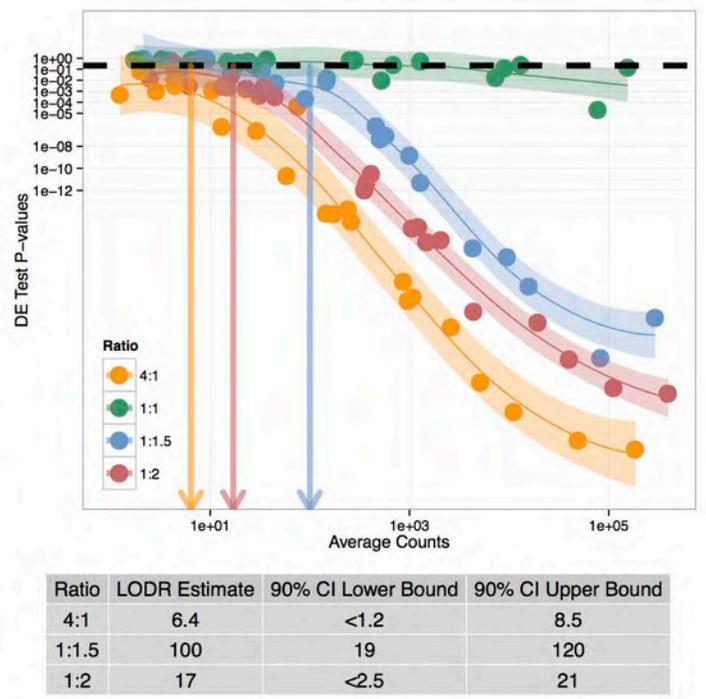

S10

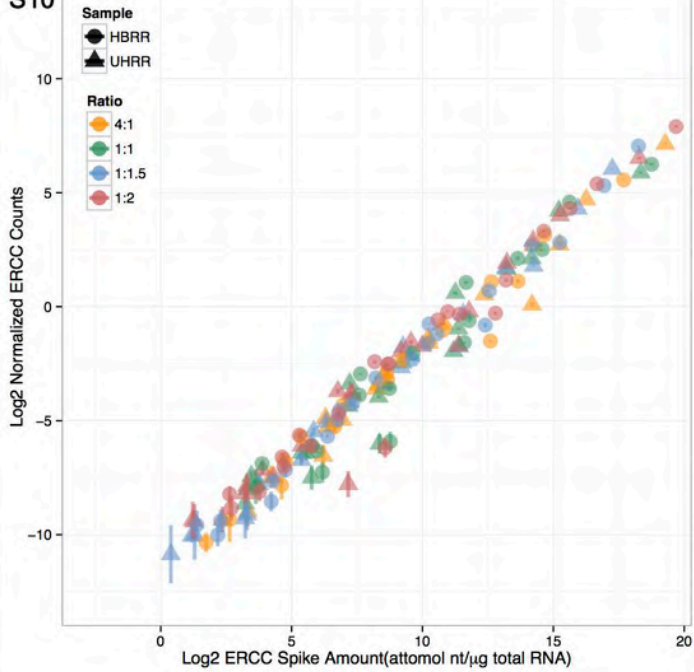
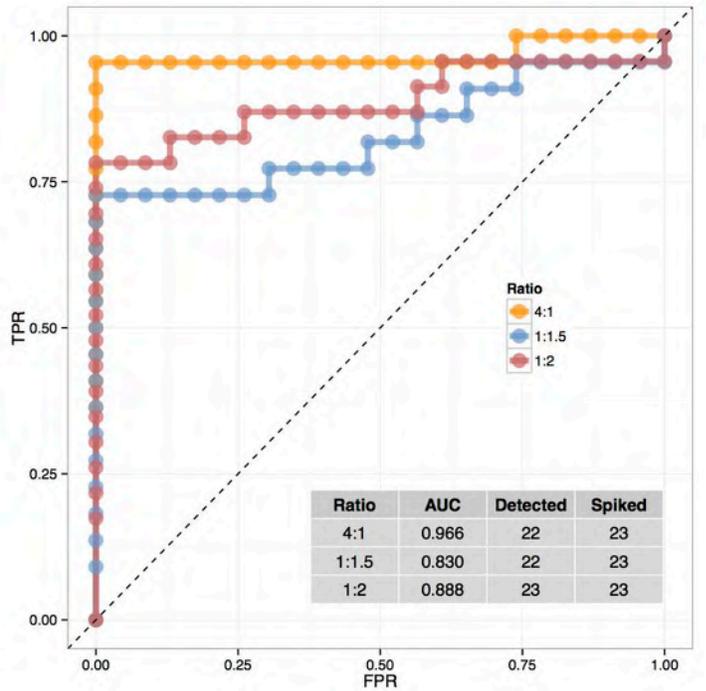
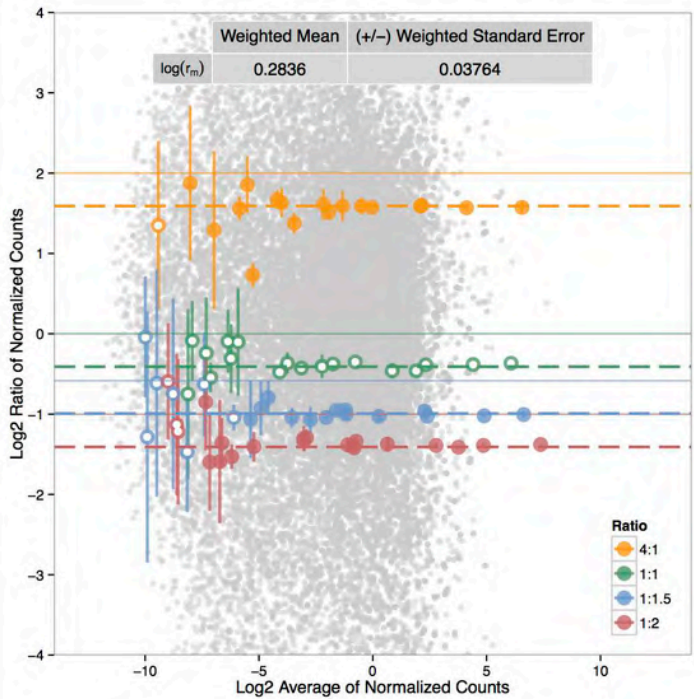
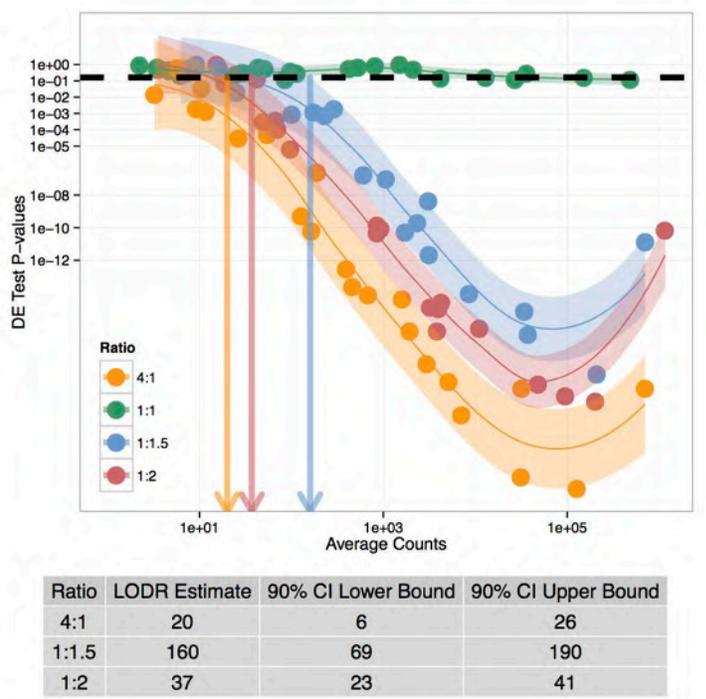

S11

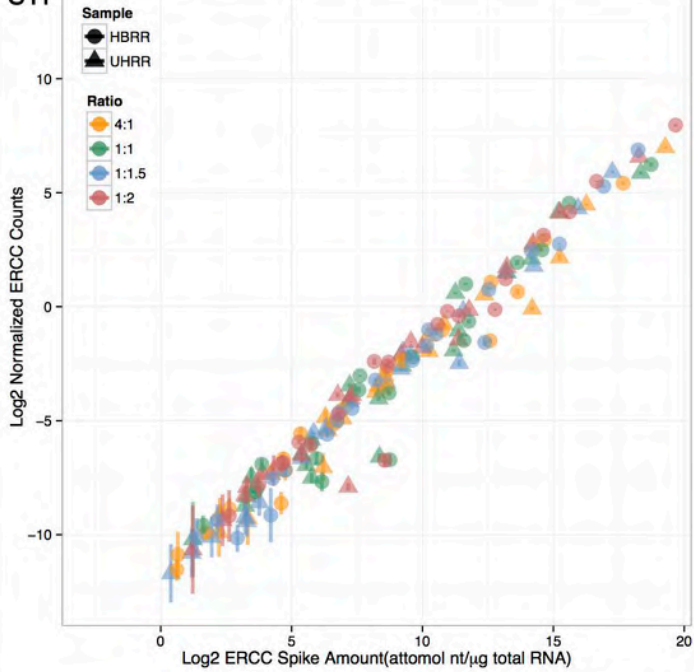
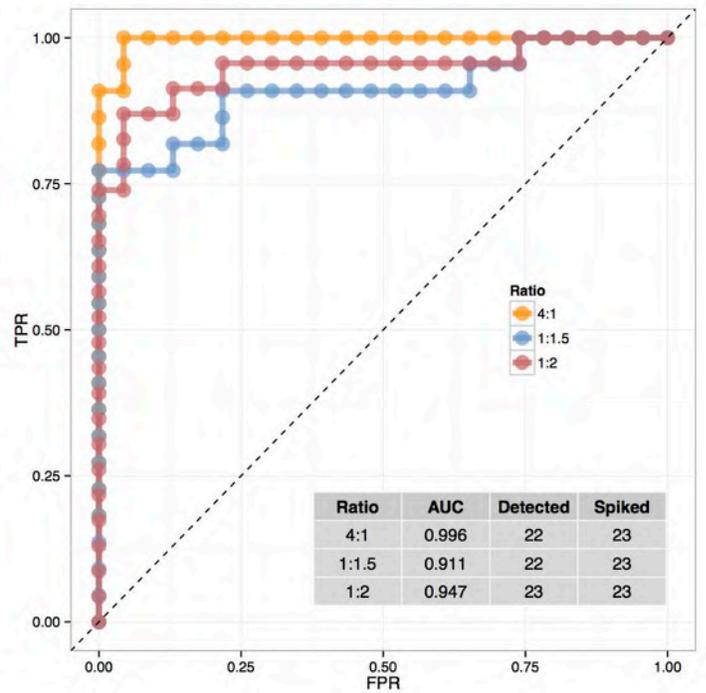
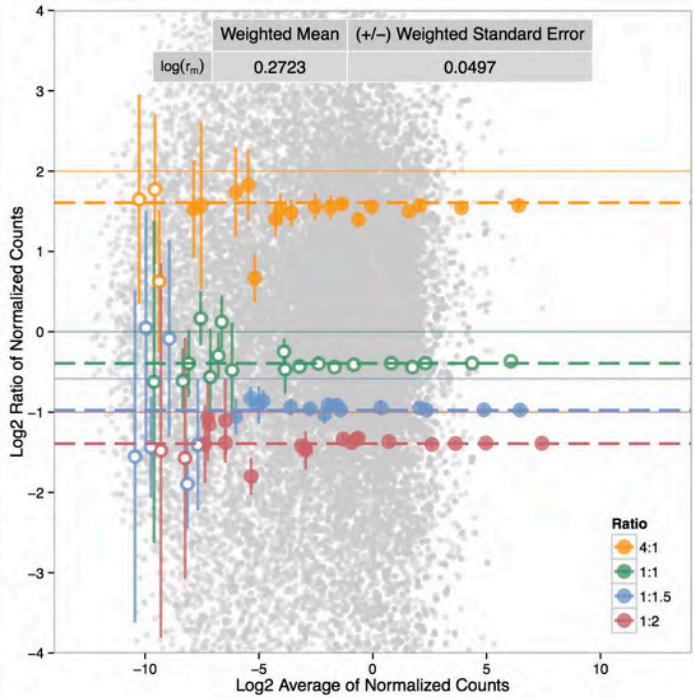
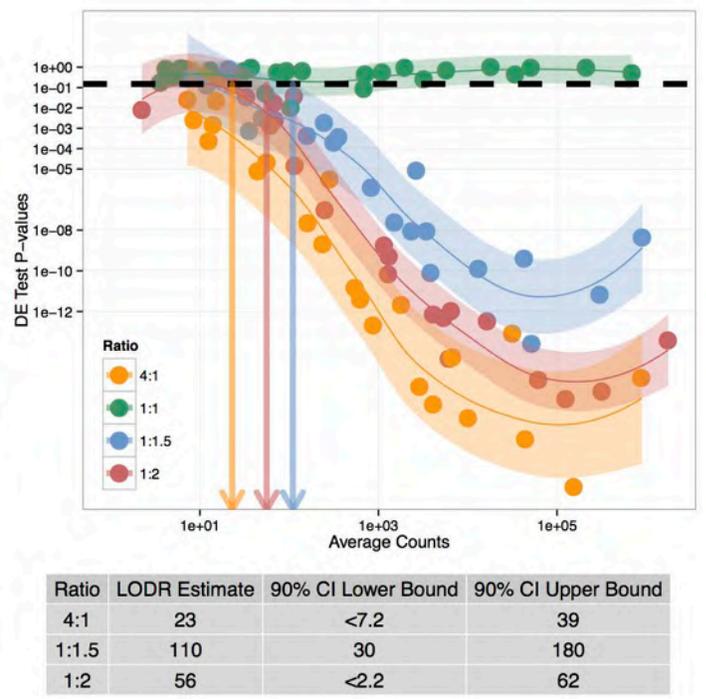

S12

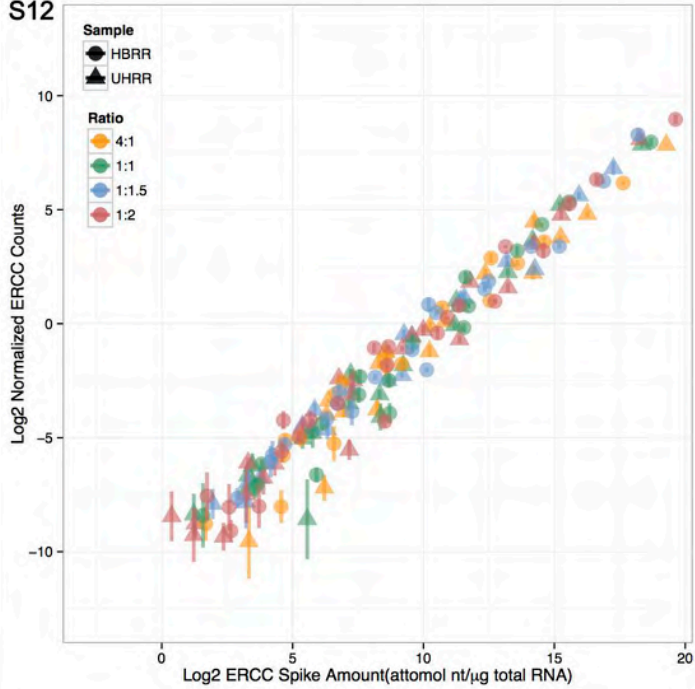
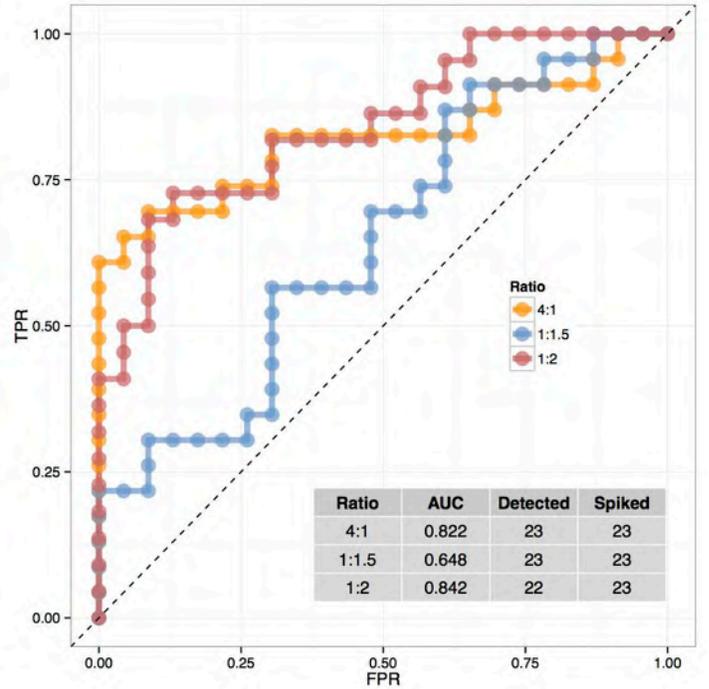
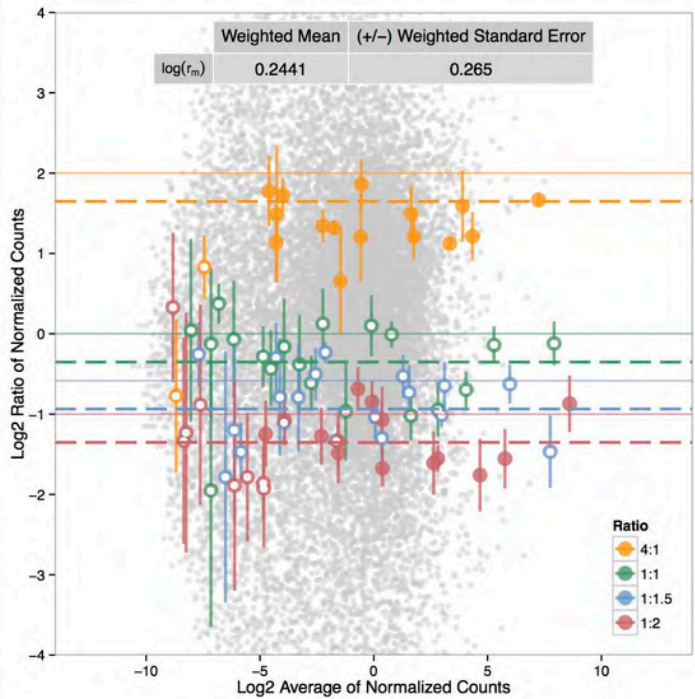
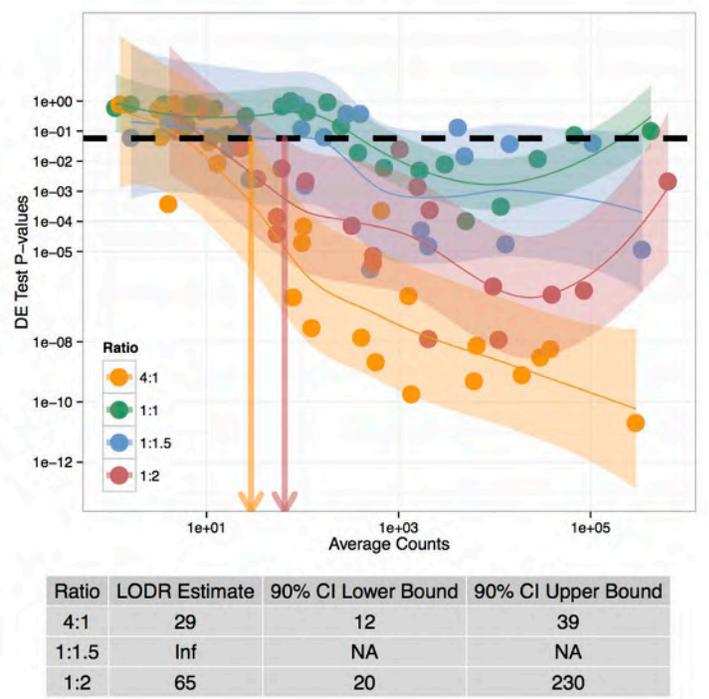

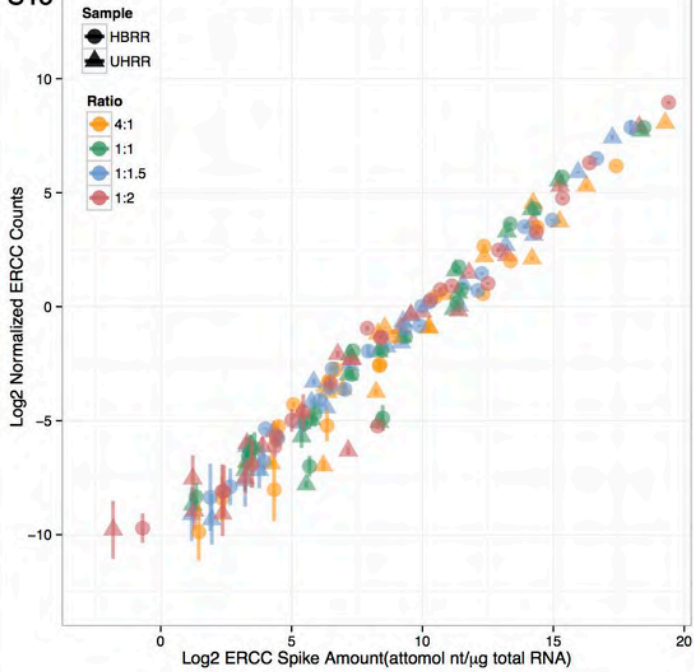
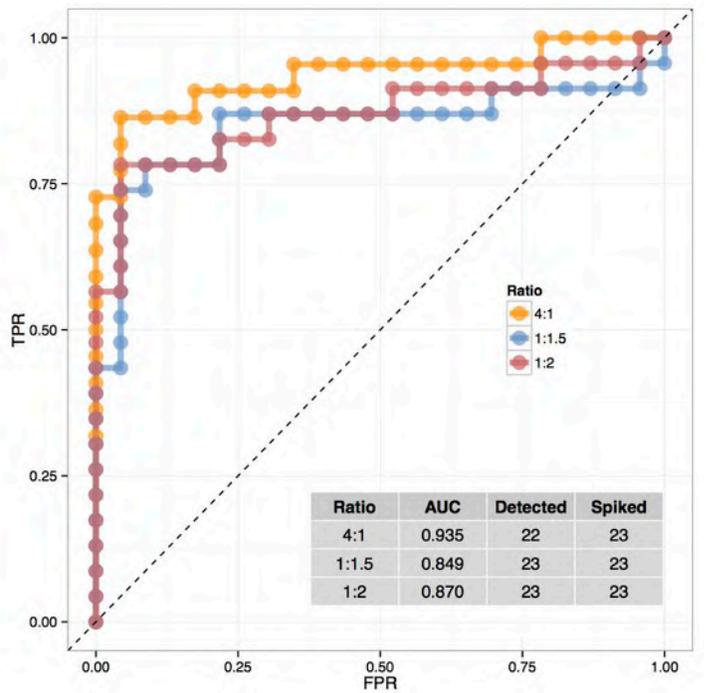
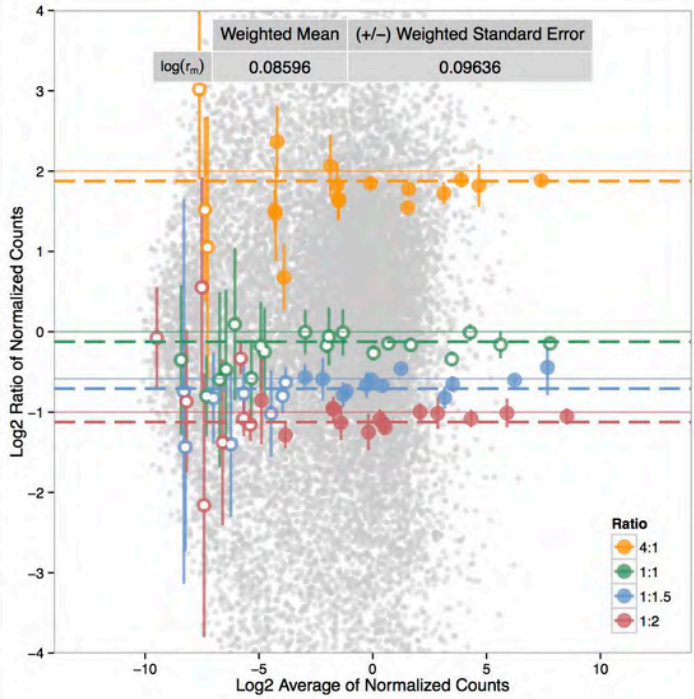
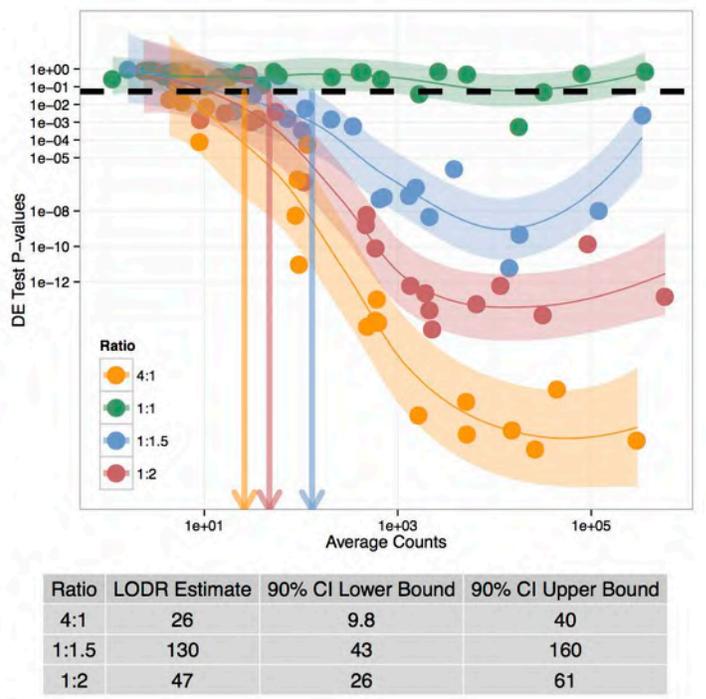

S14

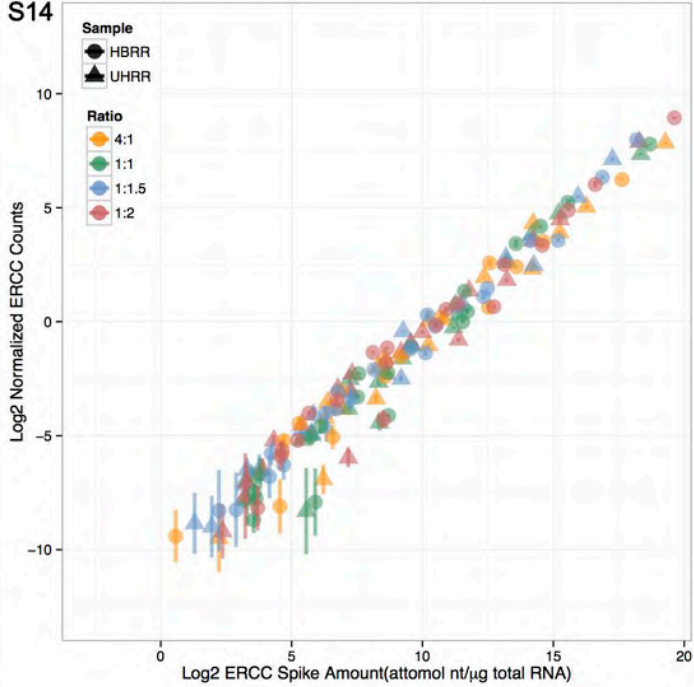
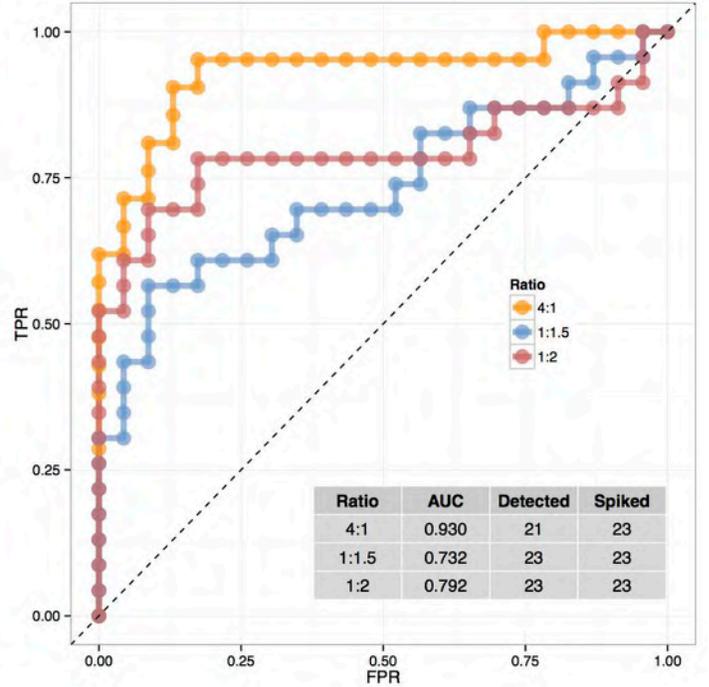
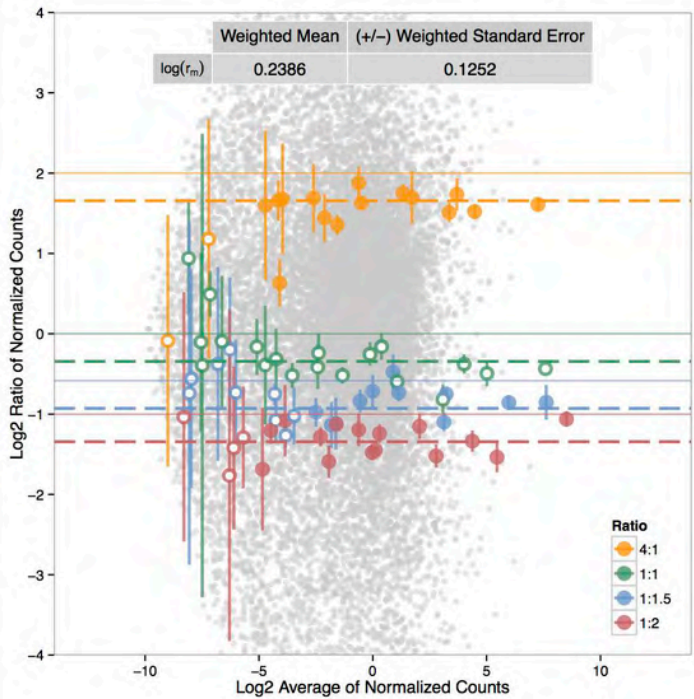
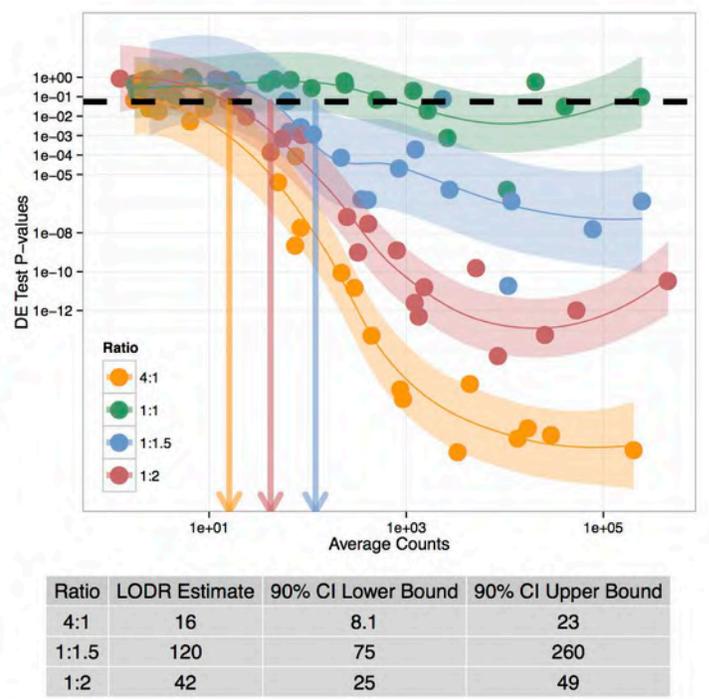

S15

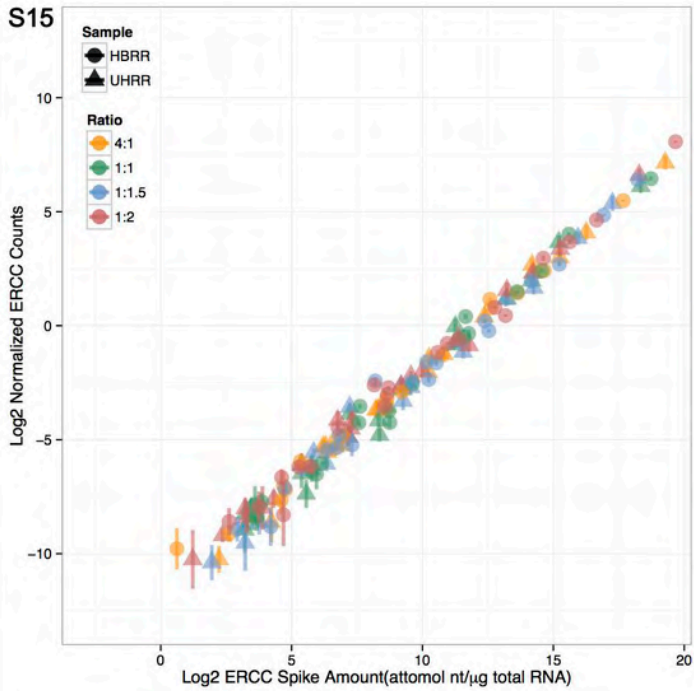
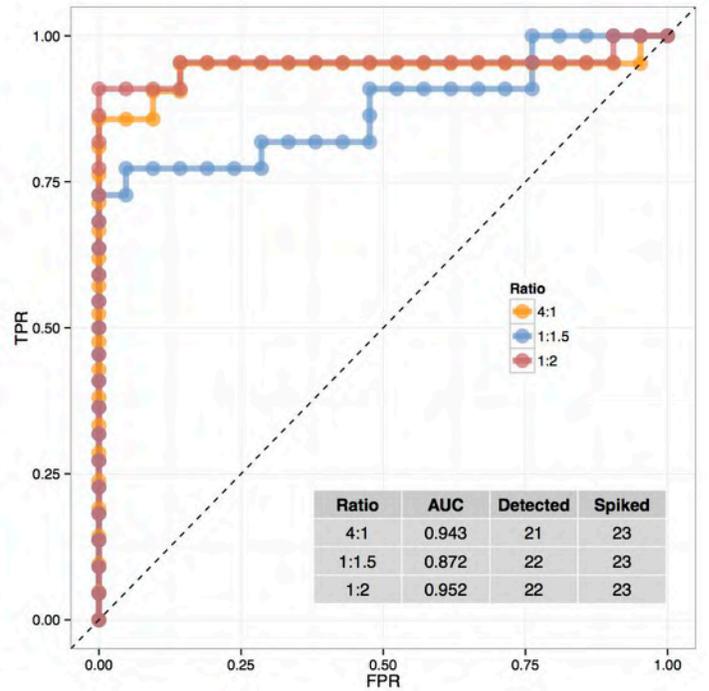
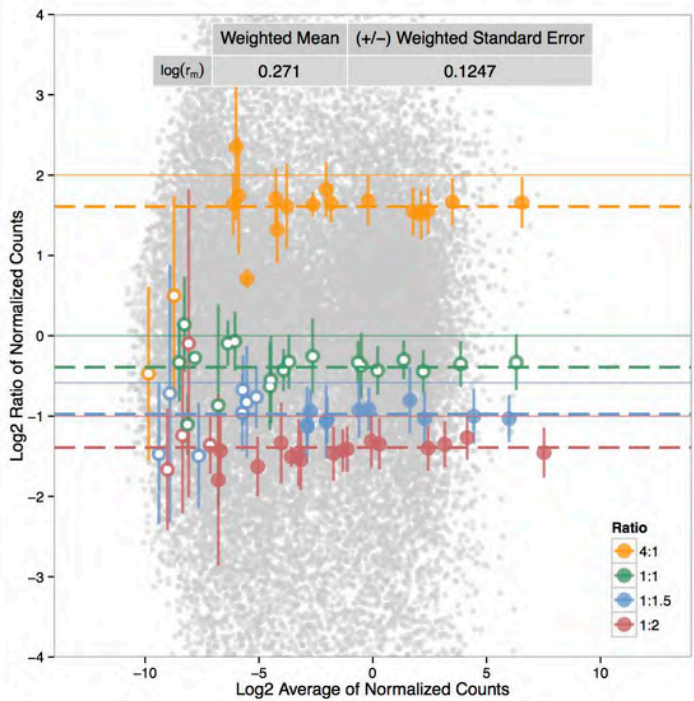
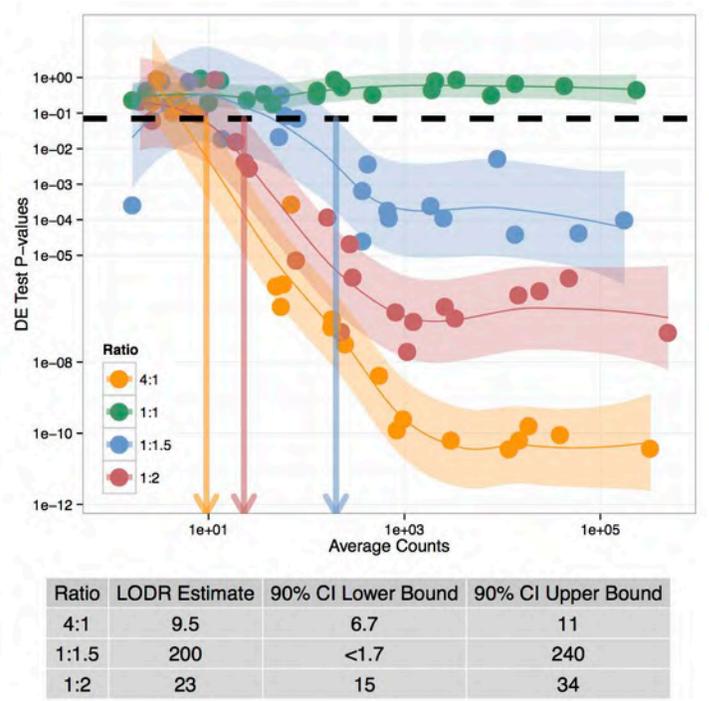

S16

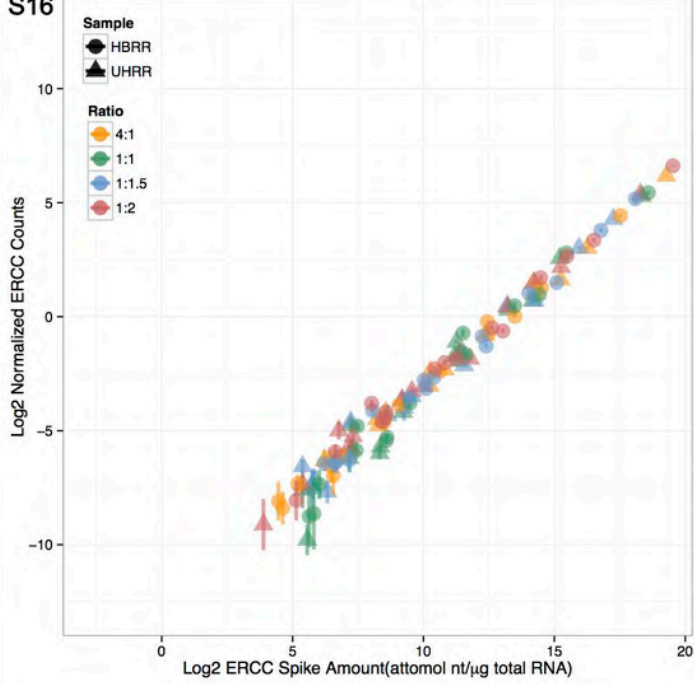
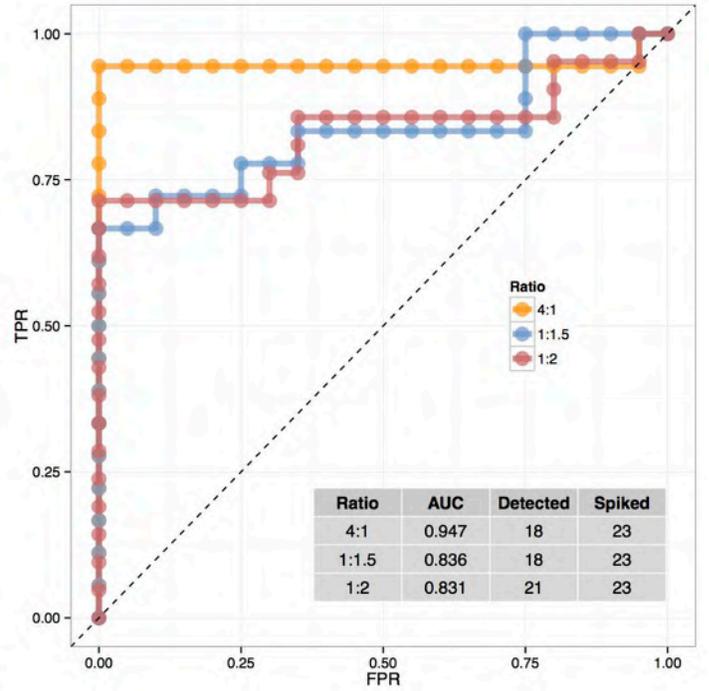
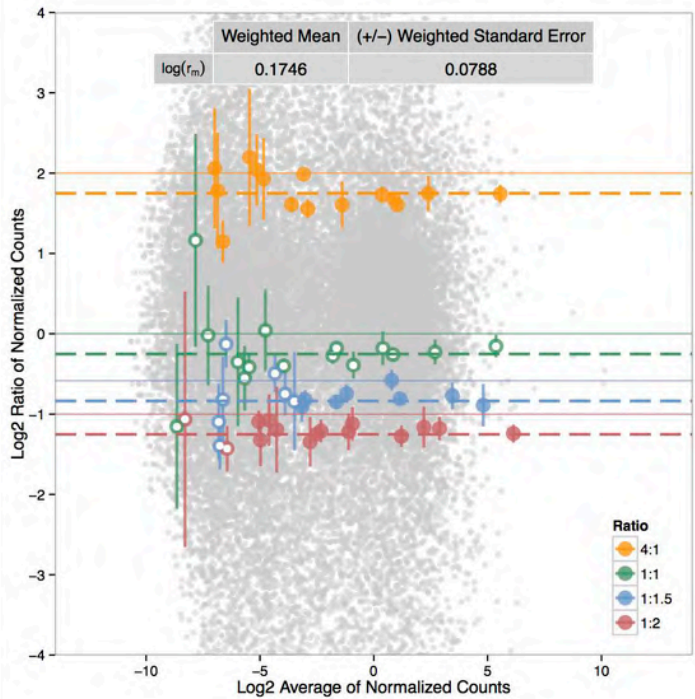
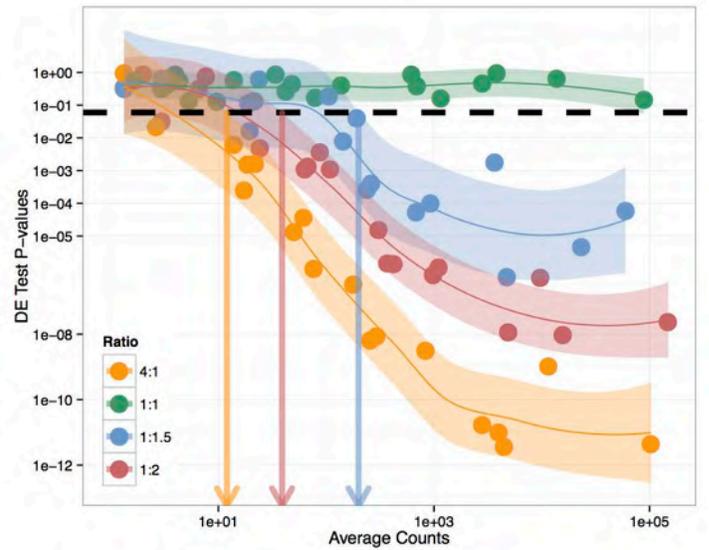

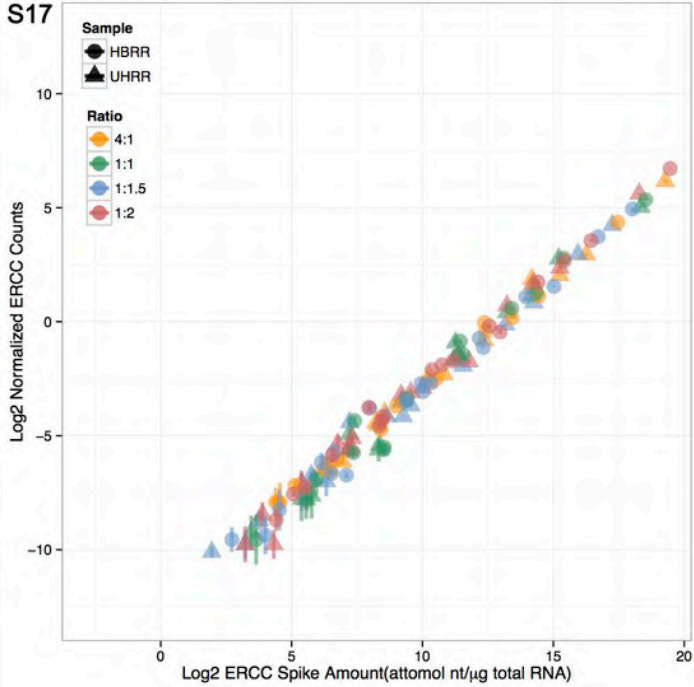
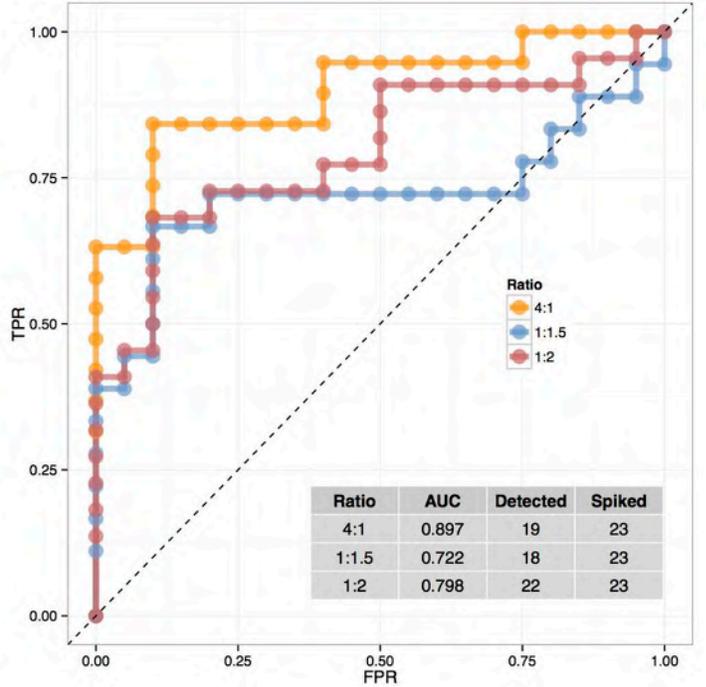
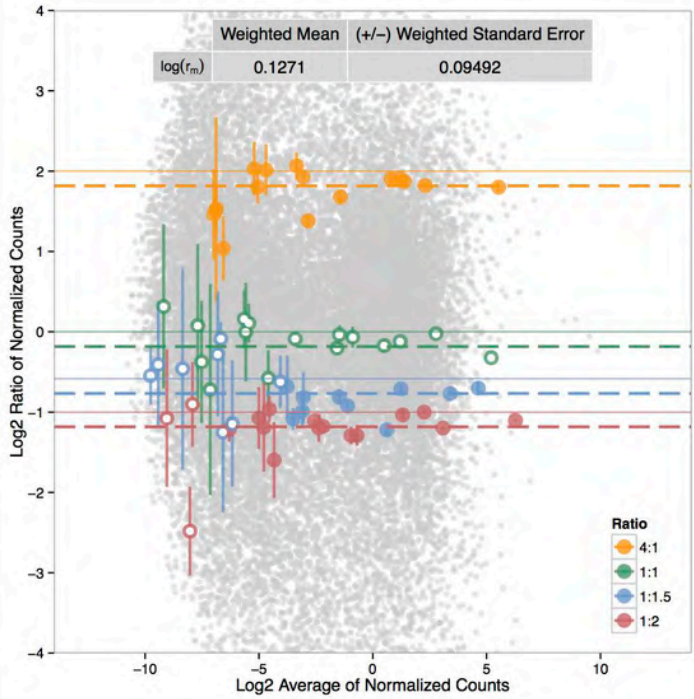
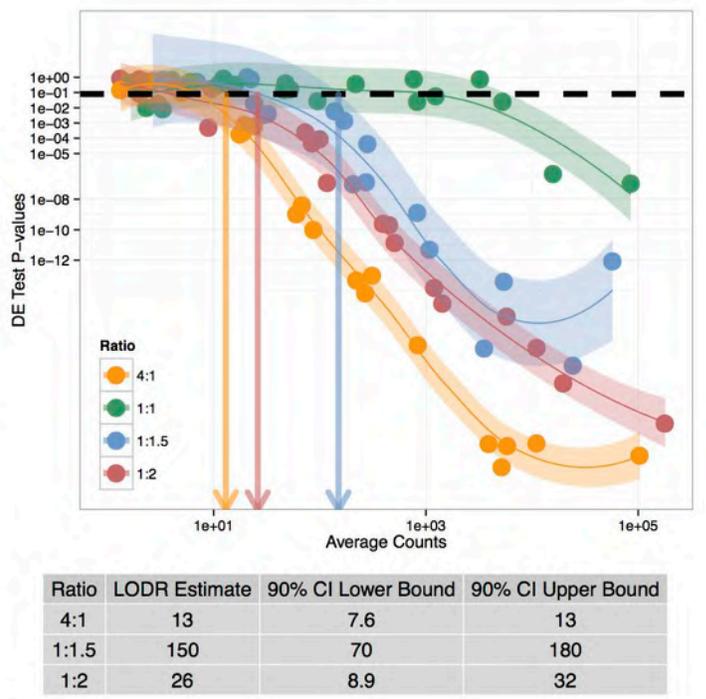

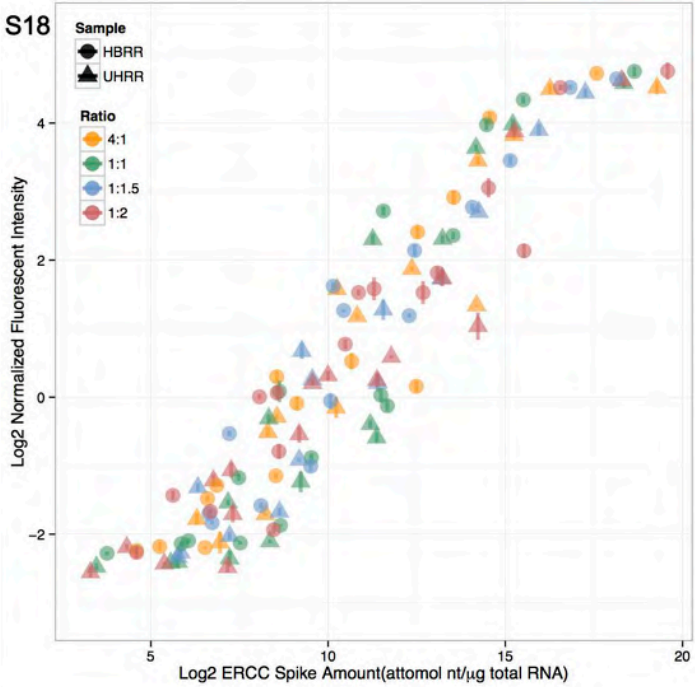
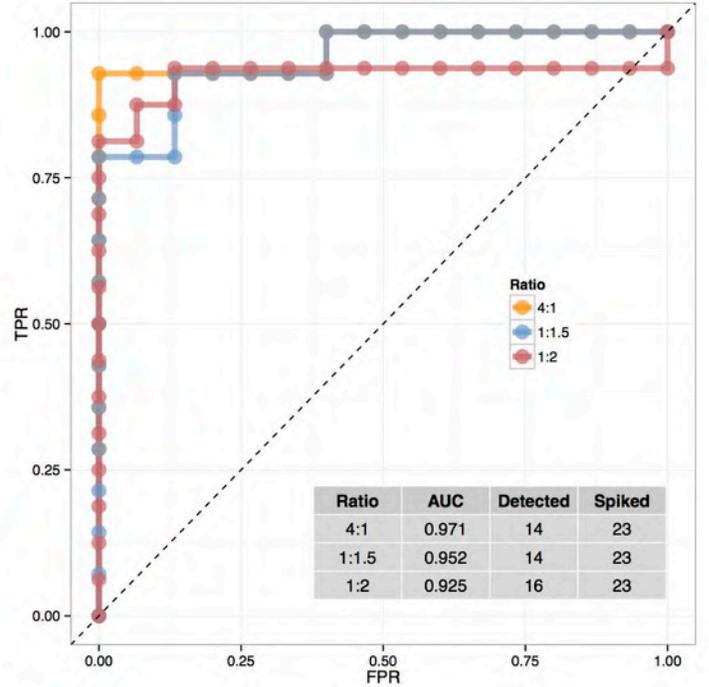
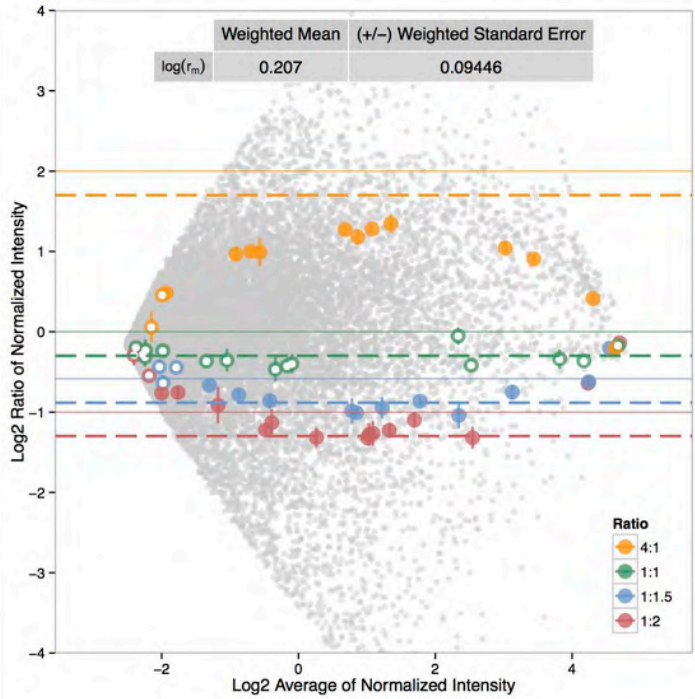
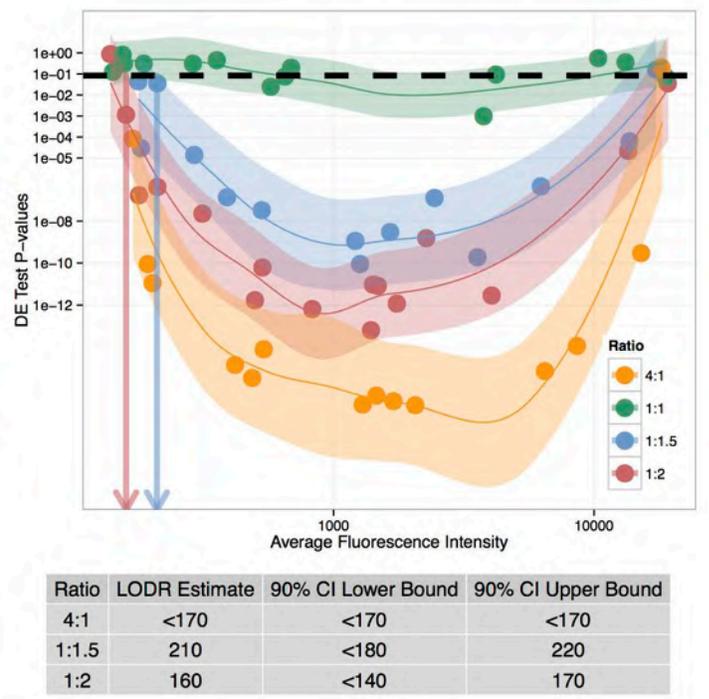

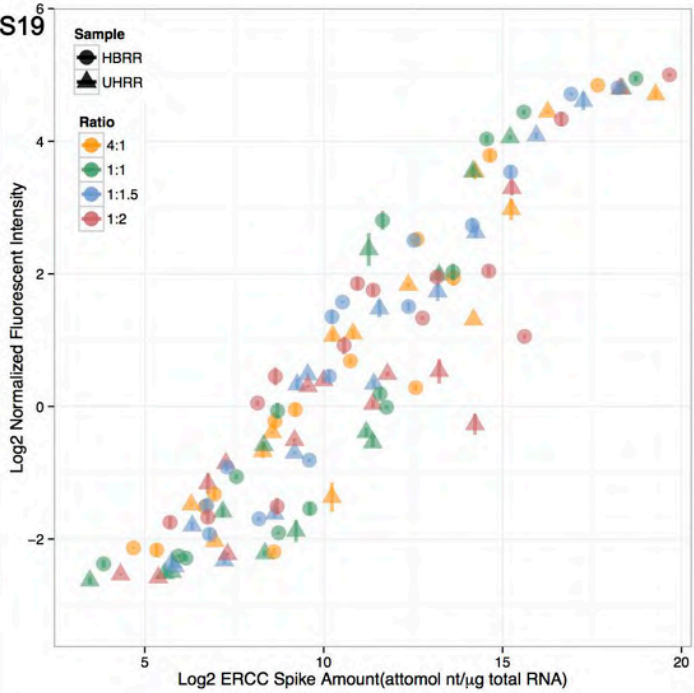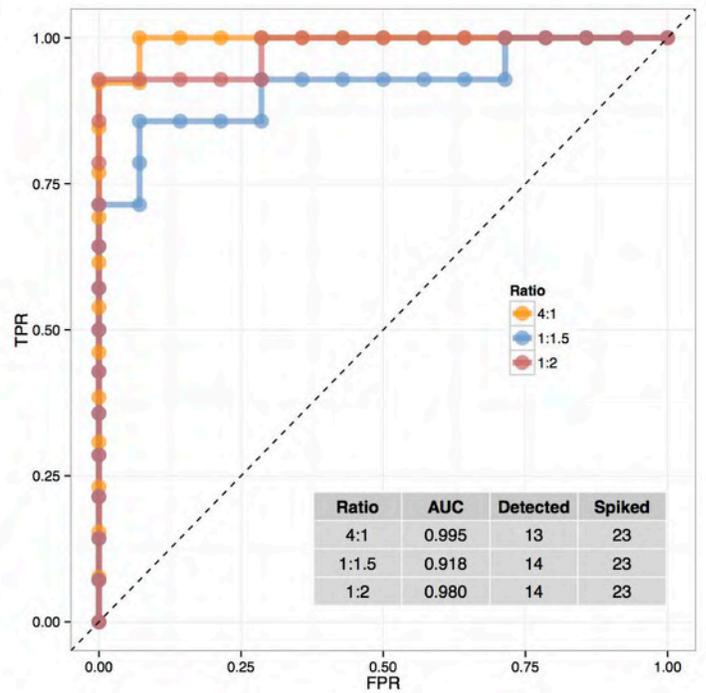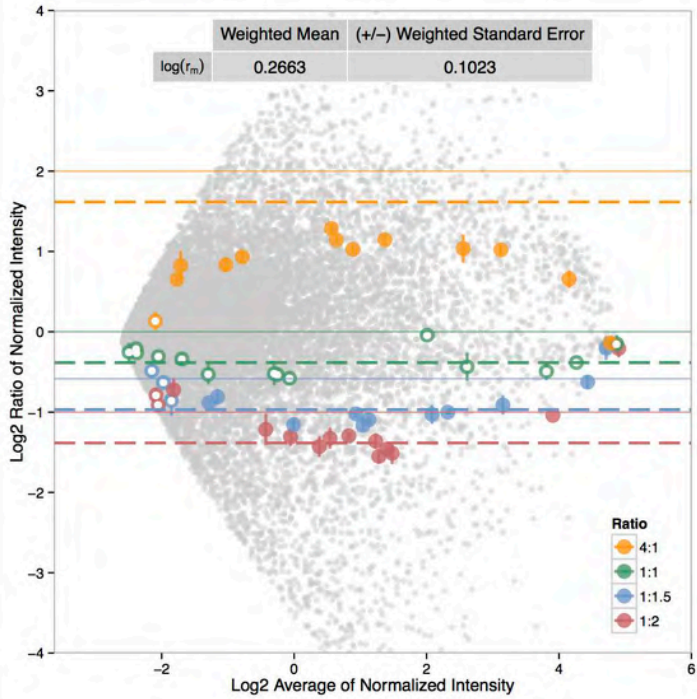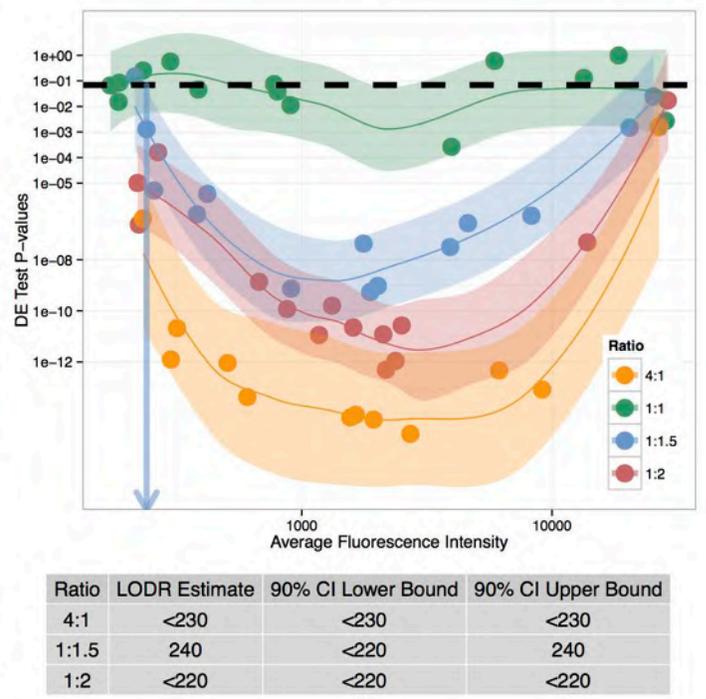

S20

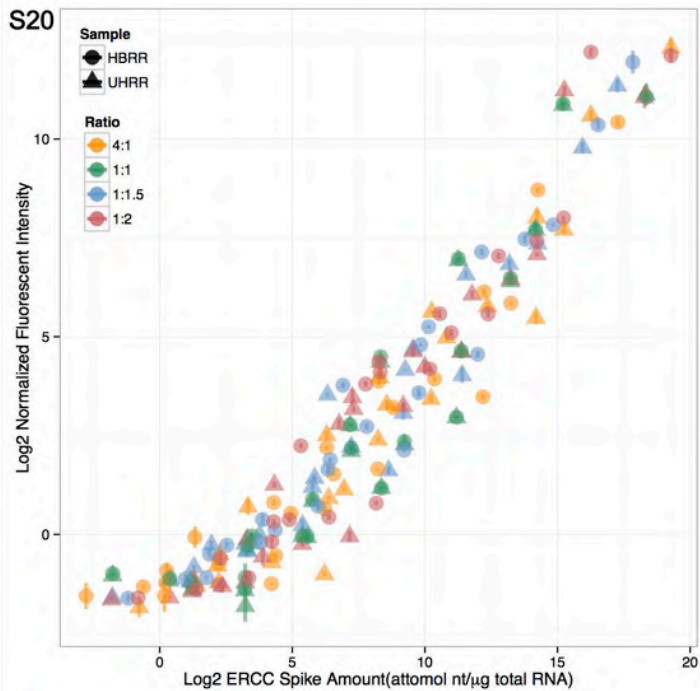
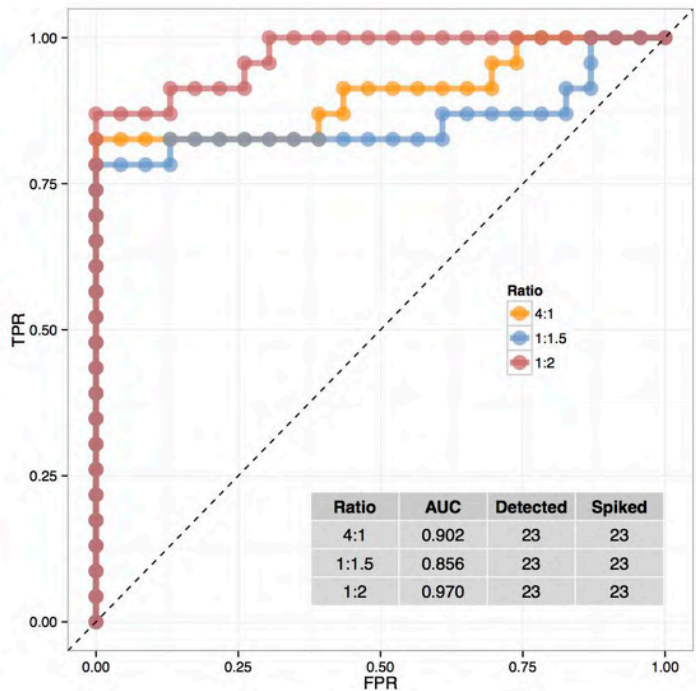
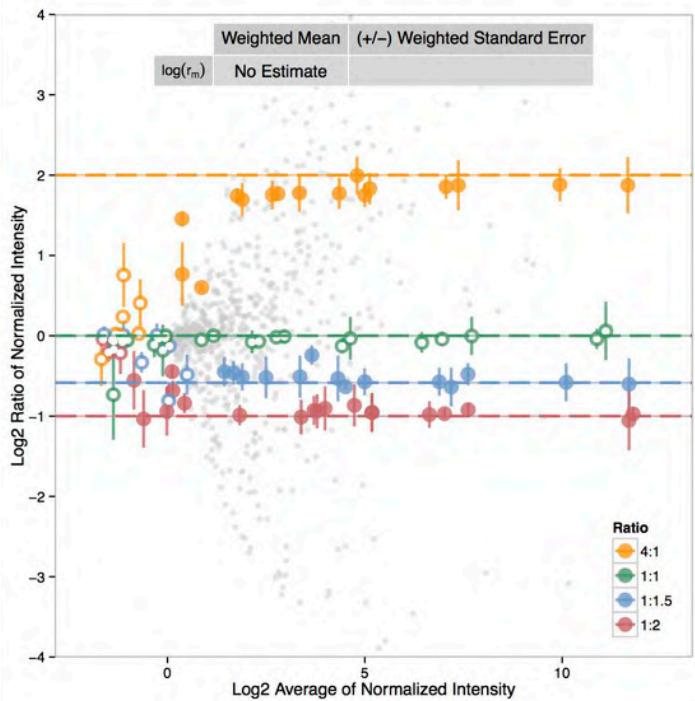
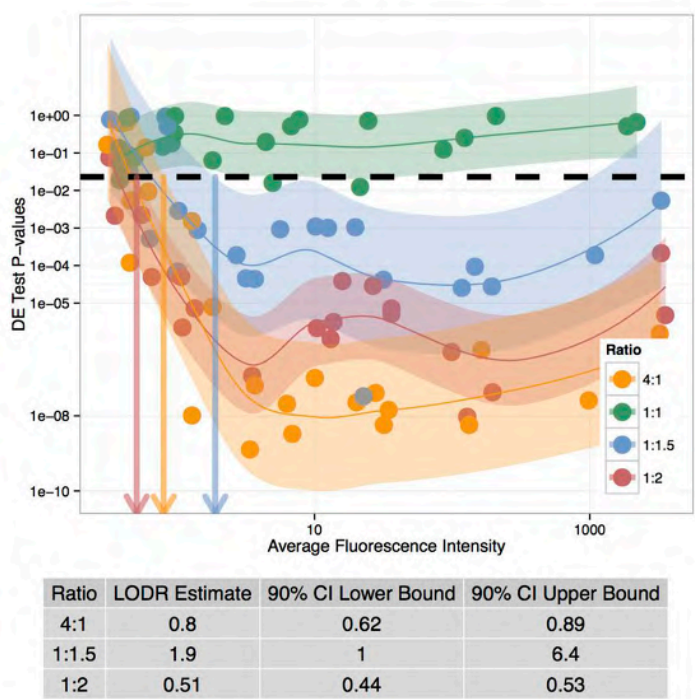

## II. ERCC specific effects across sites and platforms (Fig. S21–S22)

We looked at the deviation of ERCC controls from the expected signal-abundance relationship and found consistent ERCC-specific effects that can be attributed to polyA-selection bias. ERCC specific effects were modeled by estimating the difference between individual ERCC signal-abundance relationships and the overall model for the population of ERCC controls. These per ERCC differences are standardized by the corresponding per ERCC standard error for a unitless measure to indicate deviance from the expected signal-abundance relationship for each control. Our results show that ERCC-specific effects are consistent between laboratories for the same sample preparation protocol.

**Figure S21** ERCC-specific effects are shown for Lab 1-12. ERCC controls that are outside 1.5x the interquartile range (IQR) of the population of ERCC-specific effects at each Lab are labeled with the corresponding ERCC control code number (e.g. ERCC-00XXX).

**Figure S22** Heat map of ERCC-specific effects for ERCC controls at Lab 1–12. Only ERCC controls that were detected at all sites are included.

S21

S22

## III. Comparison of ERCC control ratio mixtures with and without poly-A selection (Fig. S23–S24)

For three of the six Illumina sequencing experiments (at Lab 2, 3, and 5) duplicate libraries were prepared with pure ERCC Mixtures (Mix 1 is sample "E" and Mix 2 is sample "F"), and barcoded and sequenced with other samples in the same experiment. One of these laboratories did perform polyA-selection on these samples (Lab 5) and the other two laboratories did not (Lab 2 and 3). ERCC controls 14, 116, and 126 had strong bias, lower signal was measured for these controls than expected in the experiment with polyA selection (Fig. S23) and this selection bias appeared to be correlated with transcript length (Fig. S24).

**Figure S23** (see next 2 pages) Signal-abundance plots and ERCC-specific effects plots of pure ERCC Mixtures (Mix 1 = E and Mix 2 = F) with duplicate library replicates at three Illumina sequencing sites At Lab 2 (a) and Lab 3 (b) poly-A selection was not performed on these sample pairs. At Lab 5 (c) poly-A selection was done on these samples. The same three ERCC controls (14, 116, and 126) are labeled in each ERCC-specific effects plot to highlight the difference between the samples for Lab 2 (d), Lab 3 (e) and Lab 5 (f).

**Figure S24** (see next 2 pages) ERCC specific effects are shown as a function of ERCC control length for the same samples shown in Fig S23, Lab 2 (a), Lab 3 (b), and Lab 5 (c). The observed polyA selection bias appears to be correlated with ERCC control length. The same three ERCC controls that were labeled in Fig. S23 are also labeled in this figure.

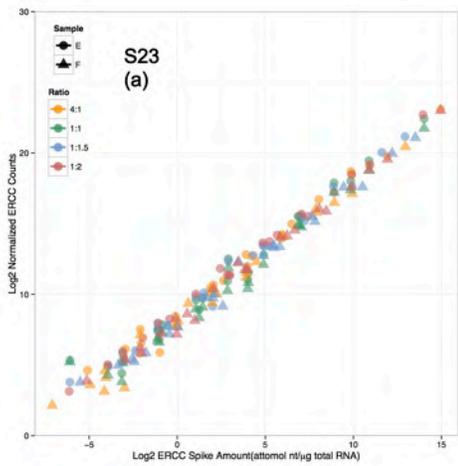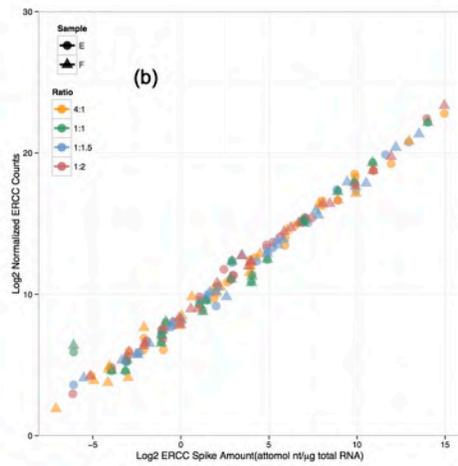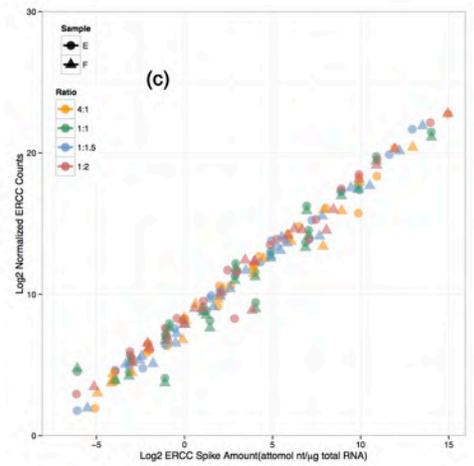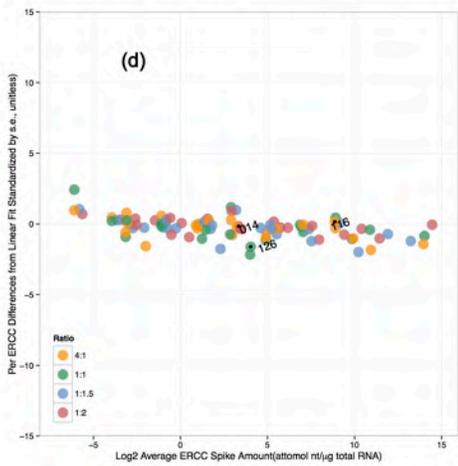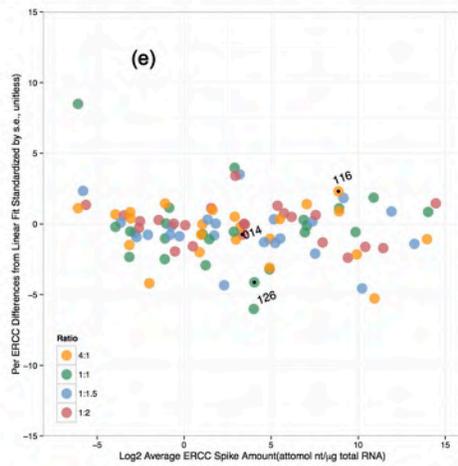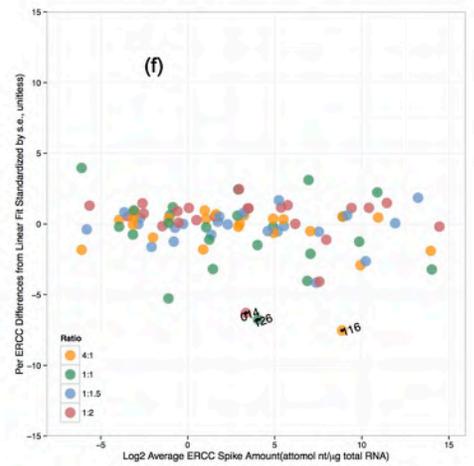

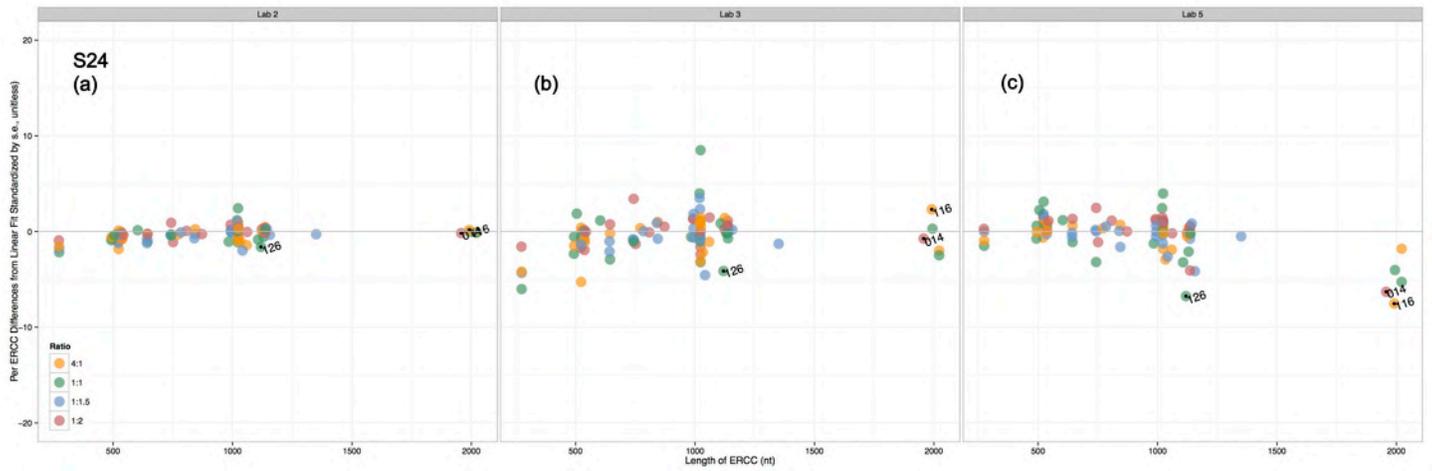

## IV. Qualitative dispersion estimate analysis to assess validity of control data (Fig. S25)

Utility of LODR estimates depends on the extent to which ERCC controls are representative of endogenous genes in a sample; before drawing strong conclusions from ERCC-derived LODR estimates, the consistency between controls and endogenous transcripts should be examined. Although control transcripts are not subject to biological variability, they are subject to the same sources of technical variation as the endogenous transcripts (e.g. library preparation, fluidics, etc.) and are also subject to additional variation from the spike-in procedure itself.

We can examine the variability of the controls and the endogenous transcripts in a sample as a function of abundance in the sequencing interlaboratory experiments (Fig. S25). Endogenous (grey) and ERCC transcript dispersion estimates (colored by ratio subpool) are shown along with a trendline fit to quasi-dispersion estimates from QuasiSeq (using edgeR dispersion trend estimates). The dispersion estimate can be used to qualitatively interpret if the variability in the control measurements is similar to variability of the endogenous transcripts. If the ERCC results are evenly spread around the trendline, then there is consistency between the dispersion of the controls and the endogenous transcripts. If a large number of controls are above the dispersion trendline then this may indicate the presences of bias, such as batch effects. Ideally the controls should be scattered around the trendline, which was the case for all experiments in the Lab 1-12 cohort except for Lab 10, which showed overdispersion. Since these samples measured at Lab 10 were shared across all of the other laboratories (one large batch was spiked prior to distribution amongst sequencing sites), the source of added variability is specific to Lab 10. This evidence of batch effects at Lab 10 is also apparent in the Lab 10 ERCC control ratio standard deviation distribution (Fig. 6d).

The qualitative assessment of ERCC control over- or under-dispersion relative to the trendline provides an indication of whether the LODR estimate using the ERCC control data is liberal or conservative. If the ERCC dispersion estimates were below the trendline, then the LODR estimates from the controls would tend to be too small and a simulation-based approach with endogenous transcripts can be implemented to obtain a more liberal LODR estimate (See supplementary material section V and Fig. S26).

**Figure S25** (next page) Endogenous (grey) and ERCC transcript dispersion estimates (colored by ratio subpool) are shown along with a trendline fit to quasi-dispersion estimates from QuasiSeq (using edgeR dispersion trend estimates).

## S25

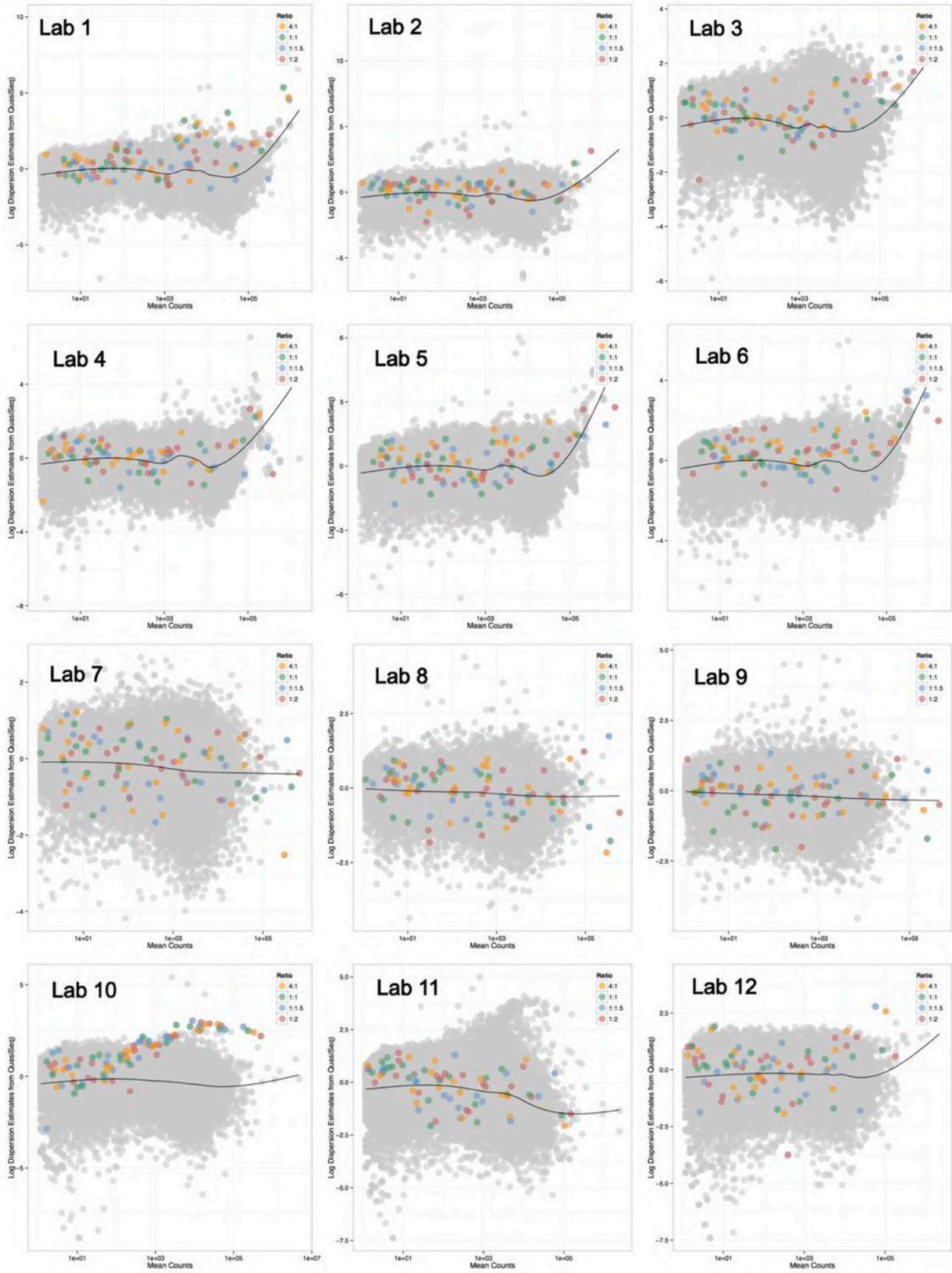

## V.    LODR Estimation from Endogenous Transcript Simulation

Both edgeR and DESeq estimate a trend describing the relationship between abundance and dispersion. This trend can be used to estimate LODR via simulation from a negative binomial model. A description of one possible approach follows.

We simulated negative binomial data using sample averages and dispersion estimates from endogenous genes. We now provide a detailed description of these simulations. Let *i* provide the index of an endogenous gene upon which simulated are to be guided. Let $y_{ijk}$ denote the observed count for gene *i* in replicate *k* of treatment *j* Let $\bar{y}_{i..} = \sum_{j,k} y_{ijk}/n$ denote the average observed count for gene *i*, where *n* is the total number of samples. Let $\hat{c}_{jk}$ denote the estimated library size factor for replicate *k* of treatment *j*. and let $\bar{y}'_{i..} = \sum_{j,k}(y_{ijk}/\hat{c}_{jk})/n$ denote the average normalized count for gene *i*. Let $\hat{\omega}_i$ denote the estimated dispersion for gene *i*, obtained from the fitted trend in edgeR).

For a given fold change, *fold*, and gene index *i*, we set $\mu'_{i1.} = 2\bar{y}'_{i..}/(1 + fold)$ and $\mu'_{i2.} = fold\ \mu'_{i1.}$, and final means used in simulation were given by $\mu_{ijk} = \mu'_{ij}.\hat{c}_{jk}$. Count data were then simulated from a negative binomial distribution such that $E(Y^{sim}_{ijk}) = \mu_{ijk}$ and $Var(Y^{sim}_{ijk}) = \mu_{ijk} + \hat{\omega}_i \mu^2_{ijk}$.

For each of the four nominal fold-changes present in the ERCCs (.5, .667, 1, and 4), we chose to simulate data for roughly every 800[th] gene (49 genes in total), when sorted by total count, to get a representative sample across the dynamic range of the endogenous genes. Simulated data was analyzed for differential expression in the same manner as the endogenous genes and ERCCs. The resulting p-values, average counts and known fold changes for the simulated data were then used to conduct an LODR analysis. Results from this analysis are shown below.

Figure S26 displays results from the LODR analysis applied to simulated data based on endogenous genes from Lab 5 of the interlaboratory study. Differential expression p-values are plotted as a function of mean counts ($\bar{y}$) for each simulated gene. The fitted trends and 90% prediction interval bounds resulting from locfit are plotted. Confidence intervals for each LODR estimate were obtained via bootstrapping (residuals from the corresponding locfit curve were repeatedly resampled to estimate LODR). These results are consistent with the ERCC LODR estimates for this experiment shown in Fig. 4b (and also Fig. S15)

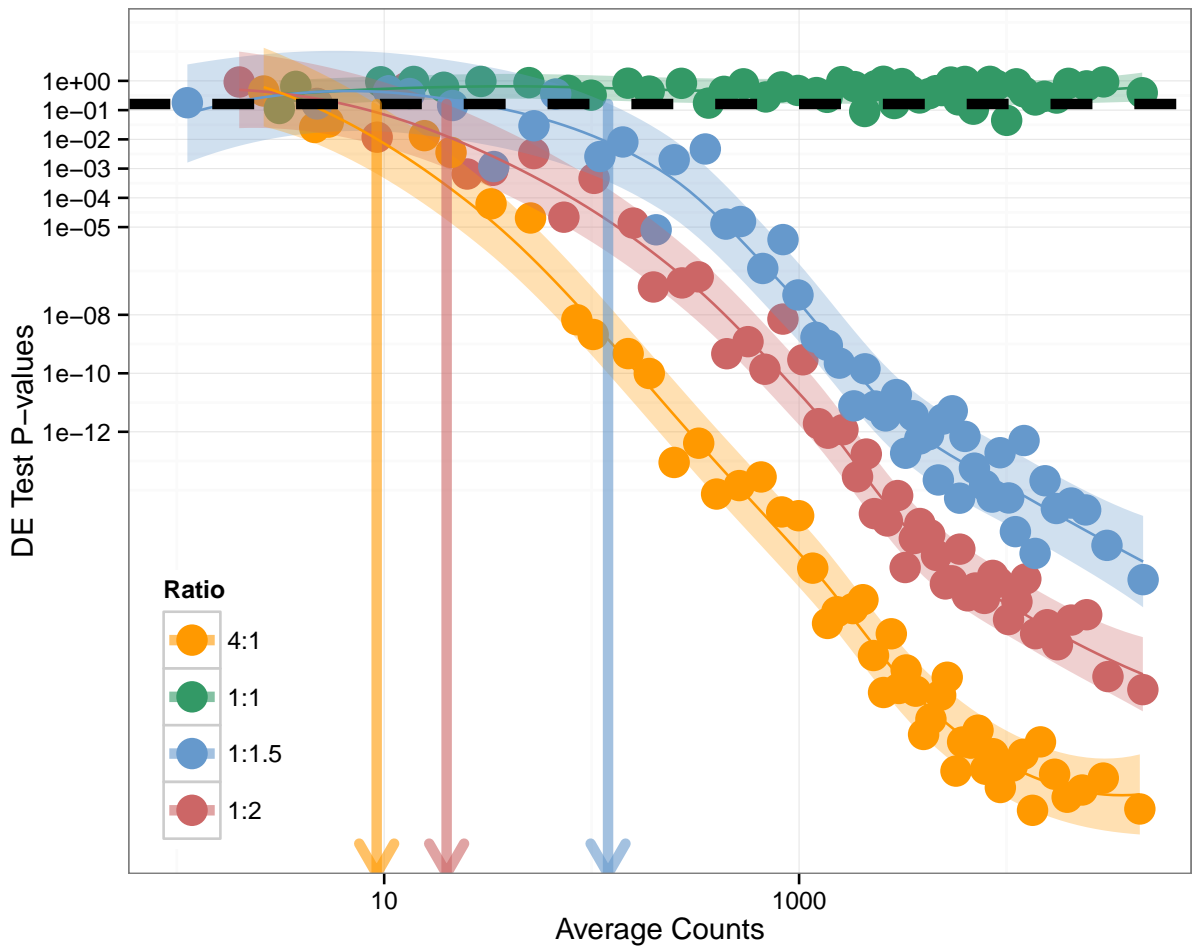

| Ratio | LODR Estimate | 90% CI Lower Bound | 90% CI Upper Bound |
|---|---|---|---|
| 4:1 | 9.2 | 6.3 | 10 |
| 1:1.5 | 120 | <1.1 | 150 |
| 1:2 | 20 | 15 | 27 |

**Figure S26.** P-values and modeled LODR as a function of abundance from simulated expression data produced from Lab 5 endogenous transcript data. The LODR results for this simulation approach are shown in the annotation table with 90% Confidence Intervals (CI)

## VI.  Figures from erccdashboard for Subread and featureCounts analysis of Lab 7-9 data (Fig. S27-S29)

These are erccdashboard analysis results for the same read data (fastq files) used in the LifeScope analysis results shown in Fig. S12-S14.

## VII.  Mapped read QC Metrics (Fig. S30–S33)

Mapped Read QC metrics from RNASeQC are shown on the next 3 pages for Lab 7-9. There were high duplication rates in the Lab 7 libraries (Fig S24) and also indication of 3' coverage bias in these samples (Fig. S30-32).

**Figure S30** The rate of duplication per library is shown for Labs 7-9 (colored by laboratory). In this faceted plot rows columns show results for different library preparations.

**Figure S31** The % of transcript coverage of 1000 medium expressed transcripts over normalized gene length is shown for Labs 7-9 (colored by laboratory). In this faceted plot rows columns show results for different library preparations.

**Figure S32** The % of transcript coverage of 1000 highest expressed transcripts over normalized gene length is shown for Labs 7-9 (colored by laboratory). In this faceted plot rows columns show results for different library preparations.

**Figure S33** Percent of rRNA reads of total reads for 18S, 28S, and 5.8S rRNA for Lab 7-9 libraries 1-4 (colored by laboratory).

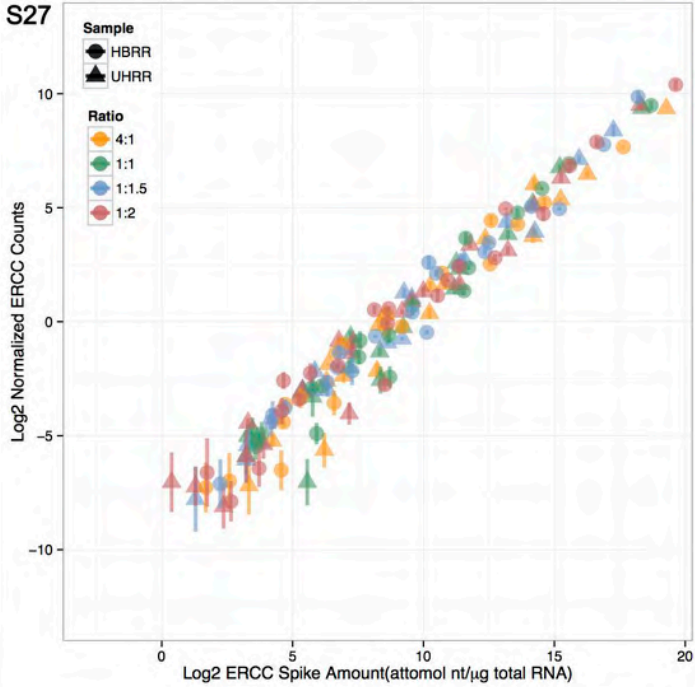
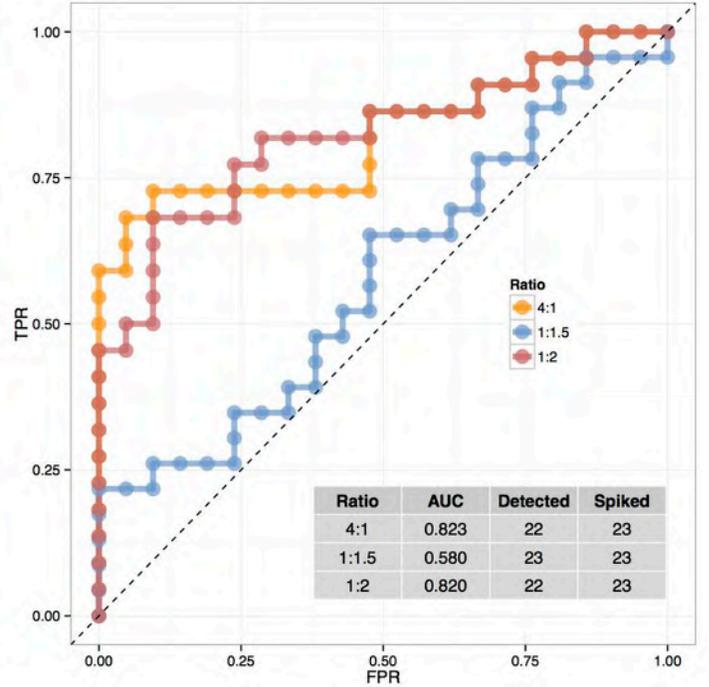
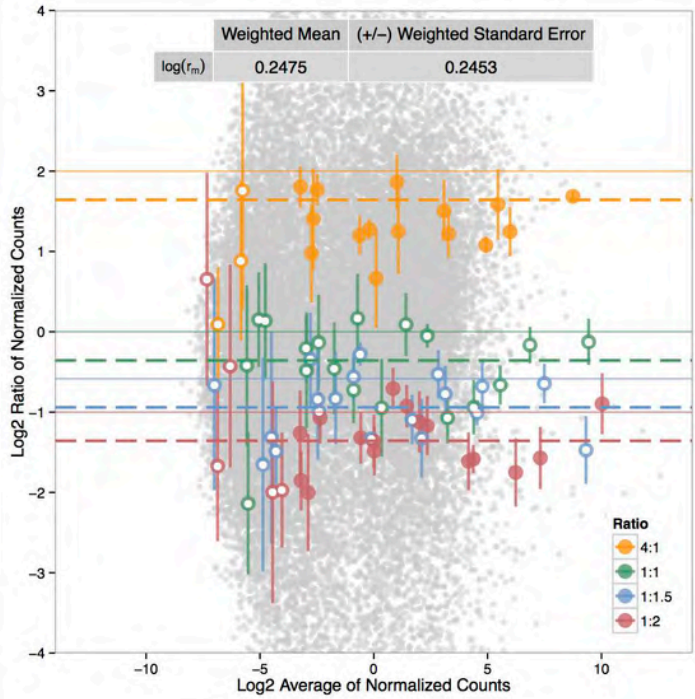
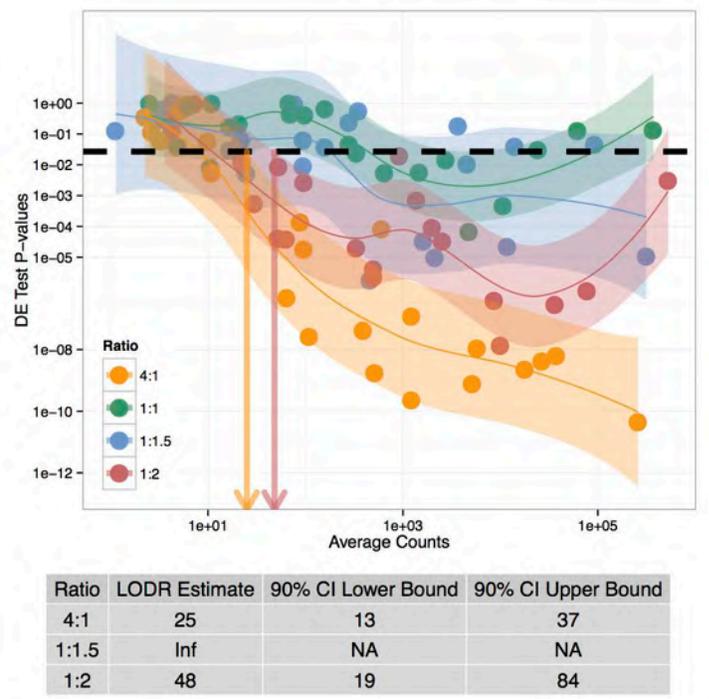

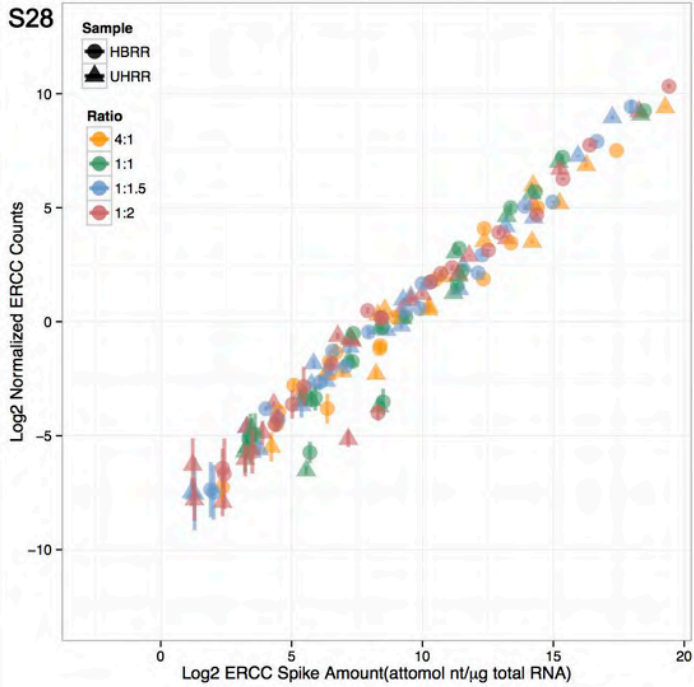
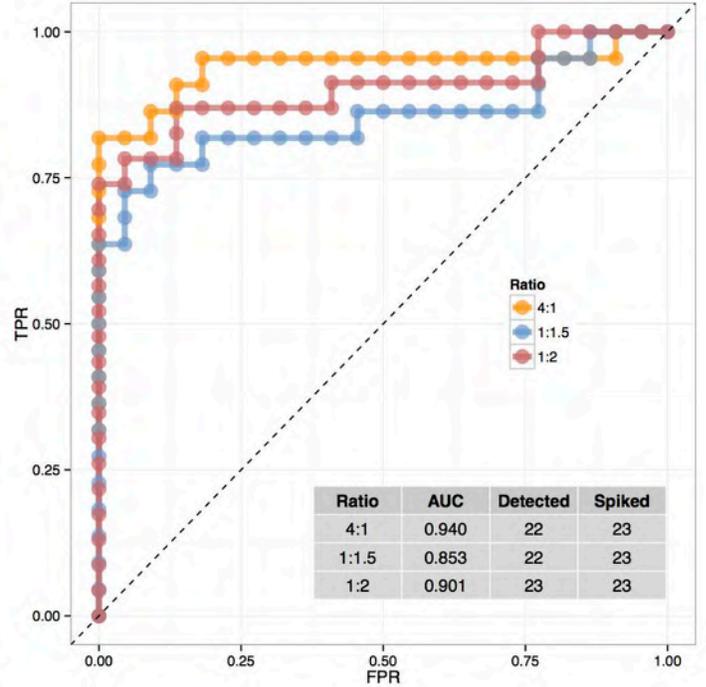
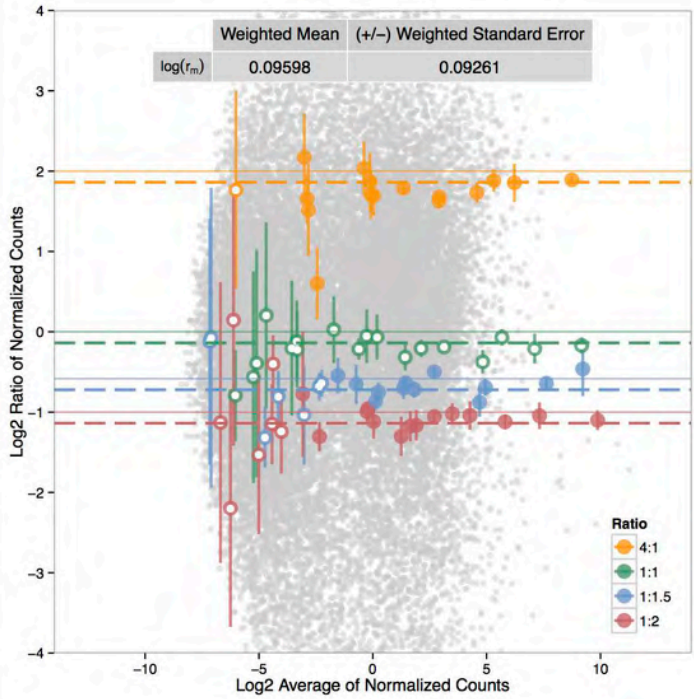
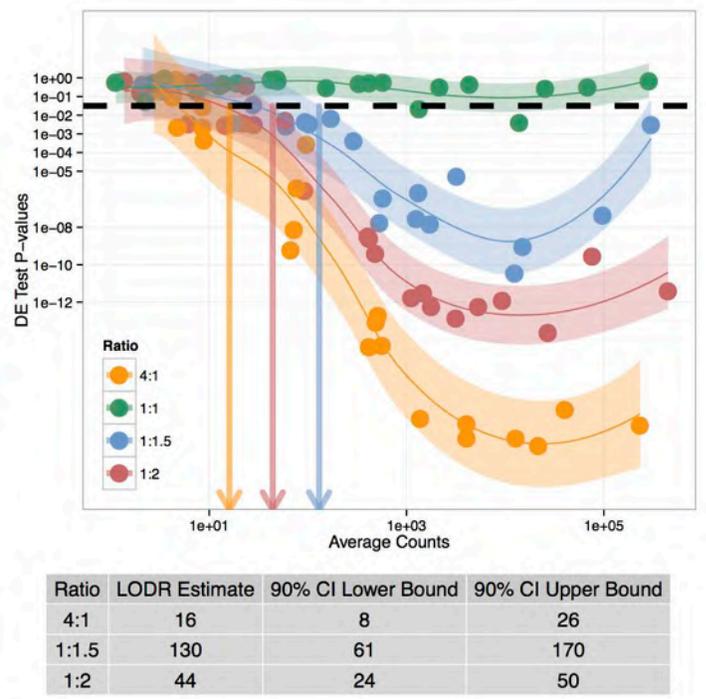

S29

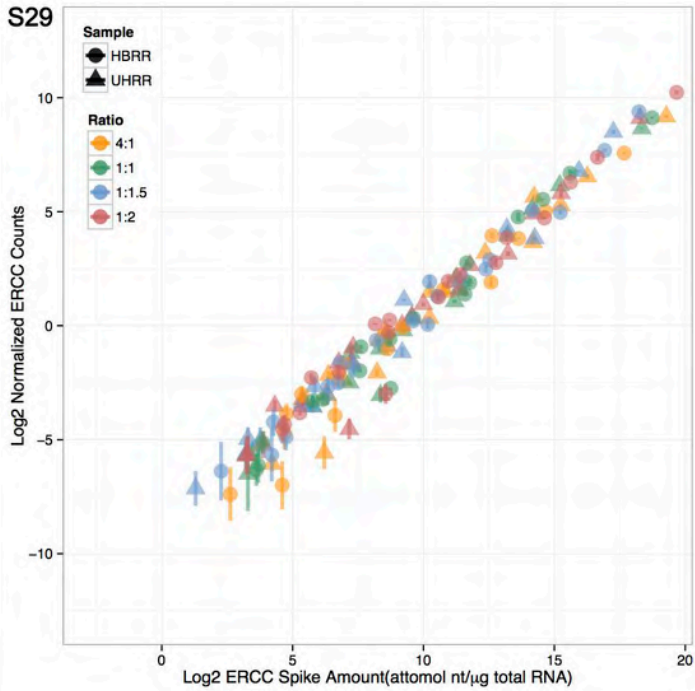
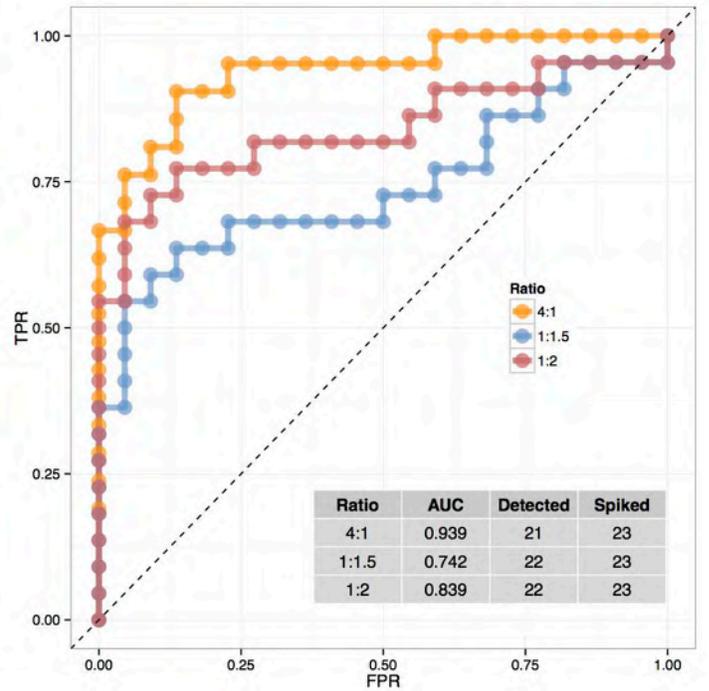
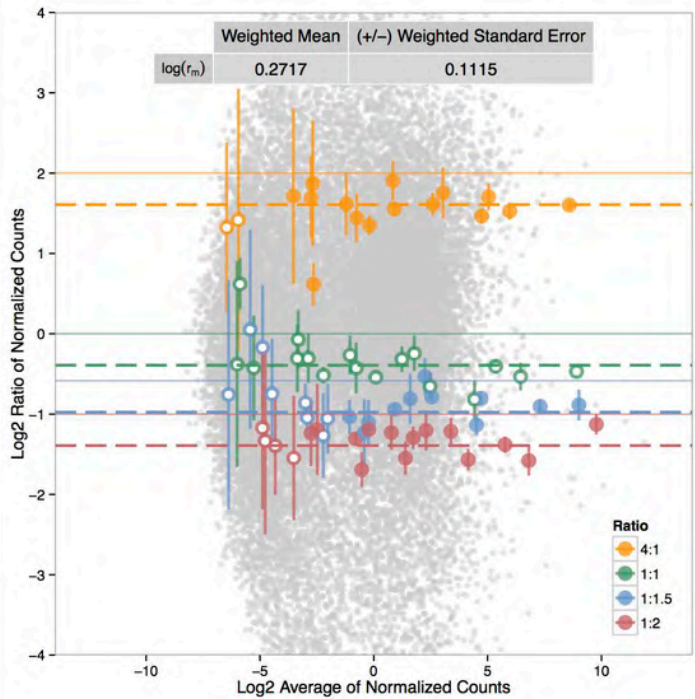
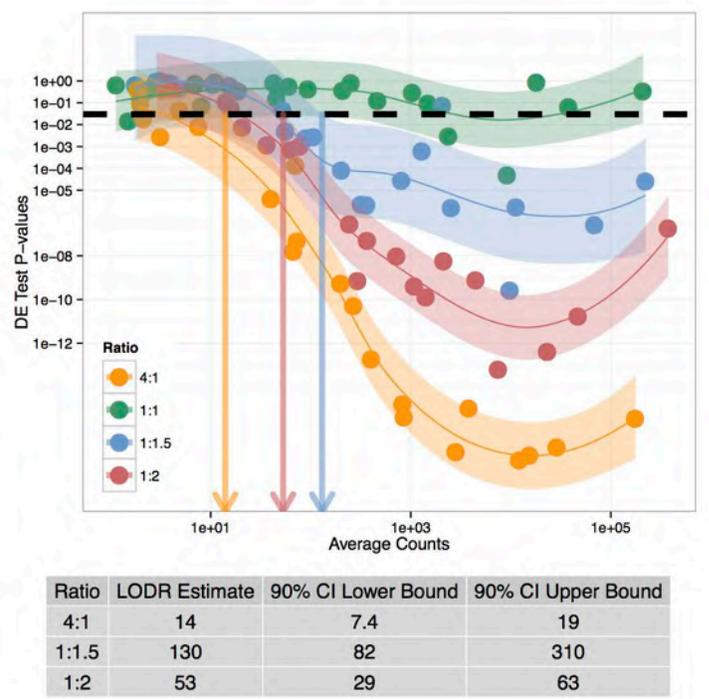

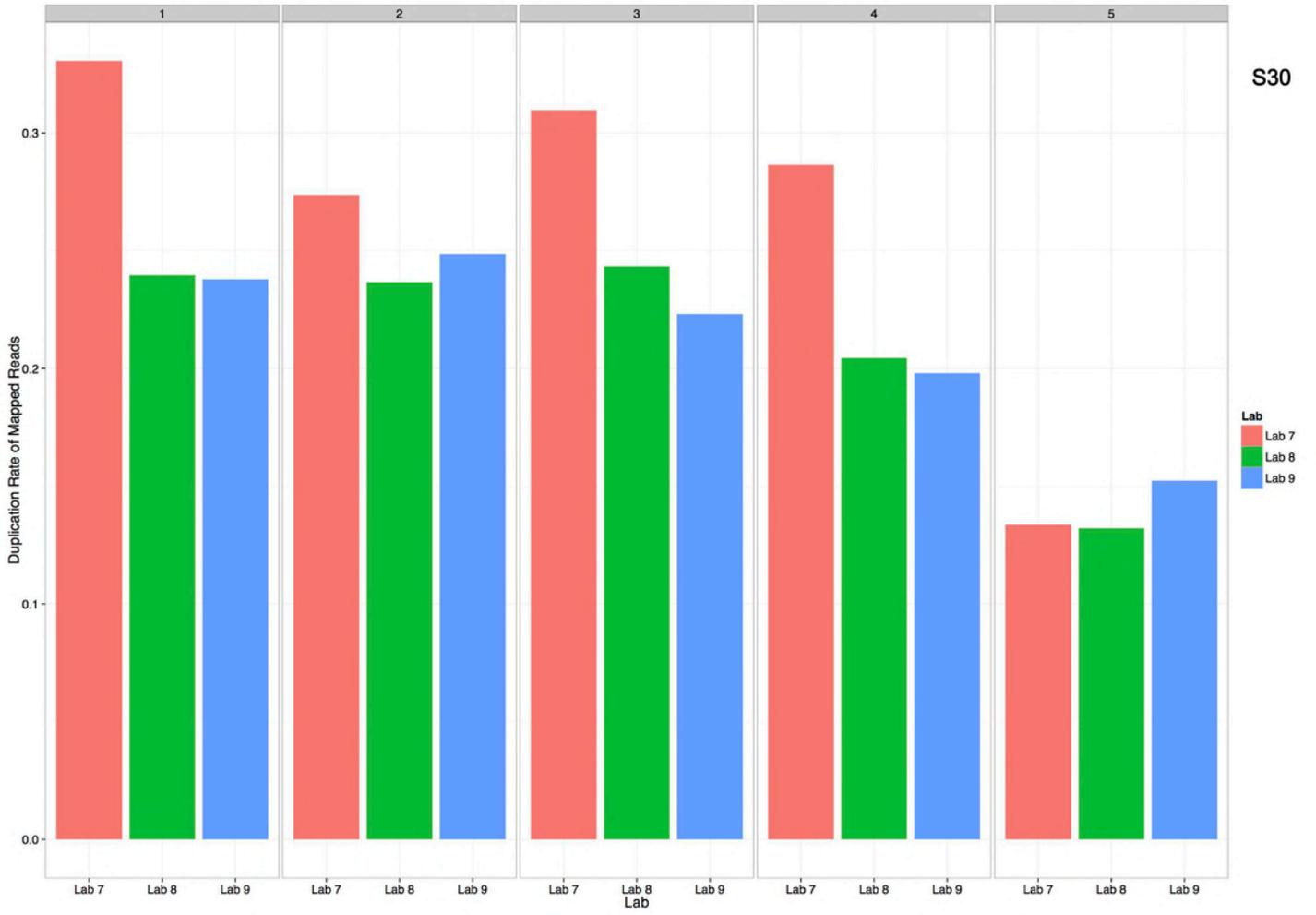

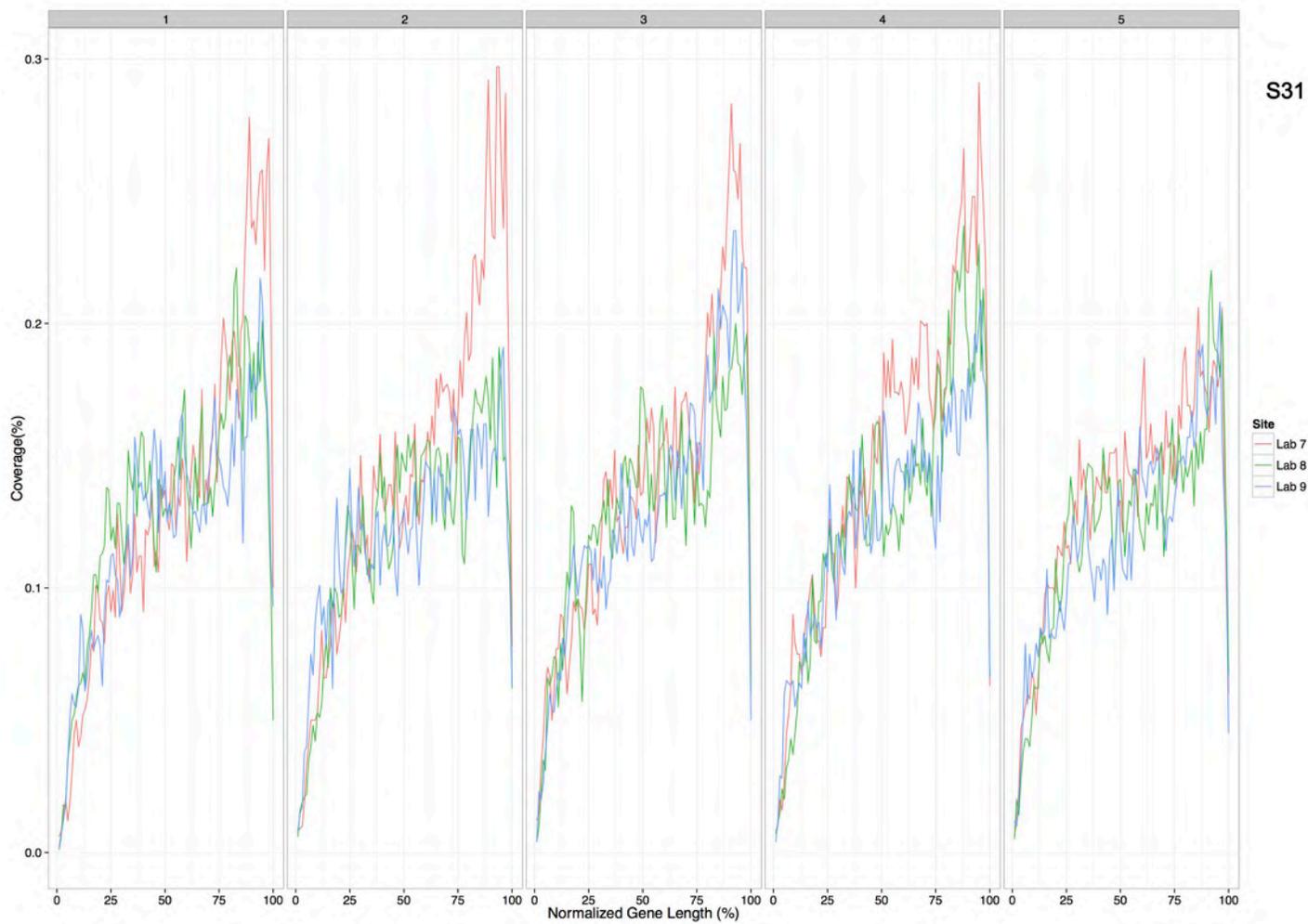

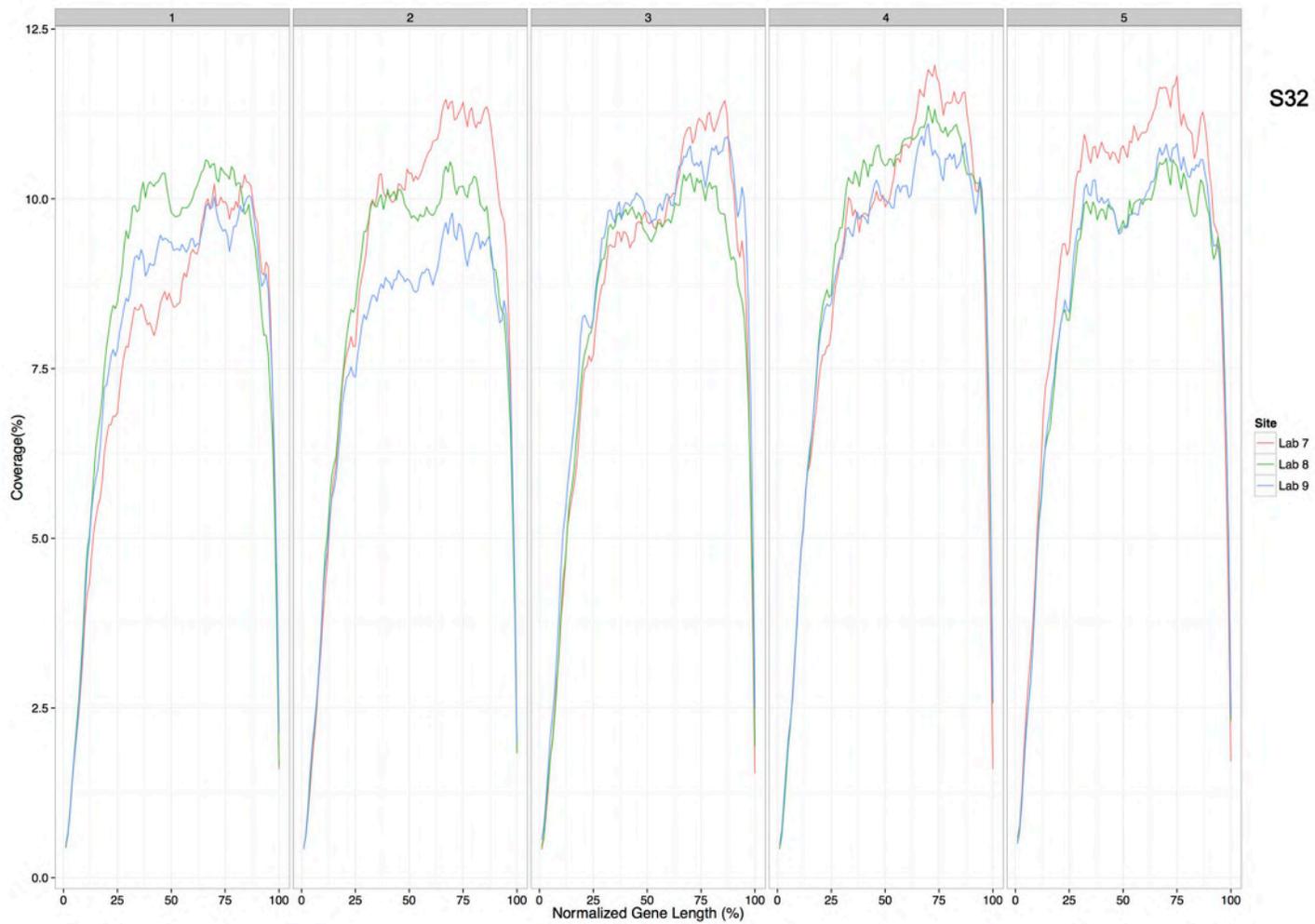

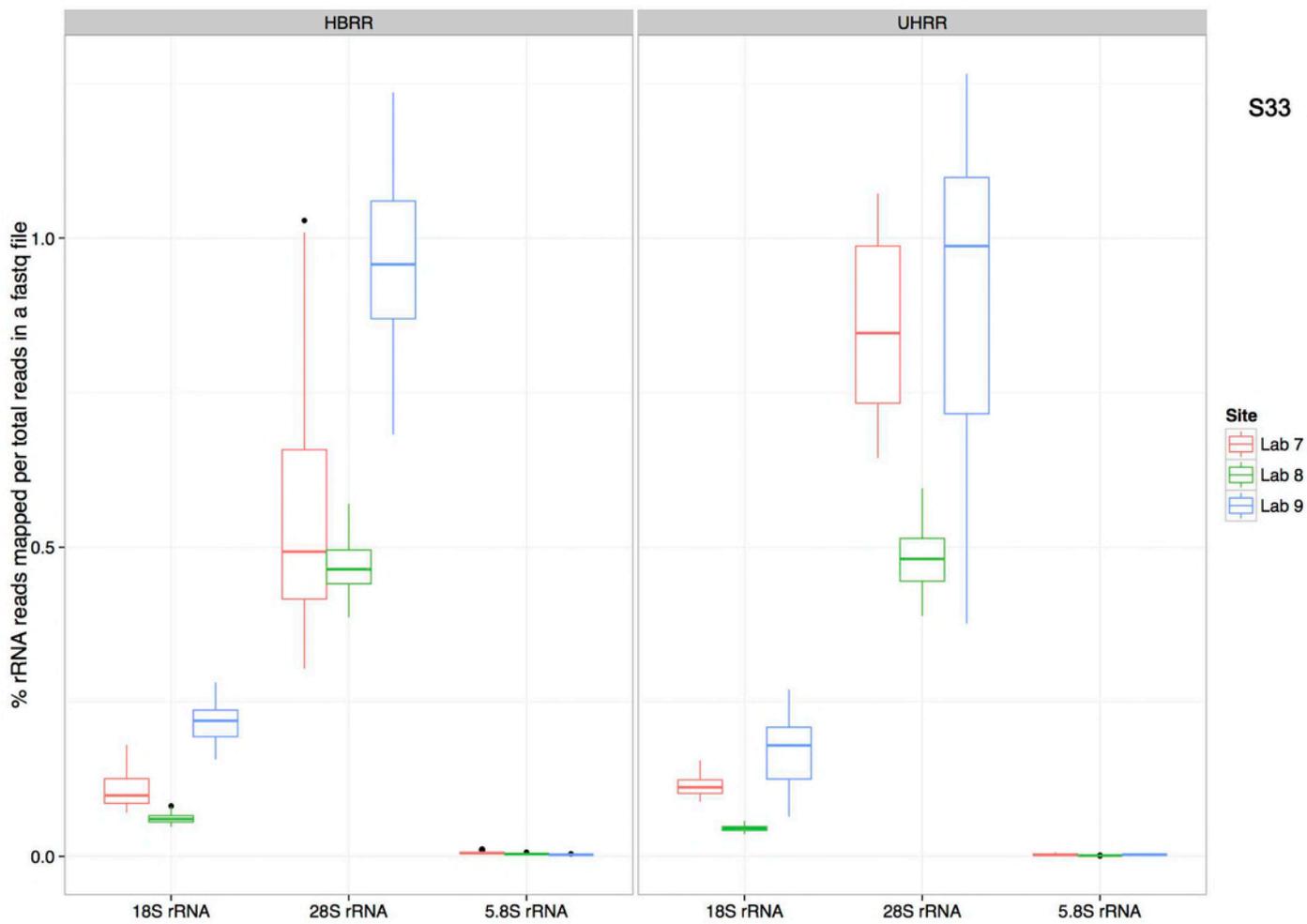